\documentclass[journal=jcisd8,manuscript=article]{achemso}

\usepackage[version=3]{mhchem}
\usepackage{amsmath}
\usepackage{amssymb}
\usepackage{array}
\usepackage{tabularx}
\usepackage{mathtools}
\usepackage[ruled]{algorithm2e}
\usepackage{setspace}
\usepackage{xcolor} 
\usepackage{hyperref}
\usepackage{listings}
\usepackage{multirow}
\usepackage{longtable}
\usepackage{booktabs}
\usepackage{comment}
\usepackage{placeins}

\hypersetup{
    urlcolor=blue,
    colorlinks=true,
    citecolor=black,
    linkcolor=black
}

\newcommand{\argmax}{\mathop{\operatorname{argmax}}\limits}
\newcommand{\defeq}{\coloneqq}

\let\tt\texttt
\let\bf\textbf

\author{David E. Graff}
\affiliation{Department of Chemistry and Chemical Biology, Harvard University, Cambridge, MA}
\author{Eugene I. Shakhnovich}
\affiliation{Department of Chemistry and Chemical Biology, Harvard University, Cambridge, MA}
\author{Connor W. Coley}
\email{ccoley@mit.edu}
\affiliation{Department of Chemical Engineering, MIT, Cambridge, MA}

\title[MolPAL]{Accelerating High-Throughput Virtual Screening Through Molecular Pool-Based Active Learning}

\abbreviations{}
\keywords{}

\begin{document}

\begin{tocentry}
\includegraphics[height=3.5cm]{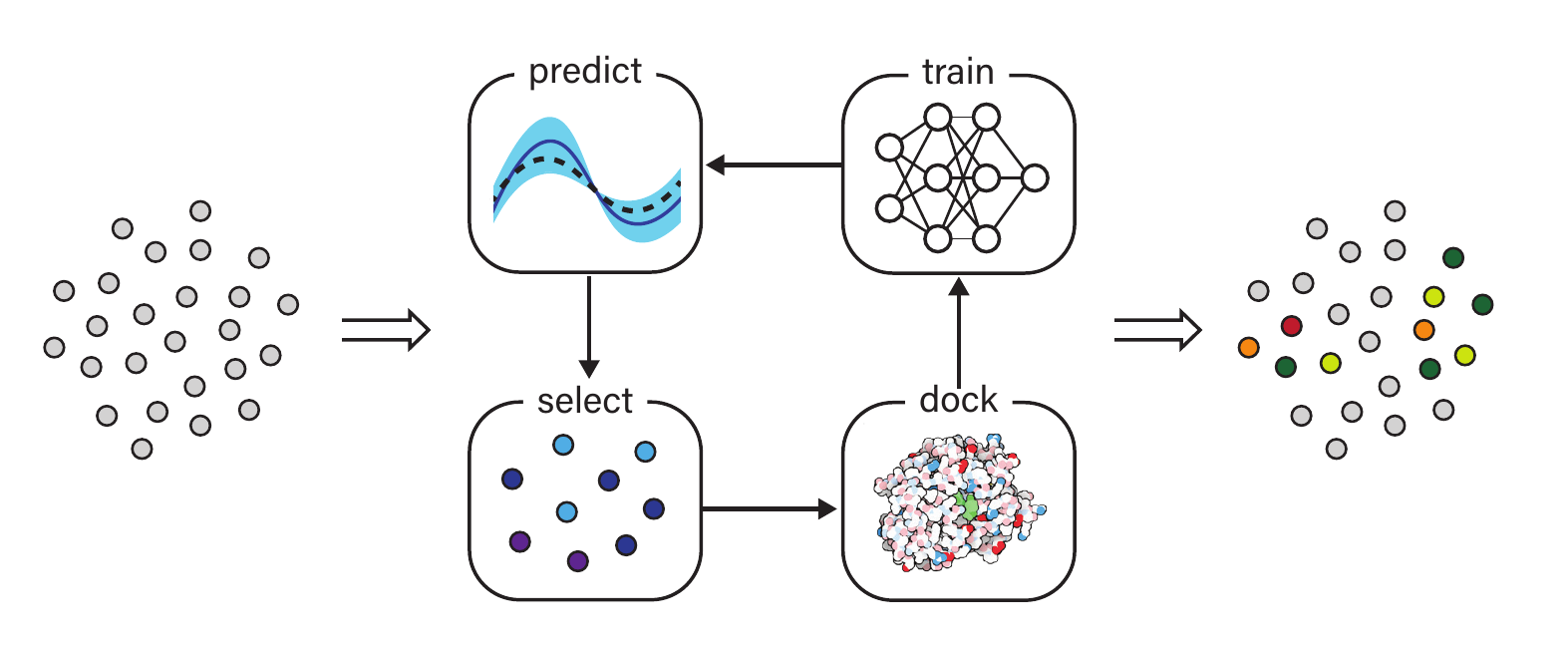}
\end{tocentry}

\begin{abstract}
    Structure-based virtual screening is an important tool in early stage drug discovery that scores the interactions between a target protein and candidate ligands. As virtual libraries continue to grow (in excess of $10^8$ molecules), so too do the resources necessary to conduct exhaustive virtual screening campaigns on these libraries. However, Bayesian optimization techniques can aid in their exploration: a surrogate structure-property relationship model trained on the predicted affinities of a subset of the library can be applied to the remaining library members, allowing the least promising compounds to be excluded from evaluation. In this study, we assess various surrogate model architectures, acquisition functions, and acquisition batch sizes as applied to several protein-ligand docking datasets and observe significant reductions in computational costs, even when using a greedy acquisition strategy; for example,  87.9\% of the top-50000 ligands can be found after testing only 2.4\% of a 100M member library. Such model-guided searches mitigate the increasing computational costs of screening increasingly large virtual libraries and can accelerate high-throughput virtual screening campaigns with applications beyond docking.
\end{abstract}

\section{Introduction}
Computer-aided drug design techniques are widely used in early stage discovery to identify small molecule ligands with affinity to a protein of interest\cite{yu_computer-aided_2017,macalino_role_2015}. Broadly speaking, these techniques fall into one of two domains: ligand-based or structure-based. Ligand-based techniques often rely on either a quantitative structure-activity relationship (QSAR) or similarity model to screen possible ligands. Both techniques require one or more previously labeled active/inactive compounds that are typically acquired through physical experimentation. In contrast to ligand-based techniques, structure-based techniques, such as computational docking and molecular dynamics, try to simulate the physical process of a ligand binding to a protein active site and assign a quantitative score intended to correlate with the free energy of binding\cite{li_overview_2019}. These techniques require a three-dimensional structure of the target protein but do not require target-specific bioactivity data. Scoring functions used in structure-based techniques are typically parameterized functions describing the intra- and intermolecular interactions at play in protein-ligand binding\cite{li_overview_2019}. As a result, structure-based methods are in principle more able to generalize to unseen protein and ligand structures compared to ligand-based methods. This advantage has been leveraged to discover novel ligand scaffolds in a number of recent virtual screening campaigns\cite{irwin_docking_2016}.

A typical virtual screening workflow will exhaustively predict the performance of ligands in a virtual chemical library. However, over the past decade, these libraries have grown so large that the computational cost of screening cannot be ignored. For example, ZINC, a popular database of commercially available compounds for virtual screening, grew from 700k to 120M structures between 2005 and 2015 and, at the time of writing, now contains roughly 1 billion molecules\cite{irwin_zinc_2005, sterling_zinc_2015}. ZINC is not unique in its gargantuan size; other enumerated virtual libraries exist that number well over one billion compounds\cite{noauthor_real_nodate}. Non-enumerated libraries contain an implicit definition of accessible molecules and can be much larger, containing anywhere from $10^{10}$ to $10^{20}$ possible compounds\cite{knehans_merck_2017, nicolaou_proximal_2016, hu_pfizer_2012}. Despite some debate around whether ``bigger is better'' in virtual screening\cite{clark_virtual_2020}, such large virtual libraries are now being used for screening in structure-based drug design workflows\cite{gorgulla_open-source_2020, lyu_ultra-large_2019, acharya_supercomputer-based_2020, mark_mcgann_gigadocking_2019}. These studies required computational resources that are inaccessible to many academic researchers (e.g., 475 CPU-years in the case of \citeauthor{gorgulla_open-source_2020}\cite{gorgulla_open-source_2020}). Moreover, this high computational cost makes such a strategy impractical to apply to many distinct protein targets. As virtual libraries grow ever larger, new strategies must be developed to mitigate the computational cost of these exhaustive screening campaigns.

\begin{figure}
    \includegraphics[width=0.6\textwidth]{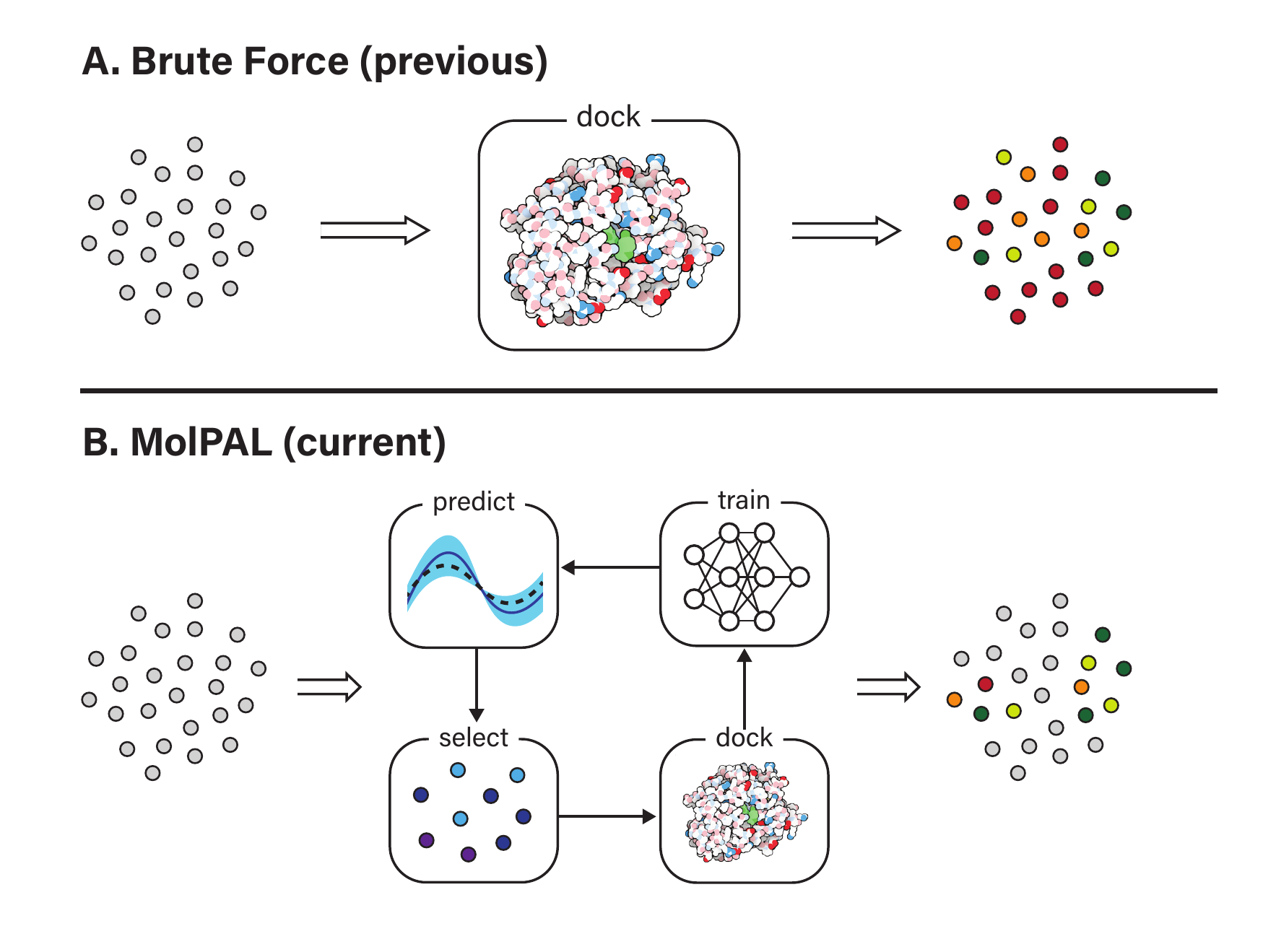}
    \caption{Overview of a computational docking screen using (\bf{A}) Brute force (exhaustive) virtual screening and (\bf{B}) Molecular pool-based active learning (\tt{MolPAL}). Grey circles depict molecules that have not been evaluated.}
    \label{fig:schematic}
\end{figure}

The goal in any virtual screening approach is to find a set of high-performing compounds---herein, computational ``hits'' with the most negative docking scores---within a significantly larger search space. To restate this formally, we are attempting to solve for the set of top-$k$ molecules ${\{x_i\}_{i=1}^k}^*$ from a virtual library $\mathcal{X}$ that maximize some black-box function of molecular identity $f : x \in \mathcal{X} \rightarrow \mathbb{R}$, i.e.,
\begin{equation}
    {\{x_i\}_{i=1}^k}^* = \argmax_{\{x_i\}_{i=1}^k \subset \mathcal{X}} \sum_{i=1}^k f(x_i).
    \label{eq:objective}
\end{equation}
In this study, the black-box function $f(x)$ is the negative docking score of a candidate ligand but other possibilities include the HOMO-LUMO gap of a candidate organic semiconductor, extinction coefficient of a candidate dye, turnover number of a candidate catalyst, etc. 
Brute force searching---screening every molecule in a library indiscriminately---is a straightforward and common strategy employed to solve this type of problem, but it necessarily spends a significant fraction of its time evaluating relatively low-performing compounds (Figure~\ref{fig:schematic}A). However, algorithmic frameworks exist that aim to solve Equation~\ref{eq:objective} with the fewest number of required evaluations.
Bayesian optimization is one such framework that uses a surrogate model trained on previously acquired data to guide the selection of subsequent experiments. We describe Bayesian optimization in more detail in the \nameref{sec:methods} section below, but we refer a reader to ref.~\citenum{frazier_tutorial_2018} for an in-depth review on the subject. The application of Bayesian optimization to physical problems is well-precedented, e.g., with applications to materials design\cite{balachandran_adaptive_2016, gubaev_accelerating_2019, xue_accelerated_2016, montoya_autonomous_2020}, bioactivity screening\cite{bilsland_yeast-based_nodate, czechtizky_integrated_2013, williams_cheaper_2015}, and molecular simulations\cite{janet_accurate_2020, ghanakota_combining_2020, konze_reaction-based_2019}, but few examples exist of its application in virtual screening for drug discovery. Furthermore, the design space is both large and discrete but not necessarily defined combinatorially, which is a relatively unexplored setting for Bayesian optimization.

Two recent examples of work that used active learning for computational drug discovery can be found in Deep Docking\cite{gentile_deep_2019} and a study from \citeauthor{pyzer-knapp_using_2020}\cite{pyzer-knapp_using_2020}. Deep Docking uses a greedy, batched optimization approach and treats docking scores as a binary classification problem during the QSAR modelling step with a fingerprint-based feed forward neural network surrogate model. \citeauthor{pyzer-knapp_using_2020} also utilized a batched optimization approach with a Gaussian process (GP) surrogate model using a parallel and distributed Thompson sampling acquisition strategy\cite{hernandez-lobato_parallel_2017}; this approach is well-suited to small design spaces (e.g., thousands of compounds) but GP training does not scale well to millions of acquired data points due to its $O(N^3)$ complexity\cite{gibbs_efficient_1997}. In contrast to Deep Docking, our work treats docking score as a continuous variable, so our surrogate model follows a regression formulation. \citeauthor{lyu_ultra-large_2019} observed correlations between the success rates of experimental validation and computed docking scores\cite{lyu_ultra-large_2019}, suggesting that preserving this information during model training will improve our ability to prioritize molecules that are more likely to be validated as active.

In this work, we leverage Bayesian optimization algorithms for docking simulations in a manner that preserves the fidelity of these structure-based methods while decreasing the computational cost needed to explore the structure-activity landscape of virtual libraries by over an order of magnitude (Figure~\ref{fig:schematic}B). We demonstrate that surrogate machine learning models can prioritize the screening of molecules that are associated with better docking scores, which are presumed to be more likely to validate experimentally\cite{lyu_ultra-large_2019}. We analyze a variety of different model architectures and acquisition functions that one can use to accelerate high-throughput virtual screening using Bayesian optimization. Specifically, we test random forest (RF), feed forward neural network (NN), and directed-message passing neural network (MPN) surrogate models along with greedy, upper confidence bound (UCB), Thompson sampling (TS), expectation of improvement (EI), and probability of improvement (PI) acquisition strategies in addition to various acquisition batch sizes. We study these optimization parameters on multiple datasets of protein-ligand docking scores for compound libraries containing roughly ten thousand, fifty thousand, two million, and one hundred million molecules. We perform these studies using \tt{MolPAL}, an open source software which we have developed and made freely available.

\section{Results}\label{sec:results}

\subsection{Small virtual libraries}\label{subsec:10k_50k}
\begin{figure}[t!]
    \includegraphics[width=\textwidth]{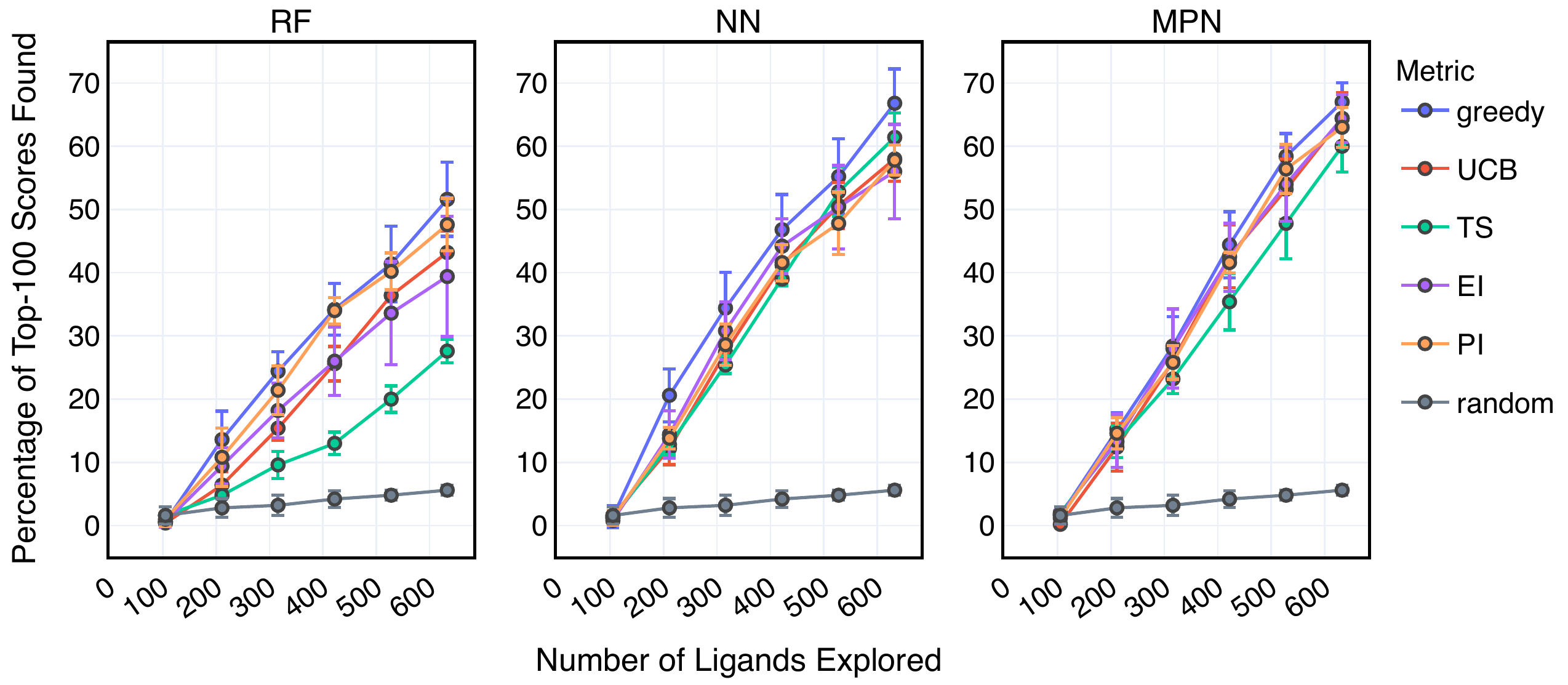}
    \caption{Bayesian optimization performance on Enamine 10k screening data as measured by the percentage of top-100 scores found as function of the number of ligands evaluated. Each trace represents the performance of the given acquisition metric with the surrogate model architecture corresponding the chart label. Each experiment began with random 1\% acquisition (ca. 100 samples) and acquired 1\% more each iteration for five iterations. Error bars reflect $\pm$ one standard deviation across five runs. RF, random forest; NN, neural network; MPN, message-passing neural network; UCB, upper confidence bound; TS, Thompson sampling; EI, expected improvement; PI, probability of improvement.}
    \label{fig:Enamine10k}
\end{figure}

As an initial evaluation of the experimental efficiency Bayesian optimization can provide, we turned to a dataset containing scores from 10,540 compounds (Enamine's smaller Discovery Diversity Collection, ``Enamine 10k'') docked against thymidylate kinase (PDB ID: 4UNN\cite{naik_structure_2015}) using AutoDock Vina\cite{trott_autodock_2010}. Data acquisition was simulated as the iterative selection of 1\% of the library (ca. 100 molecules) repeated five times after initialization with a random 1\% subset for a total acquisition of 6\% of the library. We first looked at a random forest (RF)  operating on molecular fingerprints as our surrogate model along with a greedy acquisition strategy. This combination yields clear improvement over the random baseline, representative of a brute-force search, when looking at the percentage of top-100 (ca. top-1\%) scores in the full dataset found by \tt{MolPAL} (Figure~\ref{fig:Enamine10k}, left panel). The greedy strategy finds, on average, 51.6\% ($\pm 5.9$ standard deviation over five runs) of the top-100 scores in the full dataset when exploring only 6\% of the pool. 

We define the enrichment factor (EF) as the ratio of the percentage of top-$k$ scores found by the model-guided search to the percentage of top-$k$ scores found by a random search for the same number of objective function calculations. The random baseline finds only 5.6\% of the top-100 scores in the 10k dataset, thus constituting an EF of 9.2 for the greedy random forest combination. A UCB acquisition metric, yields similar, albeit slightly lower, performance with an EF of 7.7. Surprisingly, the other optimistic acquisition metric we tested, Thompson sampling (TS), does show an improvement over the random baseline but is considerably lower than all other metrics (EF = 4.9). We attribute this lower performance to the large uncertainties in the RF model and the relatively compact distribution of predicted means, which cause the Thompson sampling strategy to behave somewhat closer to random sampling than other metrics.

We next assessed the effect of architecture for the surrogate model. Using a fully connected neural network (NN) operating on molecular fingerprints, we observed an increase in performance for all acquisition metrics (Figure~\ref{fig:Enamine10k}, middle panel). With the NN model, the least performant acquisition strategy (56.0\% with EI) was more powerful than the best performing strategy with the RF model (51.6\% with greedy acquisition.) The greedy acquisition metric remains the best performing for the NN model, achieving 66.8\% of top-100 scores found for an EF of 11.9. The third and final architecture examined is a message-passing neural network model (MPN), using the D-MPNN implementation by \citeauthor{yang_analyzing_2019}\cite{yang_analyzing_2019}. The MPN model resulted in comparable performance to the NN model (Figure~\ref{fig:Enamine10k}, right panel), with slight improvement for some metrics. However, the highest performance observed, 67.0\% (EF = 12.0) with greedy acquisition, is statistically identical to the best NN result.

\begin{figure}[b!]
    \includegraphics[width=\textwidth]{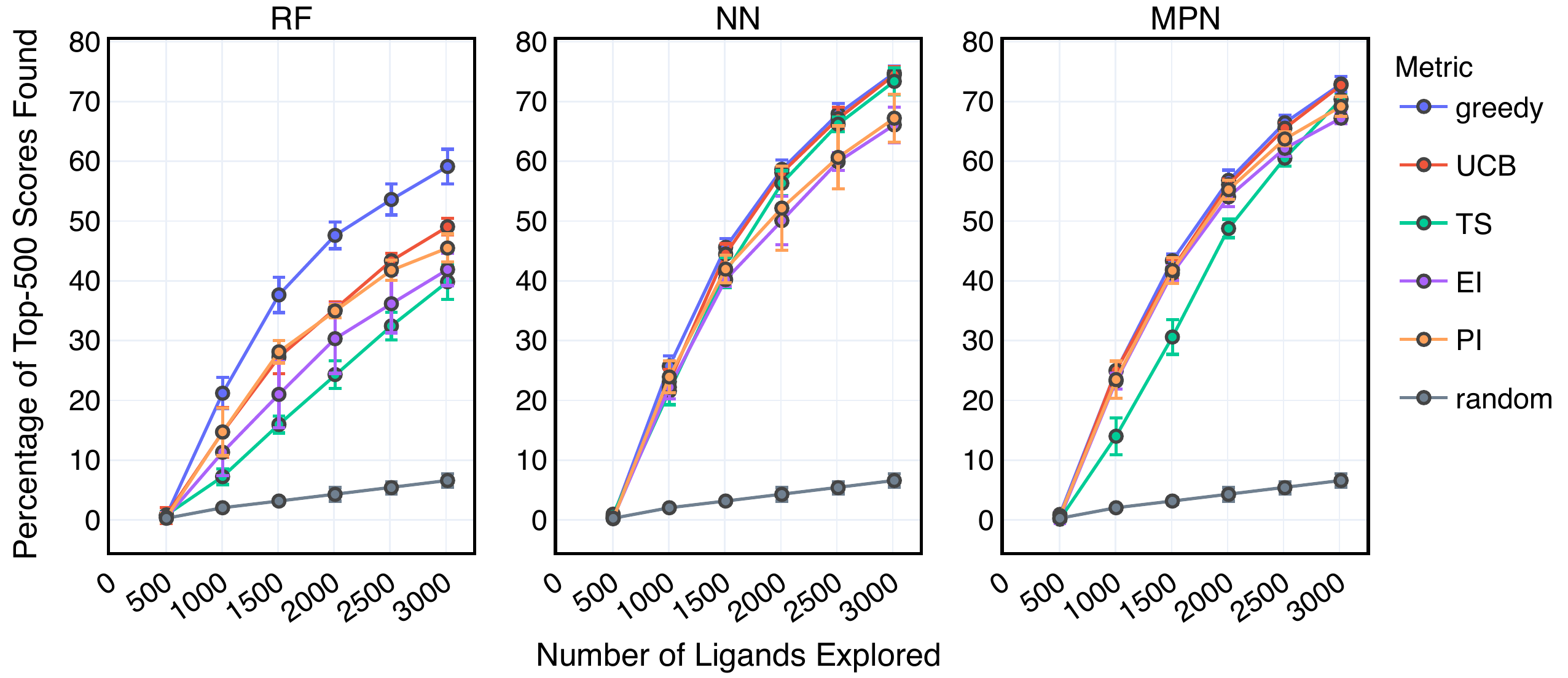}
    \caption{Bayesian optimization performance on Enamine 50k screening data as measured by the percentage of top-500 scores found as function of the number of ligands evaluated. Each trace represents the performance of the given acquisition metric with the surrogate model architecture corresponding the chart label. Each experiment began with random 1\% acquisition (ca. 500 samples) and acquired 1\% more each iteration for five iterations. Error bars reflect $\pm$ one standard deviation across five runs.}
    \label{fig:Enamine50k}
\end{figure}

These analyses were repeated for Enamine's larger Discovery Diversity Collection of 50,240 molecules (``Enamine 50k'') also against 4UNN with the same docking parameters (Figure~\ref{fig:Enamine50k}). We again took 1\% of the library as our initialization with five subsequent exploration batches of 1\% each. All of the trends remained largely the same across acquisition metrics and model architectures; we observed comparable quantitative performance for every surrogate model/acquisition metric combination as compared to the smaller library. For example, the RF model with a greedy acquisition strategy now finds 59.1\% ($\pm 2.9$) of the top-500 scores (ca. top-1\%) using the Enamine 50k library vs. the 51.6\% of the top-100 scores (ca. top-1\%) when using the Enamine 10k library. There was little difference between the results of the NN and MPN models on the Enamine 50k results, which find 74.8\% and 72.9\% of the top-500 scores after exploring just 6\% of the library, respectively. These values represent enrichment factors of 11.3 and 11.0, respectively, over the random baseline.

\subsection{Enamine HTS Collection}\label{subsec:HTS}
Encouraged by the significant enrichment observed with the 10k and 50k libraries, we next tested Enamine's 2.1 million member HTS Collection (``Enamine HTS'')--a size more typical of a high-throughput virtual screen. We again use 4UNN and Autodock Vina to define the objective function values. With this larger design space, acquisitions of 1\% represent a significant number of molecules (ca. 21,000); therefore, we also sought to reduce exploration size. Given the strong performance of the greedy acquisition metric and its simplicity (i.e., lack of a requirement of predicted variances), we focus our analysis on the this metric alone.

\begin{figure}[t!]
    \includegraphics[width=\textwidth]{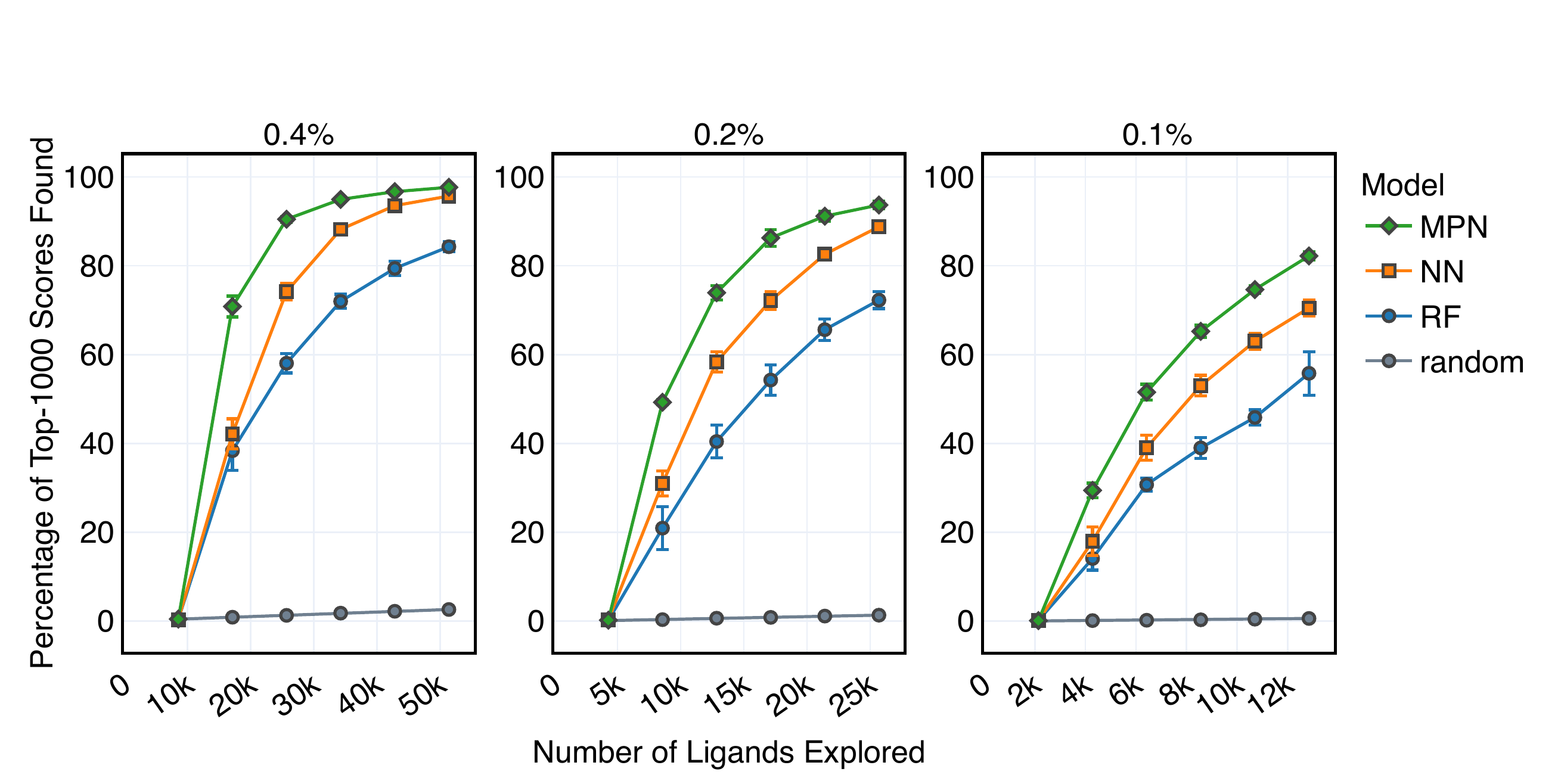}
    \caption{Bayesian optimization performance on Enamine HTS screening data as measured by the percentage of top-1000 scores found as function of the number of ligands evaluated. Each trace represents the performance of the given model with a greedy acquisition strategy. Chart labels represent the fraction of the library acquired in the random initial batch and the five subsequent exploration batches. Error bars reflect $\pm$ one standard deviation across five runs.}
    \label{fig:EnamineHTS}
\end{figure}

We tested three different batch sizes for initialization and exploration, with five exploration batches, as in our above experiments. Using a batch size of 0.4\% for a total of 2.4\% of the pool, we observed strong improvement over random exploration for all three models using the greedy metric in terms of the fraction of the top-1000 (top-0.05\%) scores found (Figure~\ref{fig:EnamineHTS}, left panel.) With a 0.4\% batch size, the random baseline finds only 2.6\% of the top-1000 scores, whereas the RF, NN, and MPN models find 84.3\% (EF = 32.4), 95.7\% (EF = 36.8), and 97.7\% (EF = 37.6) of the top-1000 scores, respectively. Lowering the total exploration size by half so that 0.2\% of the library is acquired at each iteration (a total of 1.2\% of the library) reduces the overall performance of each model, but the drop in performance is not commensurate with the decrease in performance of the random baseline (Figure~\ref{fig:EnamineHTS}, middle panel.) The MPN model is shown to be the most robust to the decrease in batch size, identifying 93.7\% of the top-1000 scores after exploring just 1.2\% of the design space for an enrichment factor of 72.0. Similarly, reducing the batch size further to 0.1\% affects the random baseline to a greater extent than any active learning strategy  (Figure~\ref{fig:EnamineHTS}, right panel.) Here, the random baseline finds only 0.6\% of the top-1000 scores but the MPN model with greedy acquisition finds 82.2\% of the top-1000 scores, representing an enrichment factor of 137. This growth in enrichment factor as exploration fraction decreases holds for other, non-greedy acquisition metrics (Tables~\ref{tbl:HTS_004_results_final}-\ref{tbl:HTS_001_results_final}).

\subsection{Ultra-large library}\label{subsec:AmpC}
One goal of the Bayesian optimization framework in our software, \tt{MolPAL}, is to scale to even larger chemical spaces. A two million member library is indeed a large collection of molecules, but screens of this size are compatible with standard high-performance computing clusters. Our final evaluation sought to demonstrate that \tt{MolPAL} can make virtual screens of $\ge 10^8$-member libraries accessible to researchers using modest, shared clusters. We turned to a recent study by \citeauthor{lyu_ultra-large_2019} that screened 99 million readily accessible molecules against AmpC $\beta$-lactamase (PDB ID: 12LS) using DOCK3.7\cite{lyu_ultra-large_2019}. The full dataset of 99.5 million molecules that were successfully docked against 12LS (``AmpC'') is used as the ground truth\cite{balius_ampc_screen_tablecsvgz_2018}. We measure our algorithm's performance as a function of the top-50000 (top-0.05\%) scores found for all three models using acquisition sizes of 0.4\%, 0.2\%, or 0.1\%.

\begin{figure}
    \includegraphics[width=\textwidth]{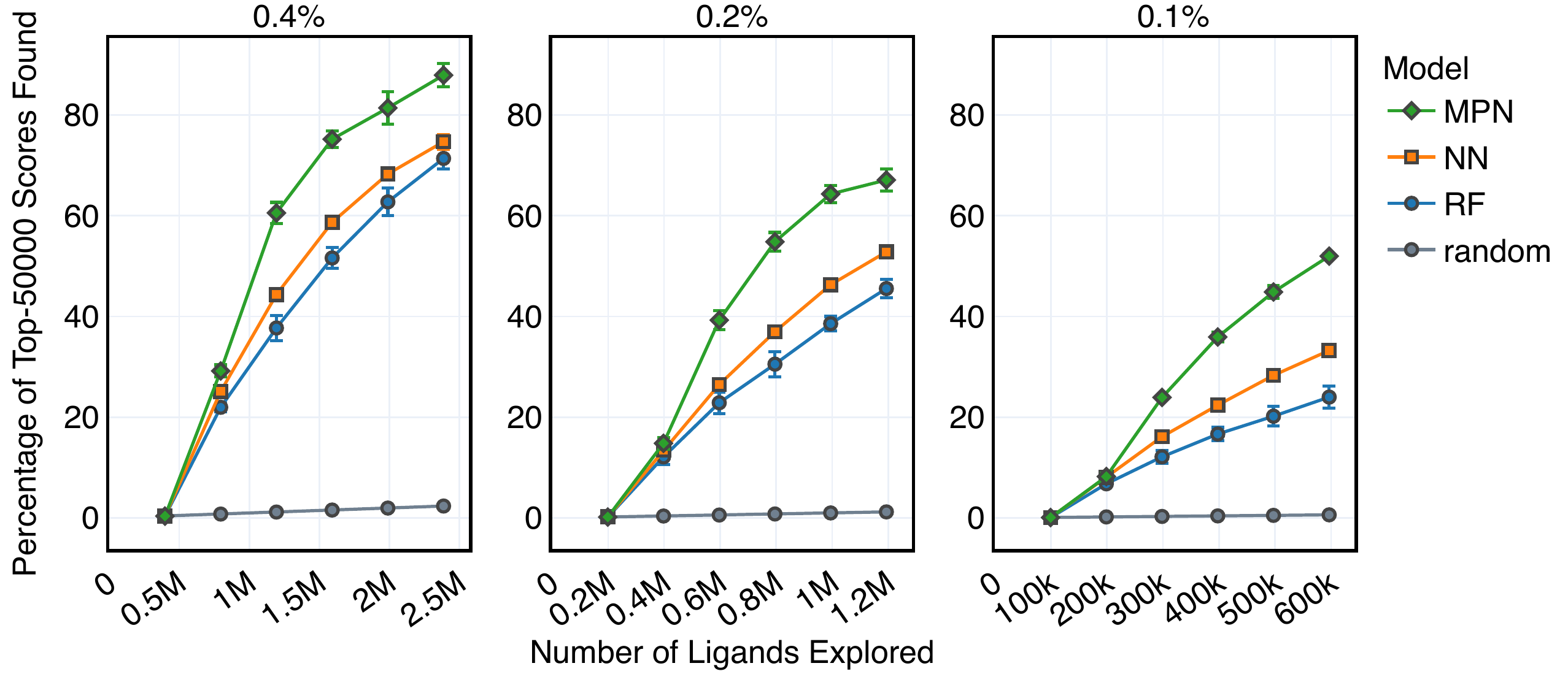}
    \caption{Bayesian optimization performance on AmpC screening data as measured by the percentage of top-50000 scores found as function of the number of ligands evaluated. Each trace represents the performance of the given model with a greedy acquisition strategy. Chart labels represent the fraction of the library taken in both the initialization batch and the five exploration batches. Error bars reflect $\pm$ one standard deviation across three runs.}
    \label{fig:AmpC100M}
\end{figure}

We see a similar qualitative trend for the AmpC dataset as for all previous experiments: namely, that the use of Bayesian optimization enables us to identify many of the top-performing compounds after exploring a small fraction of the virtual library, even when using a greedy acquisition metric (Figure~\ref{fig:AmpC100M}). For the 0.4\% batch size experiments, the MPN model finds 87.9\% of the top-50000 (ca. top-0.05\%) scores after exploring 2.4\% of the total pool (EF = 36.6). Decreasing the batch size to 0.2\% led to a recovery of 67.1\% (EF = 55.9), and further decreasing the batch size to 0.1\% resulted in 52.0\% recovery (EF = 86.7.) The NN and RF surrogate models were not as effective as the MPN, but still led to significant enrichment above the baseline. Results for additional acquisition functions can be found in Tables~\ref{tbl:AmpC_004_results_final}-\ref{tbl:AmpC_001_results_final}. The UCB acquisition metric led to notable increases in the performance of the MPN model for both a 0.4\% and 0.2\% batch size, finding 94.8\% (EF = 39.5) and 75.5\% (EF = 62.9) of the top-50000 scores, respectively; however, UCB was not consistently superior to greedy acquisition for other model architectures or datasets.

It is worth commenting on the differences in quantitative enrichment factors reported for the AmpC data and Enamine HTS data. There are at least two differences that preclude a direct comparison: (1) The top-$k$ docking scores in the AmpC data were generated by DOCK and span a range of -118.83 to -73.99. This is compared to docking scores from the Enamine HTS collection dataset calculated with AutoDock Vina, where the top-$k$ docking scores range from -12.7 to -11.0. The lower precision of AutoDock Vina scores makes the top-$k$ score metric more forgiving (discussed later in the \nameref{subsec:eval_metrics} subsection.) (2) The chemical spaces canvassed by both libraries are different. This will necessarily impact model performance and, by extension, optimization performance.

\subsection{Single-iteration active learning}\label{subsec:one_shot}
\begin{figure}[t!]
    \includegraphics[width=0.5\textwidth]{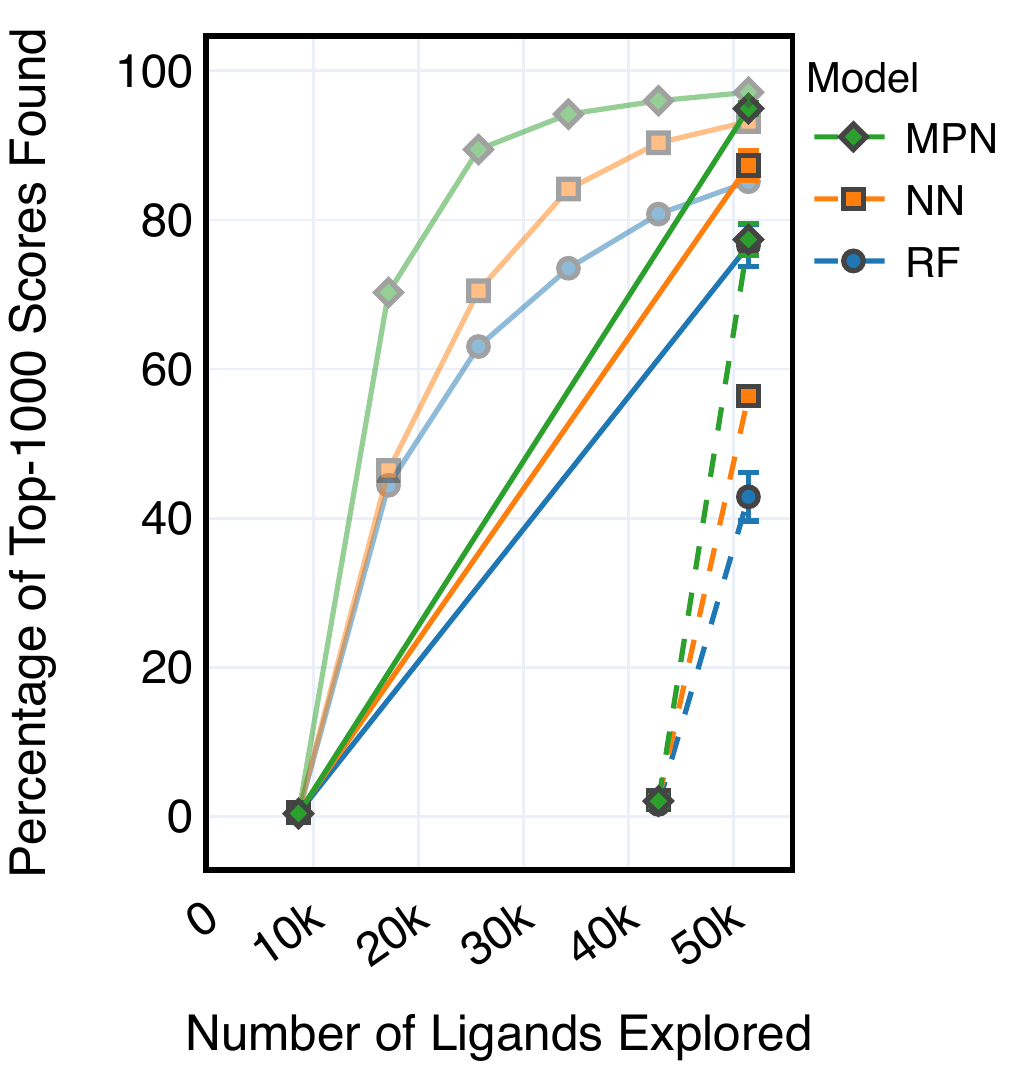}
    \caption{Single-iteration results on Enamine HTS with greedy acquisition. Solid traces: initialization batch size of 0.4\% of pool and exploration batch size of 2\% of pool. Dashed traces: initialization batch size of 2\% of pool and exploration batch size of 0.4\% of pool. Error bars reflect $\pm$ one standard deviation across three runs. Faded traces: reproduced active learning data from 0.4\% experiments (error bars omitted.)}
    \label{fig:EnamineHTS_one_shot}
\end{figure}

A critical question with these experiments is the importance of the active learning strategy. From the Enamine HTS data (Figure~\ref{fig:EnamineHTS}), we observe a sharp increase in the percentage of top-1000 scores found after the first exploration batch (e.g., from 0.4\% to 70.3\% for the MPN 0.4\% acquisition), suggesting that the randomly selected initial batch is quite informative. We look at ``single-iteration'' experiments, where the first batch is selected randomly and the second (and final) batch is selected according to our acquisition function (Figure~\ref{fig:EnamineHTS_one_shot}). Selecting all 42,000 molecules at once after training on the initial 8,400 molecules results in finding 94.9\% ($\pm$ 0.8) of the top-1000 scores  with an MPN model after exploring 2.4\% of the library (EF = 36.5). This is slightly (but significantly) lower than the active learning case with an MPN model, which finds 97.7\% of the top-1000 scores (EF = 37.6). Using an NN or an RF model, the differences in active learning versus the single-iteration case are more pronounced. We also test the setting where the initial batch consists of 42,000 molecules and the single active learning iteration only selects 8,400. The smaller acquisition size for the second batch results in considerably lower MPN performance, finding only 77.4\% of the top-1000 scores (EF = 29.8). This is worse than any model we tested with an active learning-based strategy at the same number of function evaluations. The less flexible NN and RF models suffer even more pronounced drops in performance with a large initial batch and small exploration batch.

\subsection{Dynamic convergence criterion}\label{subsec:convergence}
\begin{figure}
    \includegraphics[width=0.5\textwidth]{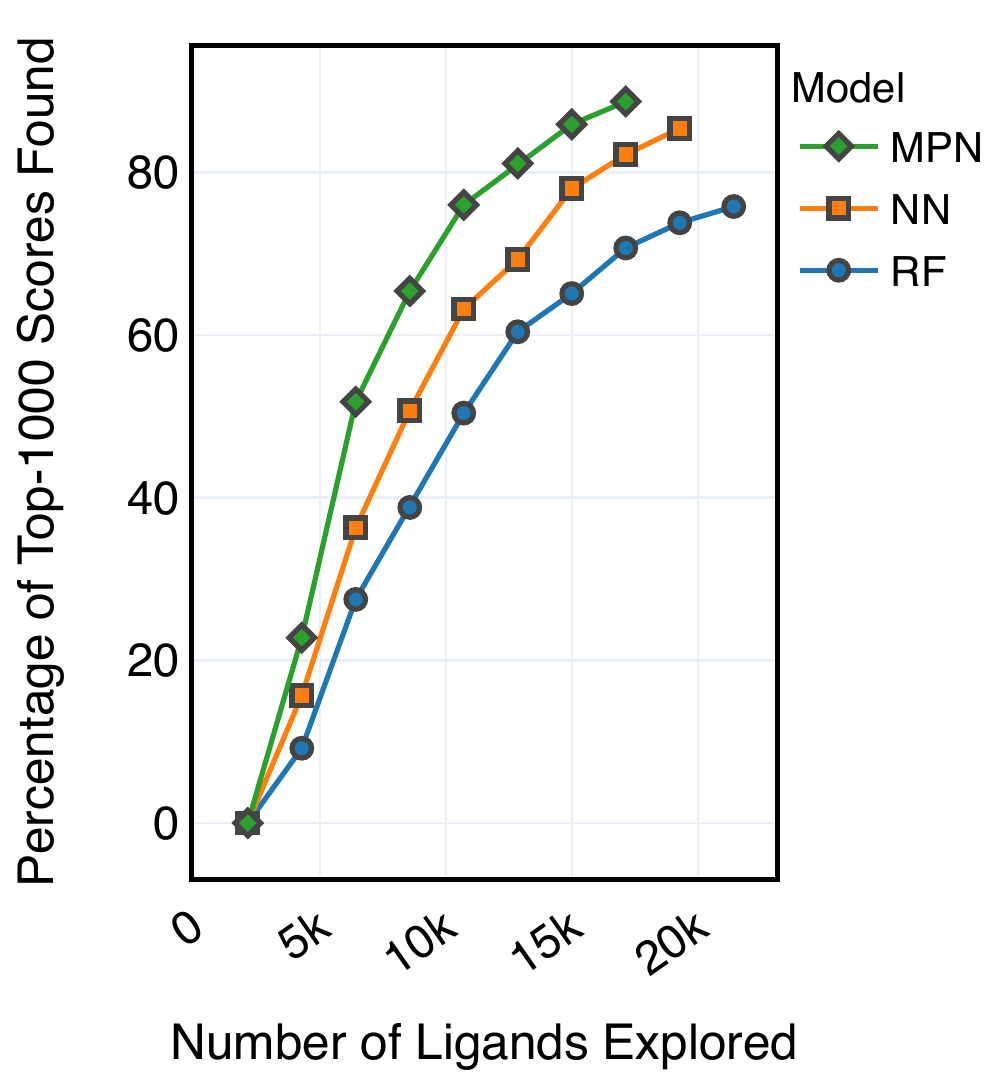}
    \caption{Convergence results on Enamine HTS collection with greedy acquisition and 0.1\% batch size for both initialization and exploration.}
    \label{fig:EnamineHTS_convergence}
\end{figure}

Our evaluations so far have demonstrated reliable performance of \tt{MolPAL} using a fixed exploration strategy (i.e., number of iterations). However, we will typically not know \textit{a priori} what an appropriate total exploration size is. We therefore define a convergence criterion for the Enamine HTS dataset that is satisfied when the fractional difference between the current average of the top-1000 scores and the rolling average of the top-1000 scores from the previous three epochs, corresponding to the top 0.05\% of compounds, is less than 0.01. Figure~\ref{fig:EnamineHTS_convergence} illustrates the use of this convergence criterion using a 0.1\% batch size (ca. 2,100 molecules) with a greedy acquisition metric. We find that not only do the higher capacity models converge sooner (MPN $>$ NN $>$ RF), but they also converge to a higher percentage of top-1000 scores found (88.7\%, 85.4\%, and 75.8\%, respectively).

\subsection{Chemical space visualization}\label{subsec:umap}
\begin{figure}
    \includegraphics[width=0.9\textwidth]{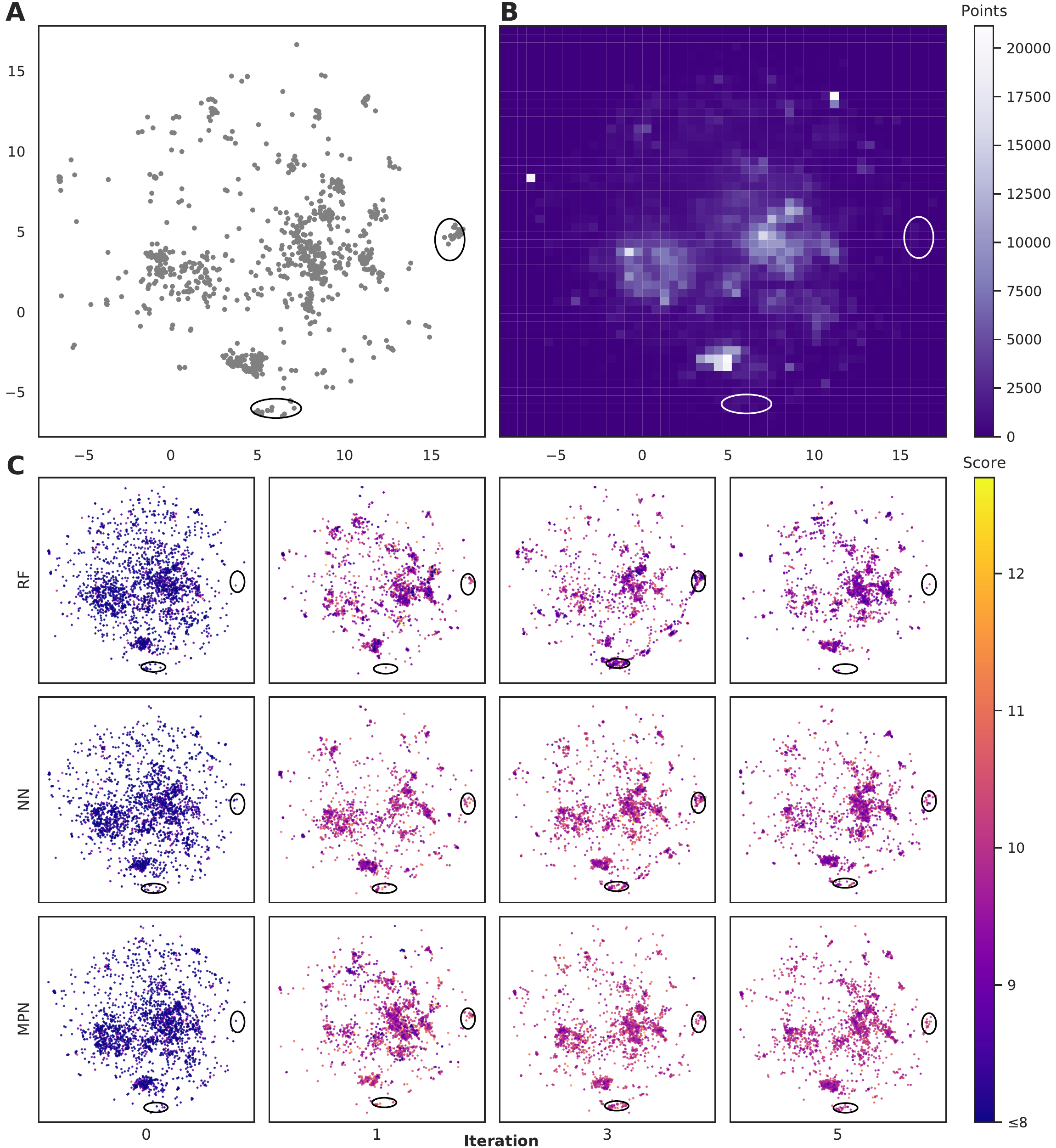}
    \caption{Visualization of the chemical space in the Enamine HTS library using UMAP embeddings of 2048-bit Atom-pair fingerprints trained on a random 10\% subset of the full library. \bf{A.} Embedded fingerprints of the top-1000 scoring molecules. \bf{B.} 2-D density plot of the embedded fingerprints of the full library. \bf{C.} Embedded fingerprints of the molecules acquired in the given iteration using a greedy acquisition metric, 0.1\% batch size, and specified surrogate model architecture. Circled regions indicate clusters of high-scoring compounds in sparse regions of chemical space. Color scale corresponds to the negative docking score (higher is better). X- and y-axes are the first and second components of the 2D UMAP embedding and range from -7.5 to 17.5}
    \label{fig:EnamineHTS_UMAP}
\end{figure}

To visualize the set of molecules acquired during the Bayesian optimization routine, we projected the 2048-bit atom-pair fingerprints of the Enamine HTS library into two dimensions using UMAP\cite{mcinnes_umap_2020} (Figures~\ref{fig:EnamineHTS_UMAP} and \ref{fig:umap_si}). The embedding of the library was trained on a random 10\% subset of the full library and then applied to the full library. Comparing the location of the top-1000 molecules (Figure~\ref{fig:EnamineHTS_UMAP}A) to the density of molecules in the entire 2M library (Figure~\ref{fig:EnamineHTS_UMAP}B) reveals several clusters (black ellipses) of high-performing compounds located in sparse regions of chemical space. To observe how the three separate surrogate models cope with this mismatch, we plot the embedded fingerprints of the molecules acquired in the zeroth (random), first, third, and fifth iterations of a 0.1\% batch size greedy search (Figure~\ref{fig:EnamineHTS_UMAP}C) While all surrogate models select candidates in these two areas, there are differences in both the speed and thoroughness with which they search them. The RF model does not begin sampling densely from this region until the third iteration and refocuses its attention elsewhere by the fifth iteration. Both the NN and MPN models acquire a larger number of points from these areas in earlier iterations. While this analysis is qualitative by nature, it illustrates how the performance differences between the three surrogate model architectures relates to the regions of chemical space they choose to explore.

\section{Discussion}\label{sec:discussion}

\subsection{Effect of acquisition strategy on performance}\label{subsec:acq_strat}
An interesting result from these experiments was the consistently strong performance of the greedy acquisition metric. This is surprising given the fact that the greedy metric is, in principle, purely exploitative and vulnerable to limiting its search to a single area of the given library's chemical space. Prior work in batched Bayesian optimization has focused on developing strategies to prevent batches from being homogeneous, including use of an inner optimization loop to construct batches one candidate at a time\cite{desautels_parallelizing_2014, tsymbalov_deeper_2019}. Despite this potential limitation of the greedy acquisition metric, it still leads to adequate exploration of these libraries and superior performance to metrics that combine some notion of exploration along with exploitation (i.e., UCB, TS, EI, PI). One confounding factor in this analysis is that methods used for uncertainty quantification in regression models are often unreliable \cite{hirschfeld_uncertainty_2020}, which may explain the poorer empirical results when acquisition depends on their predictions.

\subsection{Effect of library size}\label{subsec:lib_size}
The principal takeaway from our results on different library sizes is that Bayesian optimization is not simply a viable technique but an effective one in all of these cases. Though it is difficult to quantitatively compare algorithm performance on each dataset due to their differing chemical spaces, the impact of library size on the optimization is still worth commenting on. We observe the general trend in our data that, as library size increases, so too does top-$k$ performance given a constant fractional value for $k$, even when decreasing relative exploration size. We anticipate that this is due in part to the greater amount of training data that the surrogate model is exposed to over the course of the optimization. Despite the relative batch size decreasing, the absolute number of molecules taken in each batch and thus the number data points to train on increases from roughly 500 to nearly 8,400 when moving from the Enamine 50k dataset (1\% batch size) to the Enamine HTS dataset (0.4\% batch size). We analyzed the mean-squared error (MSE) of MPN model predictions on the entire 10k, 50k, and HTS libraries after initialization with random 1\%, 1\%, and 0.4\% batches, respectively; the MSE decreased with increasing library size: 0.3504, 0.2617, and 0.1520 (Spearman's $\rho = 0.6699$, $0.7826$, $0.9094$). This trend suggests that the overall ``diversity'' of the chemical library is not increasing at the same rate as the size, i.e., the scores of molecules in the larger chemical spaces are more easily predicted with a proportional increase in the training set size. As a result, the surrogate model is better able to aid in the acquisition of high-performing molecules.

\subsection{Consistency across repeated trials}
Bayesian optimization can be prone to large deviations across repeated trials, but our results showed consistent performance across all datasets and configurations (Tables~\ref{tbl:10k_results_final}-\ref{tbl:AmpC_001_results_final}). To investigate whether the consistency in performance is a result of consistency in the exact molecular species selected, we calculate the total number of unique SMILES strings acquired across all repeated experiments as a function of optimization iteration (Figures~\ref{fig:10k50k_smis_greedy_union} and \ref{fig:HTS_smis_greedy_union}). Comparing these results to both the theoretical maximum (each trial acquiring a completely unique subset of molecules at each iteration) and the theoretical minimum (each trial acquiring an identical subset of molecules at each iteration after the initialization) approximates the degree to which each repeat explores the same or different regions of chemical space. Traces closer to the maximum would indicate that each experiment is exploring a unique subset of highly performing molecules, and traces closer to the minimum signal the opposite: that each experiment is converging towards the same regions of chemical space. Our data are most consistent with the latter, suggesting that each trial is converging towards the same optimal subset of the library regardless of its initialization. We hypothesize that this is due to relatively smooth structure-property landscapes present in these datasets and lack of statistical uncertainty.

\subsection{Limitations of evaluation metrics}\label{subsec:eval_metrics}
In this study, three separate evaluation criteria were used to assess performance: the average top-$k$ docking score identified, the fraction of top-$k$ SMILES identified, and the fraction of top-$k$ scores identified throughout the optimization campaign (calculation details are provided in the \nameref{sec:methods} section below). The average metric is sensitive to the scale and distribution of scores, making direct comparison between datasets challenging. The top-$k$ SMILES metric asks whether a specific set of molecules is selected by the algorithm, which can be overly strict if multiple molecules have identical performance (i.e., the same score) but are not within the top-$k$ due to arbitrary data sorting (Figures~\ref{fig:10k_score_hist}-\ref{fig:HTS_score_hist}). The top-$k$ score metric overcomes this limitation by assigning equal weight to each molecule with the same score regardless of its specific position in the sorted dataset.  As a result, this makes the scores metric more forgiving for datasets with smaller ranges and lower precision (e.g., those calculated using AutoDock Vina with a precision of 0.1) than those with larger ranges and higher precision (e.g., those calculated using DOCK with a precision of 0.01). In contrast to the average top-$k$ score, however, this metric does not reward ``near-misses'', for example, identifying the $k+1$ ranked molecule with a nearly identical score to the $k$-th molecule.

\subsection{Optimal batch size for active learning}\label{subsec:batch_size}
The number of molecules selected at each iteration represents an additional hyperparameter for Bayesian optimization. In one limit, Bayesian optimization can be conducted in a purely sequential fashion, acquiring the performance of a single molecule each iteration. Fully sequential learning would offer the most up-to-date surrogate model for the acquisition of each new point but it would also be extremely costly to continually retrain the model and perform inference on the entire candidate pool. In the other limit, molecules would be selected in a single iteration, which can lead to suboptimal performance depending on the acquisition size (Figure~\ref{fig:EnamineHTS_one_shot}). Finding a principled balance between these two without resorting to empirical hyperparameter optimization is an ongoing challenge. In each of our experiments, the relative batch size was held constant at one sixth of the total exploration size. Future performance engineering work will seek to examine the effects of dynamic batch sizes in batched optimization. Note that overall batch diversity is another consideration in batched Bayesian optimization. While selected batches in this study did not appear to suffer from homogeneity, previous approaches to improve batch diversity could be explored as well\cite{wang_batched_2018, tsymbalov_deeper_2019, hernandez-lobato_parallel_2017}.

\subsection{Cost of surrogate model (re)training}\label{subsec:training_cost}
The question of optimal batch size cannot be decoupled from the computational cost of model retraining and inference. Throughout these studies, we have focused only on the number of objective function calculations necessary to achieve a given level of performance. While objective function calculation is significantly more expensive than the cost of model training and inference, inference costs scale linearly with the size of the dataset  and contribute to the overall cost of our algorithm. 

The MPN model was shown to be superior in the largest datasets (Enamine HTS and AmpC), but its costs are markedly higher than those of the fingerprint-based NN model. The tradeoff between sample efficiency (number of objective function calculations) and surrogate model costs (training and inference) should be balanced when selecting a model architecture. In our evaluation, the costs of the MPN and NN cannot be directly compared due to differences in their implementation and extent of both parallelization and precalculation. For more details, see the software design subsection in the supplementary text. An additional choice when seeking to limit surrogate model costs is whether to train the model online with newly acquired data or fully retrain the model at each iteration with all acquired data. We examined this possibility, but online learning lead to consistently lower performance in our experiments (Tables~\ref{tbl:10k_results_final}-\ref{tbl:AmpC_001_results_final}).

\section{Conclusion}
In this work, we have demonstrated the application of Bayesian optimization to the prioritization of compounds for structure-based virtual screening using chemical libraries ranging in size from 10k to 100M ligands. A thorough evaluation of different acquisition metrics and surrogate model architectures illustrates the surprisingly strong performance of a greedy acquisition strategy and the superiority of a message passing neural network over fingerprint-based feed forward neural network or random forest models. In the largest library tested, the 100M member library screened against 12LS by Lyu et al., we identify 87.9\% of the top-50000 scoring ligands with a $>$40-fold reduction in the number of docking calculations using a greedy acquisition metric and 94.8\% using the UCB acquisition metric.

We believe that this model-guided approach to compound prioritization should become standard practice as a drop-in replacement for exhaustive high-throughput virtual screening when the exact set of top-$k$ compounds is not needed. Moreover, this approach is also relevant to \emph{experimental} high-throughput screening, an expensive and important tool for challenging drug discovery problems\cite{schuffenhauer_evolution_2020}. Future work will seek to extend the open source \tt{MolPAL} software package and leverage it in a prospective manner to greatly accelerate a structure-based virtual screen of the Enamine REAL database. We also hope to expand \tt{MolPAL} beyond the initial software detailed in this report with the addition of new surrogate model architectures, the inclusion of improved uncertainty estimation techniques, and the expansion to other forms of virtual discovery, i.e., other objective functions. Finally, we envision that in addition to accelerating the practice of virtual screening, \tt{MolPAL} will increase its accessibility through the pipelining of the entire virtual screening process behind a common interface.

\section{Methods}\label{sec:methods}

\subsection{Batched Bayesian optimization}\label{para:BO}
Bayesian optimization is an active learning strategy that iteratively selects experiments to perform according to a surrogate model's predictions, often using machine learning (ML) models. In the context of this work, the Bayesian optimization was performed on a discrete set of candidate molecules, herein referred to as a ``pool'' or virtual library, and points were acquired in batches rather than one point at a time. Batched Bayesian optimization begins by first calculating the objective function $f(x)$ for a set of $n$ random points $\{x\}_{i=1}^n$ within a pool of points $\mathcal{X}$. The objective function values for these points are calculated, in this study as docking scores from AutoDock Vina, and the corresponding tuples $\{(x_i, f(x_i))\}_{i=1}^{n}$ are stored in the dataset $\mathcal{D}$. A surrogate model $\hat{f}(x)$ is then trained on these data and, along with the current maximum objective function value $f^*$, is passed to an acquisition function $\alpha(x; \hat{f}, f^*)$, which calculates the utility of evaluating the objective function at the point $x$. Utility may be measured in a number of ways: the predicted objective function value, the amount of information this new point will provide the surrogate model, the likelihood this point will improve upon the current maximum, etc\ldots. See ref.~\citenum{shahriari_taking_2016} for a detailed discussion of various acquisition functions. The set of $m$ points with the largest sum of utilities $\mathcal{S} = \argmax_{\{x_i\}_{i=1}^m \subset \mathcal{X}} \sum_{i=1}^m \alpha(x, \hat{f}, f^*)$ is then selected, or ``acquired''. The set of objective function values corresponding to these points is calculated $\{(x, f(x)): x \in \mathcal{S}\}$ and used to update the dataset $\mathcal{D}$. This process is repeated iteratively until a stopping criterion is met (e.g., a fixed number of iterations or a lack of sufficient improvement).

\begin{algorithm}[h]
    \SetAlgoNoLine
    \DontPrintSemicolon
    \setstretch{1.1}
    
    \KwIn{objective function $f(x)$, acquisition function $\alpha$, surrogate model $\hat{f}(x)$, candidate set $\mathcal{X}$}
    Select random batch $\mathcal{S} \subset \mathcal{X}$\;
    Initialize $\mathcal{D} \leftarrow \{(x, f(x)) : x \in \mathcal{S} \}$\;
    \For{$t \leftarrow 1\ \mathrm{to}\ T$}{
        Train surrogate model $\hat{f}(x)$ using $\mathcal{D}$\;
        Select batch $\displaystyle \mathcal{S} \leftarrow \argmax_{\mathcal{B} \subset \mathcal{X}, ~ |\mathcal{B}| = m} \sum_{x \in \mathcal{B}} \alpha(x; \hat{f}, f^*$)\;
        Update $\mathcal{D} \leftarrow \mathcal{D} \cup \{(x, f(x)) : x \in \mathcal{S}\}$\;
    }
    \KwResult{${\{x_i\}_{i=1}^k}^* = \argmax_{\{x_i\}_{i=1}^k \subset \mathcal{D}} \sum_{i=1}^k f(x_i)$}
    
    \caption{Batched Bayesian Optimization}
    \label{algo:BBO}
\end{algorithm}
\FloatBarrier

\subsection{Acquisition metrics}\label{para:acq_metrics}
The following acquisition functions were tested in this study:
\begingroup
{\setlength{\abovedisplayskip}{0pt}
\setlength{\belowdisplayskip}{0pt}
\allowdisplaybreaks
\begin{flalign*}
    \quad&\mathrm{Random}(x) \sim \mathcal{U}(0, 1) &\\
    &\mathrm{Greedy}(x) = \hat{\mu}(x) &\\
    &\mathrm{UCB}(x) = \hat{\mu}(x) + \beta\hat{\sigma}(x) &\\
    &\mathrm{TS}(x) \sim \mathcal{N}(\hat{\mu}(x), \hat{\sigma}^2(x)) &\\
    &\mathrm{EI}(x) =
        \begin{cases}
            \gamma(x)\Phi(z) + \hat{\sigma}(x)\phi(z), & \hat{\sigma}(x) > 0 \\
            \gamma(x), & \hat{\sigma}(x) = 0 \\
        \end{cases} &\\
    &\mathrm{PI}(x) =
        \begin{cases}
            \Phi(z), &\hat{\sigma}(x) > 0 \\
            1, &\hat{\sigma}(x) = 0\ \mathrm{and}\ \gamma(x) > 0 \\
            0, &\hat{\sigma}(x) = 0\ \mathrm{and}\ \gamma(x) \leq 0
        \end{cases} &
\end{flalign*}}
\endgroup

\noindent where: $\displaystyle \gamma(x) \defeq \hat{\mu}(x) - f^* + \xi$; $\displaystyle z(x) \defeq \frac{\gamma(x)}{\hat{\sigma}(x)}$; $\hat{\mu}(x)$ and $\hat{\sigma}^2(x)$ are the surrogate model predicted mean and uncertainty at point $x$, respectively; $\Phi$ and $\phi$ are the CDF and PDF of the standard normal distribution, respectively; and $f^*$ is the current maximum objective function value. For the experiments reported in the paper, we used $\beta=2$ and $\xi=0.01$

\subsection{Surrogate models}\label{para:models}
In the context of our study, the surrogate modelling step involved training a machine learning (ML) model and using this trained model to predict the objective function value of molecules for which the objective function has not yet been calculated. Three surrogate model architectures were investigated in these studies: random forest (RF), feed-forward neural network (NN), and directed message passing neural network (MPN) models.

\paragraph{Random forest models}
Random forest (RF) models operate by ensembling decision tree models each trained on a random subset of the training data. Broadly, a decision tree is trained on input data of the form $(\mathbf{x},y)$ where $\mathbf{x}=[x_1,\ldots,x_N]$ is a vector composed of input features $x_1,\ldots,x_N$ and $y$ is a ``target'' output value. During training, a decision tree is built by progressively partitioning the input space at decision nodes into two subspaces corresponding to the value of a given feature. This process is repeated recursively on each of the resulting subspaces until some maximum amount of partitioning is achieved and a leaf node is constructed that contains all possible values $\{y_i\}_{i=1}^M$ that correspond to the partitioning of the parent decision nodes. %
A more in-depth discussion on RF models for QSAR may be found in ref.~\citenum{svetnik2003random}. The RF surrogate model in our study was implemented using the \tt{RandomForestRegressor} class from {Scikit-Learn}\cite{pedregosa_scikit-learn_nodate} using an \tt{n\_estimators} value of 100 and a \tt{max\_depth} value of 8.

\paragraph{Feed-forward neural networks}
Feed-forward neural networks (FFNNs) comprise an input layer, an output layer, and zero or more hidden layers between the two. At each layer, the input or hidden vector is linearly transformed by a learned weight matrix and passed through an elementwise nonlinear activation function. NNs are generally trained through stochastic gradient descent to minimize a loss function, which for regression tasks is generally the mean squared error. %

The NN models in our study were implemented in {TensorFlow} \cite{abadi_tensorflow_nodate} using two fully connected hidden layers of 100 nodes each and an output size of one. Each hidden layer utilized a rectified linear unit (ReLU) activation function. The network was trained over 50 epochs with early stopping using the Adam optimizer with an initial learning rate of 0.01, a mean-squared error loss function with L2 regularization (0.01), and a batch size of 4096. Prediction uncertainties, as needed for non-greedy acquisition metrics, were estimated using Monte-Carlo Dropout via dropout layers ($p=0.2$) after each hidden layer using 10 forward passes during inference.

\paragraph{Molecular fingerprints}
Given that molecules are not naturally represented as feature vectors, inputs to both RF and NN models were generated by calculating molecular fingerprints. Fingerprint calculation algorithms vary in their implementation but can broadly be understood as encoding information about the presence or absence of substructures of the given molecule into a vector of fixed length. %
The input to both the RF and NN models used in this study is a 2048-bit Atom-pair fingerprint\cite{carhart_atom_1985} with a minimum radius of one and a maximum radius of three. A more detailed overview of molecular fingerprints is provided in ref.~\citenum{bajusz_chemical_2017}.

\paragraph{Message passing neural networks}
The third and final model architecture we tested was a directed message passing neural network (D-MPNN) model\cite{yang_analyzing_2019}, a variant of a message passing neural network (MPNN) model. In contrast to FFNNs, MPNNs operate directly on the molecular graph rather than a fixed feature vector calculated from the graph. MPNNs function in two stages, an initial message passing phase followed by a readout phase. In the message passing phase, ``messages'' are passed between atoms and/or bonds and their direct neighbors and incoming messages are used to update the ``hidden state'' of each atom and/or bond. %
The message passing phase is repeated over multiple (e.g., 3) iterations, at which point the hidden states of each atom are aggregated (e.g., summed) to produce a molecule-level feature vector. By training this model at the same time as a FFNN operating on the feature vector, MPNNs are able to learn a task-specific representation of an input molecular graph. For more details on the D-MPNN model, we refer a reader to ref.~\citenum{yang_analyzing_2019}.

Message-passing neural network models were implemented using {PyTorch}\cite{paszke_pytorch_nodate} with the \tt{MoleculeModel} class from the \tt{Chemprop} library\cite{yang_analyzing_2019} using standard settings: messages passed on directed bonds, messages subjected to ReLU activation, a learned encoded representation of dimension 300, and the output of the message-passing phase fully connected to an output layer of size 1. The model was trained using the Adam optimization algorithm, a Noam learning rate scheduler (initial, maximum, and final learning rates of $10^{-4}$, $10^{-3}$, and $10^{-4}$, respectively,) and a root mean-squared error loss function over 50 epochs with a batch size of 50. For more details on the Noam learning rate scheduler, see ref.~\citenum{vaswani_attention_2017}. The model was trained without early stopping, but the model state was reloaded from the epoch with the lowest validation score after the final training epoch. When uncertainty values were needed for metric function calculation, an MVE model based off of the work done by \citeauthor{hirschfeld_uncertainty_2020}\cite{hirschfeld_uncertainty_2020} was used. This model featured an output size of two and was trained using the loss function defined by \citeauthor{nix_estimating_1994}\cite{nix_estimating_1994}:
\begin{equation}
    \mathcal{L}(y, \hat{y}, \hat{\sigma}^2) = \frac{\log{2\pi}}{2} + \frac{\log{\hat{\sigma}^2}}{2} + \frac{({y-\hat{y}})^2}{2\hat{\sigma}^2}
    \label{eq:mve_loss}
\end{equation}

All of the surrogate models were used exactly as described above without additional hyperparameter optimization. The models were fully retrained from scratch with all acquired data at the beginning of each iteration.

\subsection{Datasets}\label{para:data}
The datasets generated for these studies were produced from docking the compounds contained in both Discovery Diversity sets and the HTS collection from Enamine against thymidylate kinase (PDB ID: 4UNN). The docking was performed using AutoDock Vina with the following command line arguments:
\begin{verbatim}
--receptor=4UNN.pdbqt --ligand=<ligand_pdbqt> --center_x=9 --center_y=20
--center_z=-5 --size_x=20 --size_y=20 --size_z=17
\end{verbatim}
All other default arguments were left as-is. The ligands were prepared from SDFs available from Enamine\cite{noauthor_diversity_nodate, noauthor_hts_nodate}, parsed into SMILES strings using RDKit \cite{noauthor_rdkit_nodate}, and processed into PDBQT files using OpenBabel\cite{oboyle_open_2011} with the \tt{--gen-3d} flag. The receptor was prepared with PDBFixer\cite{eastman_openmm_2017} using the PDB ID 4UNN, selecting only chain A, deleting all heterogens, adding all suggested missing residues and heavy atoms, and adding hydrogens for pH 7.0. The AmpC screening dataset is the publicly available dataset published by Lyu et al and was used as-is\cite{balius_ampc_screen_tablecsvgz_2018}. The score distribution of each dataset may be found in Figures~\ref{fig:10k_score_hist}-\ref{fig:AmpC_score_hist}.

\subsection{Evaluation metrics}\label{para:eval_metrics}
\tt{MolPAL} performance was judged through three evaluation metrics: (1) average top-$k$ docking score identified (``Average''), (2) the fraction of top-$k$ SMILES identified (``SMILES''), and (3) the fraction of top-$k$ scores identified (``Scores''). For (1), the average of the top-$k$ scores of molecules explored by \tt{MolPAL} was taken and divided by the true top-$k$ molecules' scores based on the full dataset. (2) was calculated by taking the intersection of the set of SMILES strings in the true top-$k$ molecules and the found top-$k$ molecules and dividing its size by $k$. (3) was calculated as the size of the intersection of the list of true top-$k$ scores and the list of observed top-$k$ scores divided by $k$.

\subsection{Hyperparameter optimization}\label{para:hyperopt}
The experiments shown in this study represent only a small fraction of the configurations in which \tt{MolPAL} may be run. A sample of settings that are supported include: various fingerprint types (e.g., RDKit, Morgan, and MACCS), input preclustering for cluster-based acquisition, different confidence estimation methods for deep learning models, etc. Given the wealth of options, an exhaustive hyperparameter optimization was outside the scope of these investigations. We looked at broad trends in both the Enamine 10k and 50k datasets and found only minor variations in performance, supporting our choice not to pursue a rigorous screening of all possible configurations.

\subsection{Software design}\label{para:software_design}
\tt{MolPAL} is built around the \tt{Explorer} class, which performs the Bayesian optimization routine shown in Algorithm \ref{algo:BBO}. The \tt{Explorer} is designed with abstraction at its core and thus relies on four helper classes that each handles an isolated element of Bayesian optimization: \tt{MoleculePool}, \tt{Model}, \tt{Acquirer}, and \tt{Objective}. In each iteration of the \tt{Explorer}'s main optimization loop, a \tt{Model} is first retrained on all observed data then applied to all molecules in the \tt{MoleculePool} to generate a predicted mean and, depending on the \tt{Model}, a predicted uncertainty for each molecule in the pool. These predictions are then passed to an \tt{Acquirer} which calculates the acquisition utility of each molecule and acquires the top-$m$ untested molecules from the \tt{MoleculePool} based on their acquisition utilities. Next, This set of candidate molecules is handed to an \tt{Objective} and the objective function value for each of these candidate molecules is calculated. Lastly, the stopping condition for the \tt{Explorer} is checked and, if satisfied, the \tt{Explorer} terminates and outputs the top-$k$ evaluated molecules. A schematic of both the design and workflow of \tt{MolPAL} may be seen in Figure~\ref{fig:molpal_schema}. The experiments performed in this study were all performed retrospectively using the \tt{LookupObjective} subclass of the \tt{Objective} with a fully generated dataset as a lookup table for objective function calculation.

\FloatBarrier
\begin{acknowledgement}
    We thank Samuel Goldman and Itai Levin for providing feedback on the manuscript and \tt{MolPAL} code. The computations in this paper were run on the FASRC Cannon cluster supported by the FAS Division of Science Research Computing Group at Harvard University. This work was funded by NIGMS RO1 068670.
\end{acknowledgement}

\begin{suppinfo}
    Additional methods and results can be found in the supporting information. All code and data needed to reproduce this study and the figures herein can be found at \url{https://github.com/coleygroup/molpal}
\end{suppinfo}

\bibliography{molpal_refs}

\providecommand{\latin}[1]{#1}
\makeatletter
\providecommand{\doi}
  {\begingroup\let\do\@makeother\dospecials
  \catcode`\{=1 \catcode`\}=2 \doi@aux}
\providecommand{\doi@aux}[1]{\endgroup\texttt{#1}}
\makeatother
\providecommand*\mcitethebibliography{\thebibliography}
\csname @ifundefined\endcsname{endmcitethebibliography}
  {\let\endmcitethebibliography\endthebibliography}{}
\begin{mcitethebibliography}{55}
\providecommand*\natexlab[1]{#1}
\providecommand*\mciteSetBstSublistMode[1]{}
\providecommand*\mciteSetBstMaxWidthForm[2]{}
\providecommand*\mciteBstWouldAddEndPuncttrue
  {\def\EndOfBibitem{\unskip.}}
\providecommand*\mciteBstWouldAddEndPunctfalse
  {\let\EndOfBibitem\relax}
\providecommand*\mciteSetBstMidEndSepPunct[3]{}
\providecommand*\mciteSetBstSublistLabelBeginEnd[3]{}
\providecommand*\EndOfBibitem{}
\mciteSetBstSublistMode{f}
\mciteSetBstMaxWidthForm{subitem}{(\alph{mcitesubitemcount})}
\mciteSetBstSublistLabelBeginEnd
  {\mcitemaxwidthsubitemform\space}
  {\relax}
  {\relax}

\bibitem[Yu and MacKerell(2017)Yu, and MacKerell]{yu_computer-aided_2017}
Yu,~W.; MacKerell,~A.~D. In \emph{Antibiotics: {Methods} and {Protocols}};
  Sass,~P., Ed.; Methods in {Molecular} {Biology}; Springer: New York, NY,
  2017; pp 85--106\relax
\mciteBstWouldAddEndPuncttrue
\mciteSetBstMidEndSepPunct{\mcitedefaultmidpunct}
{\mcitedefaultendpunct}{\mcitedefaultseppunct}\relax
\EndOfBibitem
\bibitem[Macalino \latin{et~al.}(2015)Macalino, Gosu, Hong, and
  Choi]{macalino_role_2015}
Macalino,~S. J.~Y.; Gosu,~V.; Hong,~S.; Choi,~S. Role of computer-aided drug
  design in modern drug discovery. \emph{Archives of Pharmacal Research}
  \textbf{2015}, \emph{38}, 1686--1701\relax
\mciteBstWouldAddEndPuncttrue
\mciteSetBstMidEndSepPunct{\mcitedefaultmidpunct}
{\mcitedefaultendpunct}{\mcitedefaultseppunct}\relax
\EndOfBibitem
\bibitem[Li \latin{et~al.}(2019)Li, Fu, and Zhang]{li_overview_2019}
Li,~J.; Fu,~A.; Zhang,~L. An {Overview} of {Scoring} {Functions} {Used} for
  {Protein}–{Ligand} {Interactions} in {Molecular} {Docking}.
  \emph{Interdisciplinary Sciences: Computational Life Sciences} \textbf{2019},
  \emph{11}, 320--328\relax
\mciteBstWouldAddEndPuncttrue
\mciteSetBstMidEndSepPunct{\mcitedefaultmidpunct}
{\mcitedefaultendpunct}{\mcitedefaultseppunct}\relax
\EndOfBibitem
\bibitem[Irwin and Shoichet(2016)Irwin, and Shoichet]{irwin_docking_2016}
Irwin,~J.~J.; Shoichet,~B.~K. Docking {Screens} for {Novel} {Ligands}
  {Conferring} {New} {Biology}. \emph{Journal of Medicinal Chemistry}
  \textbf{2016}, \emph{59}, 4103--4120, Publisher: American Chemical
  Society\relax
\mciteBstWouldAddEndPuncttrue
\mciteSetBstMidEndSepPunct{\mcitedefaultmidpunct}
{\mcitedefaultendpunct}{\mcitedefaultseppunct}\relax
\EndOfBibitem
\bibitem[Irwin and Shoichet(2005)Irwin, and Shoichet]{irwin_zinc_2005}
Irwin,~J.~J.; Shoichet,~B.~K. {ZINC} - {A} {Free} {Database} of {Commercially}
  {Available} {Compounds} for {Virtual} {Screening}. \emph{Journal of Chemical
  Information and Modeling} \textbf{2005}, \emph{45}, 177--182, Publisher:
  American Chemical Society\relax
\mciteBstWouldAddEndPuncttrue
\mciteSetBstMidEndSepPunct{\mcitedefaultmidpunct}
{\mcitedefaultendpunct}{\mcitedefaultseppunct}\relax
\EndOfBibitem
\bibitem[Sterling and Irwin(2015)Sterling, and Irwin]{sterling_zinc_2015}
Sterling,~T.; Irwin,~J.~J. {ZINC} 15 – {Ligand} {Discovery} for {Everyone}.
  \emph{Journal of Chemical Information and Modeling} \textbf{2015}, \emph{55},
  2324--2337, Publisher: American Chemical Society\relax
\mciteBstWouldAddEndPuncttrue
\mciteSetBstMidEndSepPunct{\mcitedefaultmidpunct}
{\mcitedefaultendpunct}{\mcitedefaultseppunct}\relax
\EndOfBibitem
\bibitem[noa()]{noauthor_real_nodate}
{REAL} {Database} - {Enamine}.
  \url{https://enamine.net/library-synthesis/real-compounds/real-database},
  Accessed 09/15/2020\relax
\mciteBstWouldAddEndPuncttrue
\mciteSetBstMidEndSepPunct{\mcitedefaultmidpunct}
{\mcitedefaultendpunct}{\mcitedefaultseppunct}\relax
\EndOfBibitem
\bibitem[Knehans \latin{et~al.}(2017)Knehans, Klingler, Kraut, Saller,
  Herrmann, Rippmann, Eiblmaier, Lemmen, and Krier]{knehans_merck_2017}
Knehans,~T.; Klingler,~F.-M.; Kraut,~H.; Saller,~H.; Herrmann,~A.;
  Rippmann,~F.; Eiblmaier,~J.; Lemmen,~C.; Krier,~M. Merck {AcceSSible}
  {InVentory} ({MASSIV}): {In} silico synthesis guided by chemical transforms
  obtained through bootstrapping reaction databases. Abstracts of {Papers} of
  the {American} {Chemical} {Society}. 2017\relax
\mciteBstWouldAddEndPuncttrue
\mciteSetBstMidEndSepPunct{\mcitedefaultmidpunct}
{\mcitedefaultendpunct}{\mcitedefaultseppunct}\relax
\EndOfBibitem
\bibitem[Nicolaou \latin{et~al.}(2016)Nicolaou, Watson, Hu, and
  Wang]{nicolaou_proximal_2016}
Nicolaou,~C.~A.; Watson,~I.~A.; Hu,~H.; Wang,~J. The {Proximal} {Lilly}
  {Collection}: {Mapping}, {Exploring} and {Exploiting} {Feasible} {Chemical}
  {Space}. \emph{Journal of Chemical Information and Modeling} \textbf{2016},
  \emph{56}, 1253--1266, Publisher: American Chemical Society\relax
\mciteBstWouldAddEndPuncttrue
\mciteSetBstMidEndSepPunct{\mcitedefaultmidpunct}
{\mcitedefaultendpunct}{\mcitedefaultseppunct}\relax
\EndOfBibitem
\bibitem[Hu \latin{et~al.}(2012)Hu, Peng, Sutton, Na, Kostrowicki, Yang,
  Thacher, Kong, Mattaparti, Zhou, Gonzalez, Ramirez-Weinhouse, and
  Kuki]{hu_pfizer_2012}
Hu,~Q.; Peng,~Z.; Sutton,~S.~C.; Na,~J.; Kostrowicki,~J.; Yang,~B.;
  Thacher,~T.; Kong,~X.; Mattaparti,~S.; Zhou,~J.~Z.; Gonzalez,~J.;
  Ramirez-Weinhouse,~M.; Kuki,~A. Pfizer {Global} {Virtual} {Library} ({PGVL}):
  {A} {Chemistry} {Design} {Tool} {Powered} by {Experimentally} {Validated}
  {Parallel} {Synthesis} {Information}. \emph{ACS Combinatorial Science}
  \textbf{2012}, \emph{14}, 579--589, Publisher: American Chemical
  Society\relax
\mciteBstWouldAddEndPuncttrue
\mciteSetBstMidEndSepPunct{\mcitedefaultmidpunct}
{\mcitedefaultendpunct}{\mcitedefaultseppunct}\relax
\EndOfBibitem
\bibitem[Clark(2020)]{clark_virtual_2020}
Clark,~D.~E. Virtual {Screening}: {Is} {Bigger} {Always} {Better}? {Or} {Can}
  {Small} {Be} {Beautiful}? \emph{Journal of Chemical Information and Modeling}
  \textbf{2020}, \emph{60}, 4120--4123, Publisher: American Chemical
  Society\relax
\mciteBstWouldAddEndPuncttrue
\mciteSetBstMidEndSepPunct{\mcitedefaultmidpunct}
{\mcitedefaultendpunct}{\mcitedefaultseppunct}\relax
\EndOfBibitem
\bibitem[Gorgulla \latin{et~al.}(2020)Gorgulla, Boeszoermenyi, Wang, Fischer,
  Coote, Padmanabha~Das, Malets, Radchenko, Moroz, Scott, Fackeldey, Hoffmann,
  Iavniuk, Wagner, and Arthanari]{gorgulla_open-source_2020}
Gorgulla,~C.; Boeszoermenyi,~A.; Wang,~Z.-F.; Fischer,~P.~D.; Coote,~P.~W.;
  Padmanabha~Das,~K.~M.; Malets,~Y.~S.; Radchenko,~D.~S.; Moroz,~Y.~S.;
  Scott,~D.~A.; Fackeldey,~K.; Hoffmann,~M.; Iavniuk,~I.; Wagner,~G.;
  Arthanari,~H. An open-source drug discovery platform enables ultra-large
  virtual screens. \emph{Nature} \textbf{2020}, \emph{580}, 663--668, Number:
  7805 Publisher: Nature Publishing Group\relax
\mciteBstWouldAddEndPuncttrue
\mciteSetBstMidEndSepPunct{\mcitedefaultmidpunct}
{\mcitedefaultendpunct}{\mcitedefaultseppunct}\relax
\EndOfBibitem
\bibitem[Lyu \latin{et~al.}(2019)Lyu, Wang, Balius, Singh, Levit, Moroz,
  O’Meara, Che, Algaa, Tolmachova, Tolmachev, Shoichet, Roth, and
  Irwin]{lyu_ultra-large_2019}
Lyu,~J.; Wang,~S.; Balius,~T.~E.; Singh,~I.; Levit,~A.; Moroz,~Y.~S.;
  O’Meara,~M.~J.; Che,~T.; Algaa,~E.; Tolmachova,~K.; Tolmachev,~A.~A.;
  Shoichet,~B.~K.; Roth,~B.~L.; Irwin,~J.~J. Ultra-large library docking for
  discovering new chemotypes. \emph{Nature} \textbf{2019}, \emph{566},
  224--229, Number: 7743 Publisher: Nature Publishing Group\relax
\mciteBstWouldAddEndPuncttrue
\mciteSetBstMidEndSepPunct{\mcitedefaultmidpunct}
{\mcitedefaultendpunct}{\mcitedefaultseppunct}\relax
\EndOfBibitem
\bibitem[Acharya \latin{et~al.}(2020)Acharya, Agarwal, Baker, Baudry, Bhowmik,
  Boehm, Byler, Coates, Chen, Cooper, Demerdash, Daidone, Eblen, R.~Ellingson,
  Forli, Glaser, Gumbart, Gunnels, Hernandez, Irle, Larkin, Lawrence, LeGrand,
  Liu, C.~Mitchell, Park, Parks, Pavlova, Petridis, Poole, Pouchard,
  Ramanathan, Rogers, Santos-Martins, Scheinberg, Sedova, Shen, Smith, Smith,
  Soto, Tsaris, Thavappiragasam, Tillack, Vermaas, Vuong, Yin, Yoo, Zahran, and
  Zanetti-Polzi]{acharya_supercomputer-based_2020}
Acharya,~A. \latin{et~al.}  Supercomputer-{Based} {Ensemble} {Docking} {Drug}
  {Discovery} {Pipeline} with {Application} to {Covid}-19. \textbf{2020},
  Publisher: ChemRxiv\relax
\mciteBstWouldAddEndPuncttrue
\mciteSetBstMidEndSepPunct{\mcitedefaultmidpunct}
{\mcitedefaultendpunct}{\mcitedefaultseppunct}\relax
\EndOfBibitem
\bibitem[{Mark McGann} and {OpenEye Scientific}(2019){Mark McGann}, and
  {OpenEye Scientific}]{mark_mcgann_gigadocking_2019}
{Mark McGann},; {OpenEye Scientific}, {GigaDocking}™ - {Structure} {Based}
  {Virtual} {Screening} of {Over} 1 {Billion} {Molecules} {Webinar}. 2019;
  \url{https://www.eyesopen.com/webinars/giga-docking-structure-based-virtual-screening},
  Accessed 09/01/2020\relax
\mciteBstWouldAddEndPuncttrue
\mciteSetBstMidEndSepPunct{\mcitedefaultmidpunct}
{\mcitedefaultendpunct}{\mcitedefaultseppunct}\relax
\EndOfBibitem
\bibitem[Frazier(2018)]{frazier_tutorial_2018}
Frazier,~P.~I. A {Tutorial} on {Bayesian} {Optimization}.
  \emph{arXiv:1807.02811 [cs, math, stat]} \textbf{2018}, arXiv:
  1807.02811\relax
\mciteBstWouldAddEndPuncttrue
\mciteSetBstMidEndSepPunct{\mcitedefaultmidpunct}
{\mcitedefaultendpunct}{\mcitedefaultseppunct}\relax
\EndOfBibitem
\bibitem[Balachandran \latin{et~al.}(2016)Balachandran, Xue, Theiler, Hogden,
  and Lookman]{balachandran_adaptive_2016}
Balachandran,~P.~V.; Xue,~D.; Theiler,~J.; Hogden,~J.; Lookman,~T. Adaptive
  {Strategies} for {Materials} {Design} using {Uncertainties}. \emph{Scientific
  Reports} \textbf{2016}, \emph{6}, 19660, Number: 1 Publisher: Nature
  Publishing Group\relax
\mciteBstWouldAddEndPuncttrue
\mciteSetBstMidEndSepPunct{\mcitedefaultmidpunct}
{\mcitedefaultendpunct}{\mcitedefaultseppunct}\relax
\EndOfBibitem
\bibitem[Gubaev \latin{et~al.}(2019)Gubaev, Podryabinkin, Hart, and
  Shapeev]{gubaev_accelerating_2019}
Gubaev,~K.; Podryabinkin,~E.~V.; Hart,~G. L.~W.; Shapeev,~A.~V. Accelerating
  high-throughput searches for new alloys with active learning of interatomic
  potentials. \emph{Computational Materials Science} \textbf{2019}, \emph{156},
  148--156\relax
\mciteBstWouldAddEndPuncttrue
\mciteSetBstMidEndSepPunct{\mcitedefaultmidpunct}
{\mcitedefaultendpunct}{\mcitedefaultseppunct}\relax
\EndOfBibitem
\bibitem[Xue \latin{et~al.}(2016)Xue, Balachandran, Hogden, Theiler, Xue, and
  Lookman]{xue_accelerated_2016}
Xue,~D.; Balachandran,~P.~V.; Hogden,~J.; Theiler,~J.; Xue,~D.; Lookman,~T.
  Accelerated search for materials with targeted properties by adaptive design.
  \emph{Nature Communications} \textbf{2016}, \emph{7}, 11241, Number: 1
  Publisher: Nature Publishing Group\relax
\mciteBstWouldAddEndPuncttrue
\mciteSetBstMidEndSepPunct{\mcitedefaultmidpunct}
{\mcitedefaultendpunct}{\mcitedefaultseppunct}\relax
\EndOfBibitem
\bibitem[Montoya \latin{et~al.}(2020)Montoya, Winther, Flores, Bligaard,
  Hummelshøj, and Aykol]{montoya_autonomous_2020}
Montoya,~J.~H.; Winther,~K.~T.; Flores,~R.~A.; Bligaard,~T.;
  Hummelshøj,~J.~S.; Aykol,~M. Autonomous intelligent agents for accelerated
  materials discovery. \emph{Chemical Science} \textbf{2020}, \emph{11},
  8517--8532, Publisher: The Royal Society of Chemistry\relax
\mciteBstWouldAddEndPuncttrue
\mciteSetBstMidEndSepPunct{\mcitedefaultmidpunct}
{\mcitedefaultendpunct}{\mcitedefaultseppunct}\relax
\EndOfBibitem
\bibitem[Bilsland \latin{et~al.}()Bilsland, Sparkes, Williams, Moss, de~Clare,
  Pir, Rowland, Aubrey, Pateman, Young, Carrington, King, and
  Oliver]{bilsland_yeast-based_nodate}
Bilsland,~E.; Sparkes,~A.; Williams,~K.; Moss,~H.~J.; de~Clare,~M.; Pir,~P.;
  Rowland,~J.; Aubrey,~W.; Pateman,~R.; Young,~M.; Carrington,~M.; King,~R.~D.;
  Oliver,~S.~G. Yeast-based automated high-throughput screens to identify
  anti-parasitic lead compounds. \emph{Open Biology} \emph{3}, 120158,
  Publisher: Royal Society\relax
\mciteBstWouldAddEndPuncttrue
\mciteSetBstMidEndSepPunct{\mcitedefaultmidpunct}
{\mcitedefaultendpunct}{\mcitedefaultseppunct}\relax
\EndOfBibitem
\bibitem[Czechtizky \latin{et~al.}(2013)Czechtizky, Dedio, Desai, Dixon,
  Farrant, Feng, Morgan, Parry, Ramjee, Selway, Schmidt, Tarver, and
  Wright]{czechtizky_integrated_2013}
Czechtizky,~W.; Dedio,~J.; Desai,~B.; Dixon,~K.; Farrant,~E.; Feng,~Q.;
  Morgan,~T.; Parry,~D.~M.; Ramjee,~M.~K.; Selway,~C.~N.; Schmidt,~T.;
  Tarver,~G.~J.; Wright,~A.~G. Integrated {Synthesis} and {Testing} of
  {Substituted} {Xanthine} {Based} {DPP4} {Inhibitors}: {Application} to {Drug}
  {Discovery}. \emph{ACS Medicinal Chemistry Letters} \textbf{2013}, \emph{4},
  768--772, Publisher: American Chemical Society\relax
\mciteBstWouldAddEndPuncttrue
\mciteSetBstMidEndSepPunct{\mcitedefaultmidpunct}
{\mcitedefaultendpunct}{\mcitedefaultseppunct}\relax
\EndOfBibitem
\bibitem[Williams \latin{et~al.}(2015)Williams, Bilsland, Sparkes, Aubrey,
  Young, Soldatova, De~Grave, Ramon, de~Clare, Sirawaraporn, Oliver, and
  King]{williams_cheaper_2015}
Williams,~K.; Bilsland,~E.; Sparkes,~A.; Aubrey,~W.; Young,~M.;
  Soldatova,~L.~N.; De~Grave,~K.; Ramon,~J.; de~Clare,~M.; Sirawaraporn,~W.;
  Oliver,~S.~G.; King,~R.~D. Cheaper faster drug development validated by the
  repositioning of drugs against neglected tropical diseases. \emph{Journal of
  The Royal Society Interface} \textbf{2015}, \emph{12}, 20141289, Publisher:
  Royal Society\relax
\mciteBstWouldAddEndPuncttrue
\mciteSetBstMidEndSepPunct{\mcitedefaultmidpunct}
{\mcitedefaultendpunct}{\mcitedefaultseppunct}\relax
\EndOfBibitem
\bibitem[Janet \latin{et~al.}(2020)Janet, Ramesh, Duan, and
  Kulik]{janet_accurate_2020}
Janet,~J.~P.; Ramesh,~S.; Duan,~C.; Kulik,~H.~J. Accurate {Multiobjective}
  {Design} in a {Space} of {Millions} of {Transition} {Metal} {Complexes} with
  {Neural}-{Network}-{Driven} {Efficient} {Global} {Optimization}. \emph{ACS
  Central Science} \textbf{2020}, acscentsci.0c00026\relax
\mciteBstWouldAddEndPuncttrue
\mciteSetBstMidEndSepPunct{\mcitedefaultmidpunct}
{\mcitedefaultendpunct}{\mcitedefaultseppunct}\relax
\EndOfBibitem
\bibitem[Ghanakota \latin{et~al.}(2020)Ghanakota, Bos, Konze, Staker, Marques,
  Marshall, Leswing, Abel, and Bhat]{ghanakota_combining_2020}
Ghanakota,~P.; Bos,~P.; Konze,~K.; Staker,~J.; Marques,~G.; Marshall,~K.;
  Leswing,~K.; Abel,~R.; Bhat,~S. Combining {Cloud}-{Based} {Free} {Energy}
  {Calculations}, {Synthetically} {Aware} {Enumerations} and {Goal}-{Directed}
  {Generative} {Machine} {Learning} for {Rapid} {Large}-{Scale} {Chemical}
  {Exploration} and {Optimization}. \textbf{2020}, Publisher: ChemRxiv\relax
\mciteBstWouldAddEndPuncttrue
\mciteSetBstMidEndSepPunct{\mcitedefaultmidpunct}
{\mcitedefaultendpunct}{\mcitedefaultseppunct}\relax
\EndOfBibitem
\bibitem[Konze \latin{et~al.}(2019)Konze, Bos, Dahlgren, Leswing,
  Tubert-Brohman, Bortolato, Robbason, Abel, and
  Bhat]{konze_reaction-based_2019}
Konze,~K.~D.; Bos,~P.~H.; Dahlgren,~M.~K.; Leswing,~K.; Tubert-Brohman,~I.;
  Bortolato,~A.; Robbason,~B.; Abel,~R.; Bhat,~S. Reaction-{Based}
  {Enumeration}, {Active} {Learning}, and {Free} {Energy} {Calculations} {To}
  {Rapidly} {Explore} {Synthetically} {Tractable} {Chemical} {Space} and
  {Optimize} {Potency} of {Cyclin}-{Dependent} {Kinase} 2 {Inhibitors}.
  \emph{Journal of Chemical Information and Modeling} \textbf{2019}, \emph{59},
  3782--3793\relax
\mciteBstWouldAddEndPuncttrue
\mciteSetBstMidEndSepPunct{\mcitedefaultmidpunct}
{\mcitedefaultendpunct}{\mcitedefaultseppunct}\relax
\EndOfBibitem
\bibitem[Gentile \latin{et~al.}(2019)Gentile, Agrawal, Hsing, Ban, Norinder,
  Gleave, and Cherkasov]{gentile_deep_2019}
Gentile,~F.; Agrawal,~V.; Hsing,~M.; Ban,~F.; Norinder,~U.; Gleave,~M.~E.;
  Cherkasov,~A. Deep {Docking} - a {Deep} {Learning} {Approach} for {Virtual}
  {Screening} of {Big} {Chemical} {Datasets}. \emph{bioRxiv} \textbf{2019},
  2019.12.15.877316, Publisher: Cold Spring Harbor Laboratory Section: New
  Results\relax
\mciteBstWouldAddEndPuncttrue
\mciteSetBstMidEndSepPunct{\mcitedefaultmidpunct}
{\mcitedefaultendpunct}{\mcitedefaultseppunct}\relax
\EndOfBibitem
\bibitem[Pyzer-Knapp(2020)]{pyzer-knapp_using_2020}
Pyzer-Knapp,~E.~O. Using {Bayesian} {Optimization} to {Accelerate} {Virtual}
  {Screening} for the {Discovery} of {Therapeutics} {Appropriate} for
  {Repurposing} for {COVID}-19. \emph{arXiv:2005.07121 [cs, q-bio]}
  \textbf{2020}, arXiv: 2005.07121\relax
\mciteBstWouldAddEndPuncttrue
\mciteSetBstMidEndSepPunct{\mcitedefaultmidpunct}
{\mcitedefaultendpunct}{\mcitedefaultseppunct}\relax
\EndOfBibitem
\bibitem[Hernández-Lobato \latin{et~al.}(2017)Hernández-Lobato, Requeima,
  Pyzer-Knapp, and Aspuru-Guzik]{hernandez-lobato_parallel_2017}
Hernández-Lobato,~J.~M.; Requeima,~J.; Pyzer-Knapp,~E.~O.; Aspuru-Guzik,~A.
  Parallel and {Distributed} {Thompson} {Sampling} for {Large}-scale
  {Accelerated} {Exploration} of {Chemical} {Space}. \emph{arXiv:1706.01825
  [stat]} \textbf{2017}, arXiv: 1706.01825\relax
\mciteBstWouldAddEndPuncttrue
\mciteSetBstMidEndSepPunct{\mcitedefaultmidpunct}
{\mcitedefaultendpunct}{\mcitedefaultseppunct}\relax
\EndOfBibitem
\bibitem[Gibbs and MacKay(1997)Gibbs, and MacKay]{gibbs_efficient_1997}
Gibbs,~M.; MacKay,~D. J.~C. \emph{Efficient {Implementation} of {Gaussian}
  {Processes}}; 1997\relax
\mciteBstWouldAddEndPuncttrue
\mciteSetBstMidEndSepPunct{\mcitedefaultmidpunct}
{\mcitedefaultendpunct}{\mcitedefaultseppunct}\relax
\EndOfBibitem
\bibitem[Naik \latin{et~al.}(2015)Naik, Raichurkar, Bandodkar, Varun, Bhat,
  Kalkhambkar, Murugan, Menon, Bhat, Paul, Iyer, Hussein, Tucker, Vogtherr,
  Embrey, McMiken, Prasad, Gill, Ugarkar, Venkatraman, Read, and
  Panda]{naik_structure_2015}
Naik,~M. \latin{et~al.}  Structure {Guided} {Lead} {Generation} for {M}.
  tuberculosis {Thymidylate} {Kinase} ({Mtb} {TMK}): {Discovery} of
  3-{Cyanopyridone} and 1,6-{Naphthyridin}-2-one as {Potent} {Inhibitors}.
  \emph{Journal of Medicinal Chemistry} \textbf{2015}, \emph{58}, 753--766,
  Publisher: American Chemical Society\relax
\mciteBstWouldAddEndPuncttrue
\mciteSetBstMidEndSepPunct{\mcitedefaultmidpunct}
{\mcitedefaultendpunct}{\mcitedefaultseppunct}\relax
\EndOfBibitem
\bibitem[Trott and Olson(2010)Trott, and Olson]{trott_autodock_2010}
Trott,~O.; Olson,~A.~J. {AutoDock} {Vina}: {Improving} the speed and accuracy
  of docking with a new scoring function, efficient optimization, and
  multithreading. \emph{Journal of Computational Chemistry} \textbf{2010},
  \emph{31}, 455--461, \_eprint:
  https://onlinelibrary.wiley.com/doi/pdf/10.1002/jcc.21334\relax
\mciteBstWouldAddEndPuncttrue
\mciteSetBstMidEndSepPunct{\mcitedefaultmidpunct}
{\mcitedefaultendpunct}{\mcitedefaultseppunct}\relax
\EndOfBibitem
\bibitem[Yang \latin{et~al.}(2019)Yang, Swanson, Jin, Coley, Eiden, Gao,
  Guzman-Perez, Hopper, Kelley, Mathea, Palmer, Settels, Jaakkola, Jensen, and
  Barzilay]{yang_analyzing_2019}
Yang,~K.; Swanson,~K.; Jin,~W.; Coley,~C.; Eiden,~P.; Gao,~H.;
  Guzman-Perez,~A.; Hopper,~T.; Kelley,~B.; Mathea,~M.; Palmer,~A.;
  Settels,~V.; Jaakkola,~T.; Jensen,~K.; Barzilay,~R. Analyzing {Learned}
  {Molecular} {Representations} for {Property} {Prediction}. \emph{Journal of
  Chemical Information and Modeling} \textbf{2019}, \emph{59}, 3370--3388\relax
\mciteBstWouldAddEndPuncttrue
\mciteSetBstMidEndSepPunct{\mcitedefaultmidpunct}
{\mcitedefaultendpunct}{\mcitedefaultseppunct}\relax
\EndOfBibitem
\bibitem[Balius \latin{et~al.}(2018)Balius, Lyu, Shoichet, and
  Irwin]{balius_ampc_screen_tablecsvgz_2018}
Balius,~T.; Lyu,~J.; Shoichet,~B.~K.; Irwin,~J.~J.
  {AmpC}\_screen\_table.csv.gz. 2018;
  \url{https://figshare.com/articles/AmpC_screen_table_csv_gz/7359626},
  Accessed 06/01/2020\relax
\mciteBstWouldAddEndPuncttrue
\mciteSetBstMidEndSepPunct{\mcitedefaultmidpunct}
{\mcitedefaultendpunct}{\mcitedefaultseppunct}\relax
\EndOfBibitem
\bibitem[McInnes \latin{et~al.}(2020)McInnes, Healy, and
  Melville]{mcinnes_umap_2020}
McInnes,~L.; Healy,~J.; Melville,~J. {UMAP}: {Uniform} {Manifold}
  {Approximation} and {Projection} for {Dimension} {Reduction}.
  \emph{arXiv:1802.03426 [cs, stat]} \textbf{2020}, arXiv: 1802.03426\relax
\mciteBstWouldAddEndPuncttrue
\mciteSetBstMidEndSepPunct{\mcitedefaultmidpunct}
{\mcitedefaultendpunct}{\mcitedefaultseppunct}\relax
\EndOfBibitem
\bibitem[Desautels \latin{et~al.}(2014)Desautels, Krause, and
  Burdick]{desautels_parallelizing_2014}
Desautels,~T.; Krause,~A.; Burdick,~J.~W. Parallelizing
  {Exploration}-{Exploitation} {Tradeoffs} in {Gaussian} {Process} {Bandit}
  {Optimization}. \emph{Journal of Machine Learning Research} \textbf{2014},
  \emph{15}, 4053--4103\relax
\mciteBstWouldAddEndPuncttrue
\mciteSetBstMidEndSepPunct{\mcitedefaultmidpunct}
{\mcitedefaultendpunct}{\mcitedefaultseppunct}\relax
\EndOfBibitem
\bibitem[Tsymbalov \latin{et~al.}(2019)Tsymbalov, Makarychev, Shapeev, and
  Panov]{tsymbalov_deeper_2019}
Tsymbalov,~E.; Makarychev,~S.; Shapeev,~A.; Panov,~M. Deeper {Connections}
  between {Neural} {Networks} and {Gaussian} {Processes} {Speed}-up {Active}
  {Learning}. \emph{Proceedings of the Twenty-Eighth International Joint
  Conference on Artificial Intelligence} \textbf{2019}, 3599--3605, arXiv:
  1902.10350\relax
\mciteBstWouldAddEndPuncttrue
\mciteSetBstMidEndSepPunct{\mcitedefaultmidpunct}
{\mcitedefaultendpunct}{\mcitedefaultseppunct}\relax
\EndOfBibitem
\bibitem[Hirschfeld \latin{et~al.}(2020)Hirschfeld, Swanson, Yang, Barzilay,
  and Coley]{hirschfeld_uncertainty_2020}
Hirschfeld,~L.; Swanson,~K.; Yang,~K.; Barzilay,~R.; Coley,~C.~W. Uncertainty
  {Quantification} {Using} {Neural} {Networks} for {Molecular} {Property}
  {Prediction}. \emph{Journal of Chemical Information and Modeling}
  \textbf{2020}, \emph{60}, 3770--3780\relax
\mciteBstWouldAddEndPuncttrue
\mciteSetBstMidEndSepPunct{\mcitedefaultmidpunct}
{\mcitedefaultendpunct}{\mcitedefaultseppunct}\relax
\EndOfBibitem
\bibitem[Wang \latin{et~al.}(2018)Wang, Gehring, Kohli, and
  Jegelka]{wang_batched_2018}
Wang,~Z.; Gehring,~C.; Kohli,~P.; Jegelka,~S. Batched {Large}-scale {Bayesian}
  {Optimization} in {High}-dimensional {Spaces}. \emph{arXiv:1706.01445 [cs,
  math, stat]} \textbf{2018}, arXiv: 1706.01445\relax
\mciteBstWouldAddEndPuncttrue
\mciteSetBstMidEndSepPunct{\mcitedefaultmidpunct}
{\mcitedefaultendpunct}{\mcitedefaultseppunct}\relax
\EndOfBibitem
\bibitem[Schuffenhauer \latin{et~al.}(2020)Schuffenhauer, Schneider,
  Hintermann, Auld, Blank, Cotesta, Engeloch, Fechner, Gaul, Giovannoni,
  Jansen, Joslin, Krastel, Lounkine, Manchester, Monovich, Pelliccioli,
  Schwarze, Shultz, Stiefl, and Baeschlin]{schuffenhauer_evolution_2020}
Schuffenhauer,~A. \latin{et~al.}  Evolution of {Novartis}’ {Small} {Molecule}
  {Screening} {Deck} {Design}. \emph{Journal of Medicinal Chemistry}
  \textbf{2020}, Publisher: American Chemical Society\relax
\mciteBstWouldAddEndPuncttrue
\mciteSetBstMidEndSepPunct{\mcitedefaultmidpunct}
{\mcitedefaultendpunct}{\mcitedefaultseppunct}\relax
\EndOfBibitem
\bibitem[Shahriari \latin{et~al.}(2016)Shahriari, Swersky, Wang, Adams, and
  de~Freitas]{shahriari_taking_2016}
Shahriari,~B.; Swersky,~K.; Wang,~Z.; Adams,~R.~P.; de~Freitas,~N. Taking the
  {Human} {Out} of the {Loop}: {A} {Review} of {Bayesian} {Optimization}.
  \emph{Proceedings of the IEEE} \textbf{2016}, \emph{104}, 148--175\relax
\mciteBstWouldAddEndPuncttrue
\mciteSetBstMidEndSepPunct{\mcitedefaultmidpunct}
{\mcitedefaultendpunct}{\mcitedefaultseppunct}\relax
\EndOfBibitem
\bibitem[Svetnik \latin{et~al.}(2003)Svetnik, Liaw, Tong, Culberson, Sheridan,
  and Feuston]{svetnik2003random}
Svetnik,~V.; Liaw,~A.; Tong,~C.; Culberson,~J.~C.; Sheridan,~R.~P.;
  Feuston,~B.~P. Random forest: a classification and regression tool for
  compound classification and QSAR modeling. \emph{Journal of chemical
  information and computer sciences} \textbf{2003}, \emph{43}, 1947--1958\relax
\mciteBstWouldAddEndPuncttrue
\mciteSetBstMidEndSepPunct{\mcitedefaultmidpunct}
{\mcitedefaultendpunct}{\mcitedefaultseppunct}\relax
\EndOfBibitem
\bibitem[Pedregosa \latin{et~al.}(2011)Pedregosa, Varoquaux, Gramfort, Michel,
  Thirion, Grisel, Blondel, Prettenhofer, Weiss, Dubourg, \latin{et~al.}
  others]{pedregosa_scikit-learn_nodate}
Pedregosa,~F.; Varoquaux,~G.; Gramfort,~A.; Michel,~V.; Thirion,~B.;
  Grisel,~O.; Blondel,~M.; Prettenhofer,~P.; Weiss,~R.; Dubourg,~V.,
  \latin{et~al.}  Scikit-learn: Machine learning in Python. \emph{the Journal
  of machine Learning research} \textbf{2011}, \emph{12}, 2825--2830\relax
\mciteBstWouldAddEndPuncttrue
\mciteSetBstMidEndSepPunct{\mcitedefaultmidpunct}
{\mcitedefaultendpunct}{\mcitedefaultseppunct}\relax
\EndOfBibitem
\bibitem[Abadi \latin{et~al.}(2016)Abadi, Agarwal, Barham, Brevdo, Chen, Citro,
  Corrado, Davis, Dean, Devin, \latin{et~al.} others]{abadi_tensorflow_nodate}
Abadi,~M.; Agarwal,~A.; Barham,~P.; Brevdo,~E.; Chen,~Z.; Citro,~C.;
  Corrado,~G.~S.; Davis,~A.; Dean,~J.; Devin,~M., \latin{et~al.}  Tensorflow:
  Large-scale machine learning on heterogeneous distributed systems.
  \emph{arXiv preprint arXiv:1603.04467} \textbf{2016}, \relax
\mciteBstWouldAddEndPunctfalse
\mciteSetBstMidEndSepPunct{\mcitedefaultmidpunct}
{}{\mcitedefaultseppunct}\relax
\EndOfBibitem
\bibitem[Carhart \latin{et~al.}(1985)Carhart, Smith, and
  Venkataraghavan]{carhart_atom_1985}
Carhart,~R.~E.; Smith,~D.~H.; Venkataraghavan,~R. Atom pairs as molecular
  features in structure-activity studies: definition and applications.
  \emph{Journal of Chemical Information and Computer Sciences} \textbf{1985},
  \emph{25}, 64--73, Publisher: American Chemical Society\relax
\mciteBstWouldAddEndPuncttrue
\mciteSetBstMidEndSepPunct{\mcitedefaultmidpunct}
{\mcitedefaultendpunct}{\mcitedefaultseppunct}\relax
\EndOfBibitem
\bibitem[Bajusz \latin{et~al.}(2017)Bajusz, Rácz, and
  Héberger]{bajusz_chemical_2017}
Bajusz,~D.; Rácz,~A.; Héberger,~K. \emph{Reference {Module} in {Chemistry},
  {Molecular} {Sciences} and {Chemical} {Engineering}}; 2017; Journal
  Abbreviation: Reference Module in Chemistry, Molecular Sciences and Chemical
  Engineering\relax
\mciteBstWouldAddEndPuncttrue
\mciteSetBstMidEndSepPunct{\mcitedefaultmidpunct}
{\mcitedefaultendpunct}{\mcitedefaultseppunct}\relax
\EndOfBibitem
\bibitem[Paszke \latin{et~al.}(2019)Paszke, Gross, Massa, Lerer, Bradbury,
  Chanan, Killeen, Lin, Gimelshein, Antiga, \latin{et~al.}
  others]{paszke_pytorch_nodate}
Paszke,~A.; Gross,~S.; Massa,~F.; Lerer,~A.; Bradbury,~J.; Chanan,~G.;
  Killeen,~T.; Lin,~Z.; Gimelshein,~N.; Antiga,~L., \latin{et~al.}  Pytorch: An
  imperative style, high-performance deep learning library. Advances in neural
  information processing systems. 2019; pp 8026--8037\relax
\mciteBstWouldAddEndPuncttrue
\mciteSetBstMidEndSepPunct{\mcitedefaultmidpunct}
{\mcitedefaultendpunct}{\mcitedefaultseppunct}\relax
\EndOfBibitem
\bibitem[Vaswani \latin{et~al.}(2017)Vaswani, Shazeer, Parmar, Uszkoreit,
  Jones, Gomez, Kaiser, and Polosukhin]{vaswani_attention_2017}
Vaswani,~A.; Shazeer,~N.; Parmar,~N.; Uszkoreit,~J.; Jones,~L.; Gomez,~A.~N.;
  Kaiser,~L.; Polosukhin,~I. Attention {Is} {All} {You} {Need}.
  \emph{arXiv:1706.03762 [cs]} \textbf{2017}, arXiv: 1706.03762\relax
\mciteBstWouldAddEndPuncttrue
\mciteSetBstMidEndSepPunct{\mcitedefaultmidpunct}
{\mcitedefaultendpunct}{\mcitedefaultseppunct}\relax
\EndOfBibitem
\bibitem[Nix and Weigend(1994)Nix, and Weigend]{nix_estimating_1994}
Nix,~D.~A.; Weigend,~A.~S. Estimating the mean and variance of the target
  probability distribution. Proceedings of 1994 {IEEE} {International}
  {Conference} on {Neural} {Networks} ({ICNN}'94). 1994; pp 55--60 vol.1\relax
\mciteBstWouldAddEndPuncttrue
\mciteSetBstMidEndSepPunct{\mcitedefaultmidpunct}
{\mcitedefaultendpunct}{\mcitedefaultseppunct}\relax
\EndOfBibitem
\bibitem[noa()]{noauthor_diversity_nodate}
Diversity {Libraries} - {Enamine}.
  \url{https://enamine.net/hit-finding/diversity-libraries}, Accessed
  04/01/2020\relax
\mciteBstWouldAddEndPuncttrue
\mciteSetBstMidEndSepPunct{\mcitedefaultmidpunct}
{\mcitedefaultendpunct}{\mcitedefaultseppunct}\relax
\EndOfBibitem
\bibitem[noa()]{noauthor_hts_nodate}
{HTS} {Collection} - {Enamine}.
  \url{https://enamine.net/hit-finding/compound-collections/screening-collection/hts-collection},
  Accessed 04/01/2020\relax
\mciteBstWouldAddEndPuncttrue
\mciteSetBstMidEndSepPunct{\mcitedefaultmidpunct}
{\mcitedefaultendpunct}{\mcitedefaultseppunct}\relax
\EndOfBibitem
\bibitem[noa()]{noauthor_rdkit_nodate}
{RDKit}. \url{http://rdkit.org/}, Accessed 10/20/2020\relax
\mciteBstWouldAddEndPuncttrue
\mciteSetBstMidEndSepPunct{\mcitedefaultmidpunct}
{\mcitedefaultendpunct}{\mcitedefaultseppunct}\relax
\EndOfBibitem
\bibitem[O'Boyle \latin{et~al.}(2011)O'Boyle, Banck, James, Morley,
  Vandermeersch, and Hutchison]{oboyle_open_2011}
O'Boyle,~N.~M.; Banck,~M.; James,~C.~A.; Morley,~C.; Vandermeersch,~T.;
  Hutchison,~G.~R. Open {Babel}: {An} open chemical toolbox. \emph{Journal of
  Cheminformatics} \textbf{2011}, \emph{3}, 33\relax
\mciteBstWouldAddEndPuncttrue
\mciteSetBstMidEndSepPunct{\mcitedefaultmidpunct}
{\mcitedefaultendpunct}{\mcitedefaultseppunct}\relax
\EndOfBibitem
\bibitem[Eastman \latin{et~al.}(2017)Eastman, Swails, Chodera, McGibbon, Zhao,
  Beauchamp, Wang, Simmonett, Harrigan, Stern, Wiewiora, Brooks, and
  Pande]{eastman_openmm_2017}
Eastman,~P.; Swails,~J.; Chodera,~J.~D.; McGibbon,~R.~T.; Zhao,~Y.;
  Beauchamp,~K.~A.; Wang,~L.-P.; Simmonett,~A.~C.; Harrigan,~M.~P.;
  Stern,~C.~D.; Wiewiora,~R.~P.; Brooks,~B.~R.; Pande,~V.~S. {OpenMM} 7:
  {Rapid} development of high performance algorithms for molecular dynamics.
  \emph{PLOS Computational Biology} \textbf{2017}, \emph{13}, 1--17, Publisher:
  Public Library of Science\relax
\mciteBstWouldAddEndPuncttrue
\mciteSetBstMidEndSepPunct{\mcitedefaultmidpunct}
{\mcitedefaultendpunct}{\mcitedefaultseppunct}\relax
\EndOfBibitem
\end{mcitethebibliography}

\clearpage

\renewcommand{\thefigure}{S\arabic{figure}}
\renewcommand{\thetable}{S\arabic{table}}
\setcounter{figure}{0} 
\setcounter{table}{0} 
\setcounter{page}{1}

\begin{center}
    \textsf{\LARGE{\textbf{Supporting Information}}} \\
    \vspace{0.5cm}
    \textsf{\Large{\textbf{Accelerating High-Throughput Virtual Screening Through Molecular Pool-Based Active Learning}}} \\
    \vspace{0.6cm}
    \textsf{\large{David E. Graff,$^\dagger$ Eugene Shakhnovich,$^\dagger$ Connor W. Coley$^{\ast,\ddagger}$}}\\
    \vspace{0.6cm}
    $\dagger$\textit{Department of Chemistry and Chemical Biology, Harvard University, Cambridge, MA} \\
    \vspace{0.3cm}
    $\ddagger$\textit{Department of Chemical Engineering, MIT, Cambridge, MA} \\
    \vspace{0.3cm}
    \textsf{E-mail: ccoley@mit.edu}
\end{center}

\vspace{0.5cm}

\normalsize

\section{Additional Methods}\label{sec:addl_methods}

\subsection{Elaboration on software design}\label{subsec:software_design}
The design choices detailed in the \nameref{para:software_design} paragraph of the \nameref{sec:methods} section are critical to both the testing and extension of the \tt{MolPAL} software. Namely, the decision to rely on the \tt{MoleculePool}, \tt{Model}, \tt{Acquirer}, and \tt{Objective} helper classes enables the rapid and facile testing of different combinations of model architectures and acquisition strategies for a given objective optimization. This choice also enables the straightforward extension of \tt{MolPAL} with new surrogate model architectures, acquisition metrics, and objective functions. The \tt{Model} and \tt{Objective} are both defined as minimal abstract base classes built around an adapter design pattern. This enables the simple interfacing of popular machine learning learning libraries (e.g., {Scikit-Learn}, {PyTorch}, and {TensorFlow}) via the \tt{Model} and virtual screening software via the \tt{Objective} with the \tt{Explorer} class. The \tt{MoleculePool} is primarily an abstraction of a list of molecules stored. This class stores both a molecule's SMILES string and, if necessary, its precalculated fingerprint. The fingerprint is used for clustering, if desired, and as input to models expecting vectors as inputs (e.g., RF and NN models) during the model inference step. The data stored by the \tt{MoleculePool} is all stored on disk to enable the seamless application of \tt{MolPAL} to all-sizes of virtual libraries. Molecular graphs, the input to the MPN model, are not capable of being stored either in memory or on disk due to their large memory footprint in their current implementation. As such, they are recalculated as necessary.

\begin{figure}
    \centering
    \includegraphics[width=0.7\textwidth]{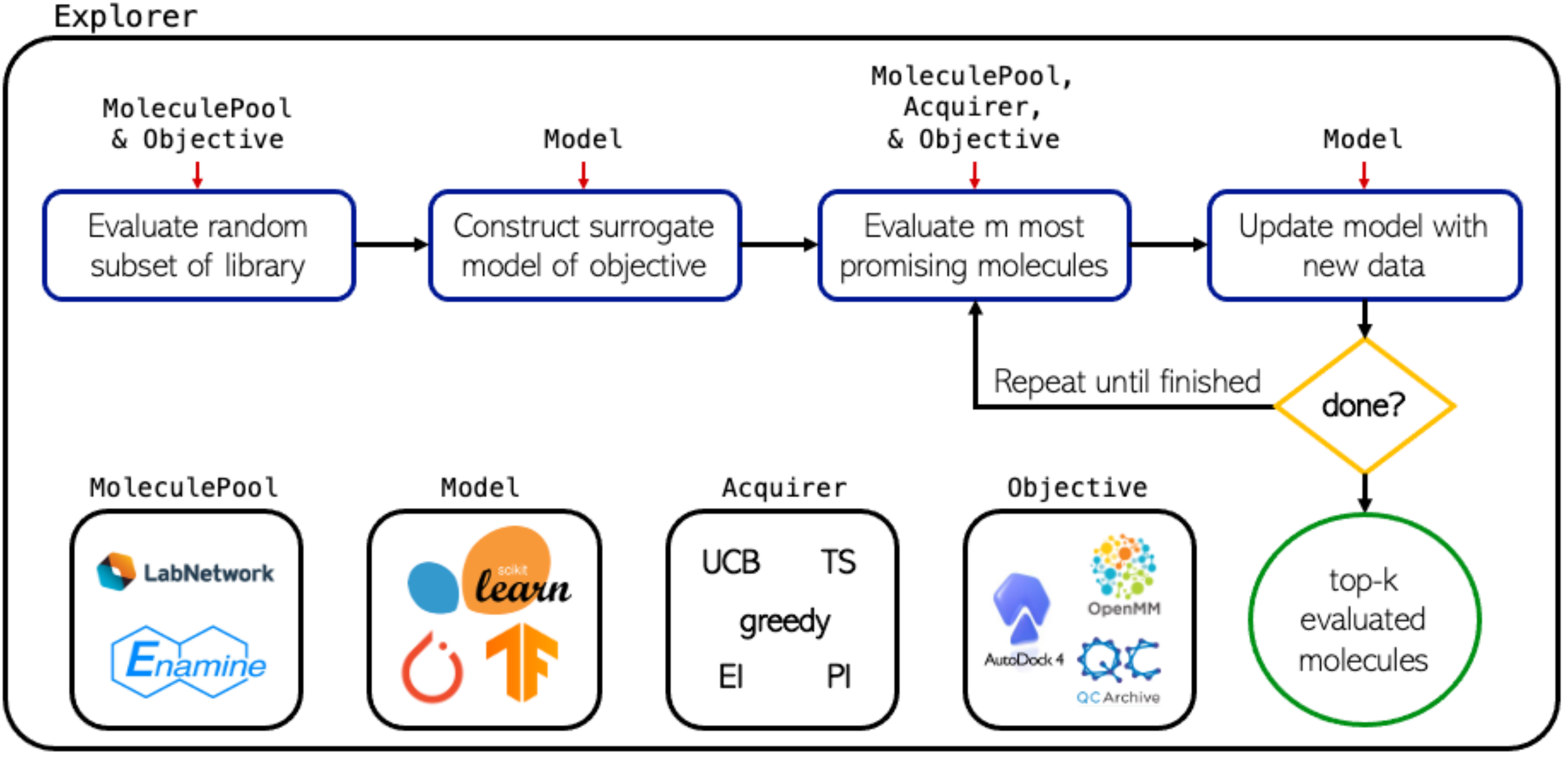}
    \caption{Overview of the MolPAL software structure and workflow.}
    \label{fig:molpal_schema}
\end{figure}

\subsection{Alternative surrogate models}
\paragraph{Feed forward neural network models}
Two alternative NN models were defined for confidence estimation purposes: an ensemble model and a mean-variance estimation (MVE) model. The ensemble model was the same as the base model, with the only difference being that an ensemble of five models was trained. Each of the trained models was used for inference, and these five separate predictions were averaged and a variance taken to produce both a mean predicted value and an uncertainty estimate, respectively. The mean-variance estimation model used an output layer size of two, the learning rate was increased to 0.05 from 0.01, and the same loss function from the MPN-MVE  was used (Equation~\ref{eq:mve_loss}). Neither of these alternate models was used for experiments due to their lower performance as compared to the dropout model.

\paragraph{Directed-message passing neural network models}
An MPN dropout model was also defined for confidence estimation purposes. This model was built similar to the NN dropout model, with the key difference being that the dropout layer was prepended to the hidden layer. Again, a dropout probability of 0.2 was used and dropout was performed during model inference. Mean predicted values were calculated by averaging 10 forward passes through the model and the variance of these predictions was used to as the predicted uncertainty. This alternate model was not used in experiments due to its significantly higher inference costs.

\subsection{Retraining strategy}
In addition to fully retraining the surrogate model from scratch using all acquired data, we tested an online training strategy. For online training, the trained surrogate model from the previous epoch was trained only on newly acquired data. Note that online training applies only to the NN and MPN models, as the RF is reinitialized each time it is fit.

\newpage
\section{Additional Results}

\subsection{Dataset score distributions}
\begin{figure}
    \centering
    \includegraphics[width=0.9\textwidth]{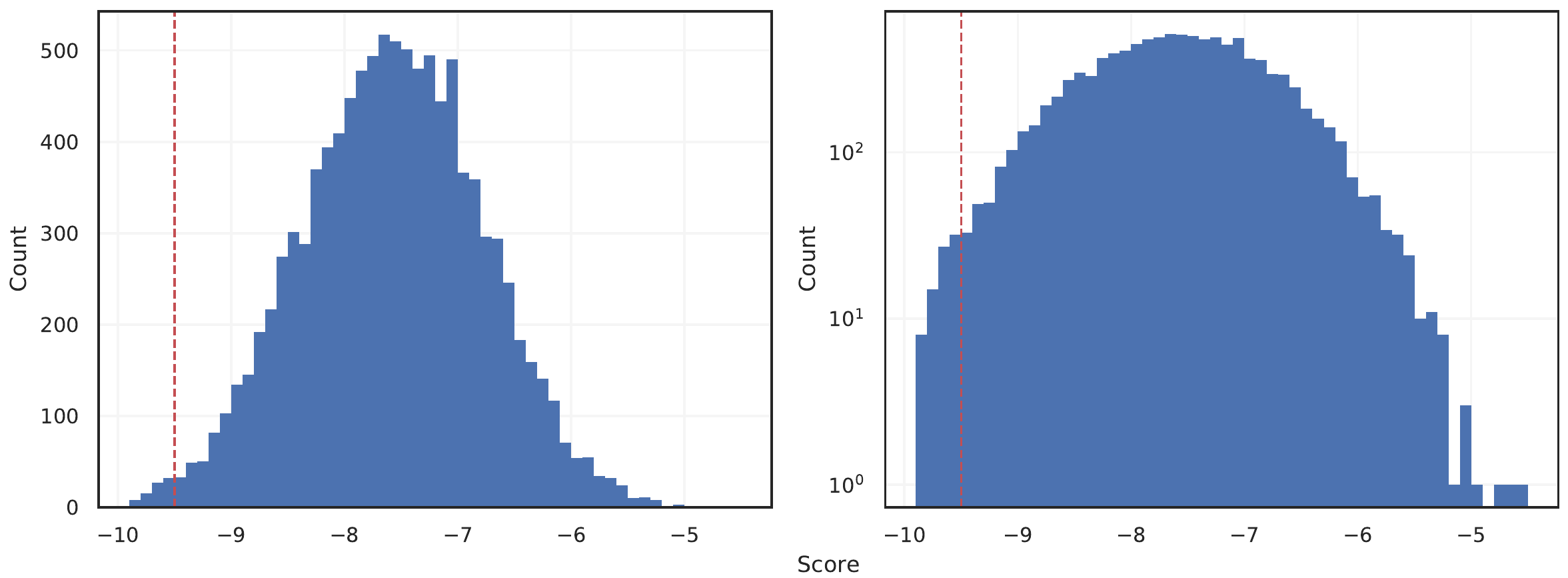}
    \caption{Distribution of docking scores in the Enamine 10k dataset with a bin size of 0.1. Red, dashed line corresponds to the $k$\textsuperscript{th} best score ($k=100$).}
    \label{fig:10k_score_hist}
\end{figure}

\begin{figure}
    \centering
    \includegraphics[width=0.9\textwidth]{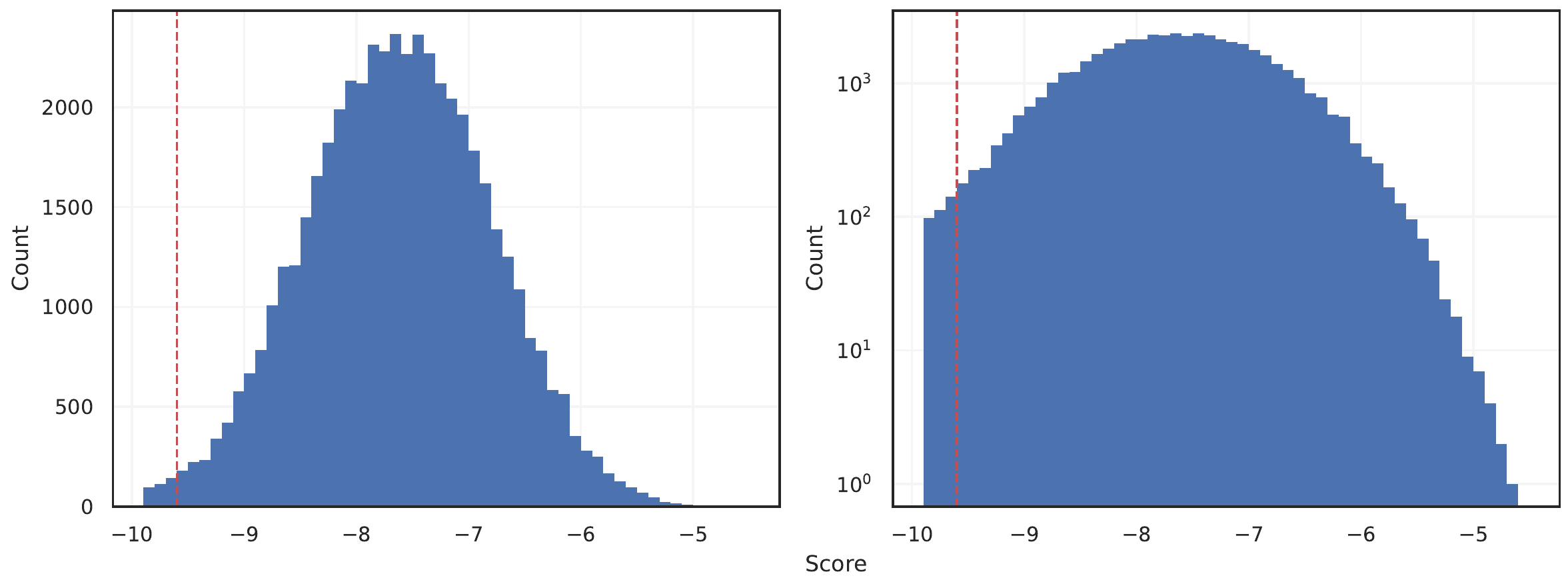}
    \caption{Distribution of docking scores in the Enamine 50k dataset with a bin size of 0.1. Red, dashed line corresponds to the $k$\textsuperscript{th} best score ($k=500$).}
    \label{fig:50k_score_hist}
\end{figure}

\begin{figure}
    \centering
    \includegraphics[width=0.9\textwidth]{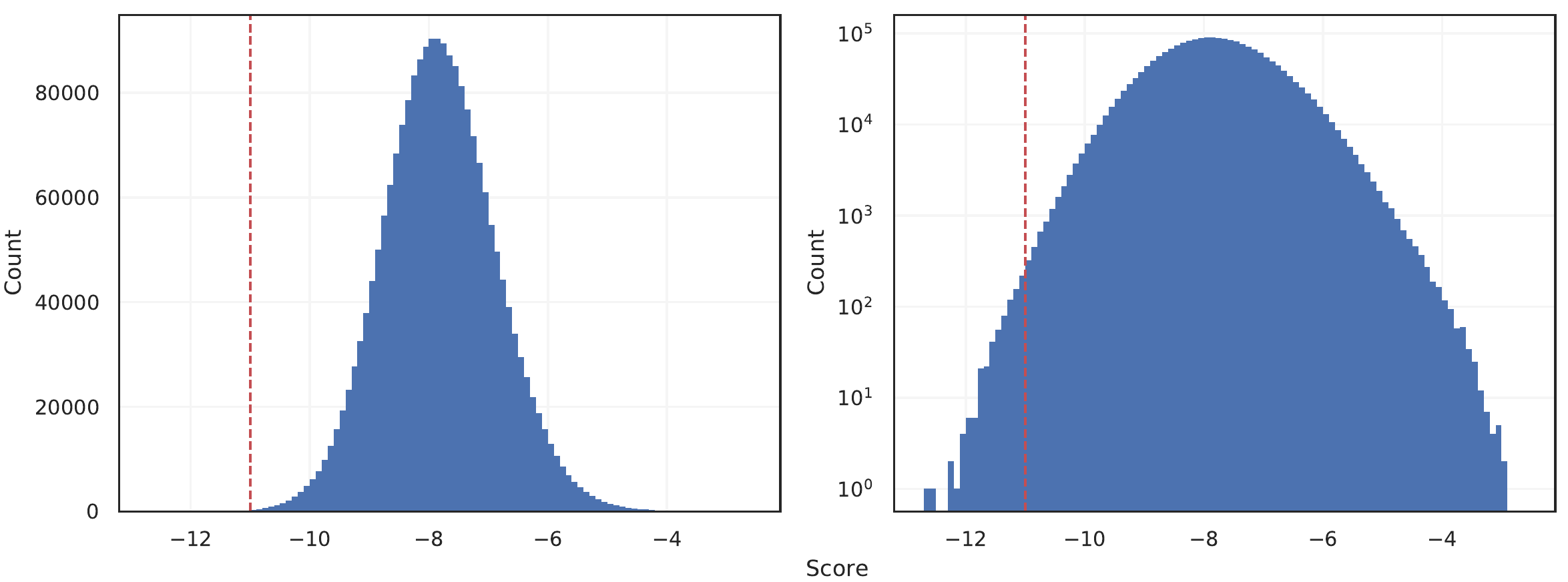}
    \caption{Distribution of docking scores in the Enamine HTS dataset with a bin size of 0.1. Red, dashed line corresponds to the $k$\textsuperscript{th} best score ($k=1000$).}
    \label{fig:HTS_score_hist}
\end{figure}

\begin{figure}
    \centering
    \includegraphics[width=0.9\textwidth]{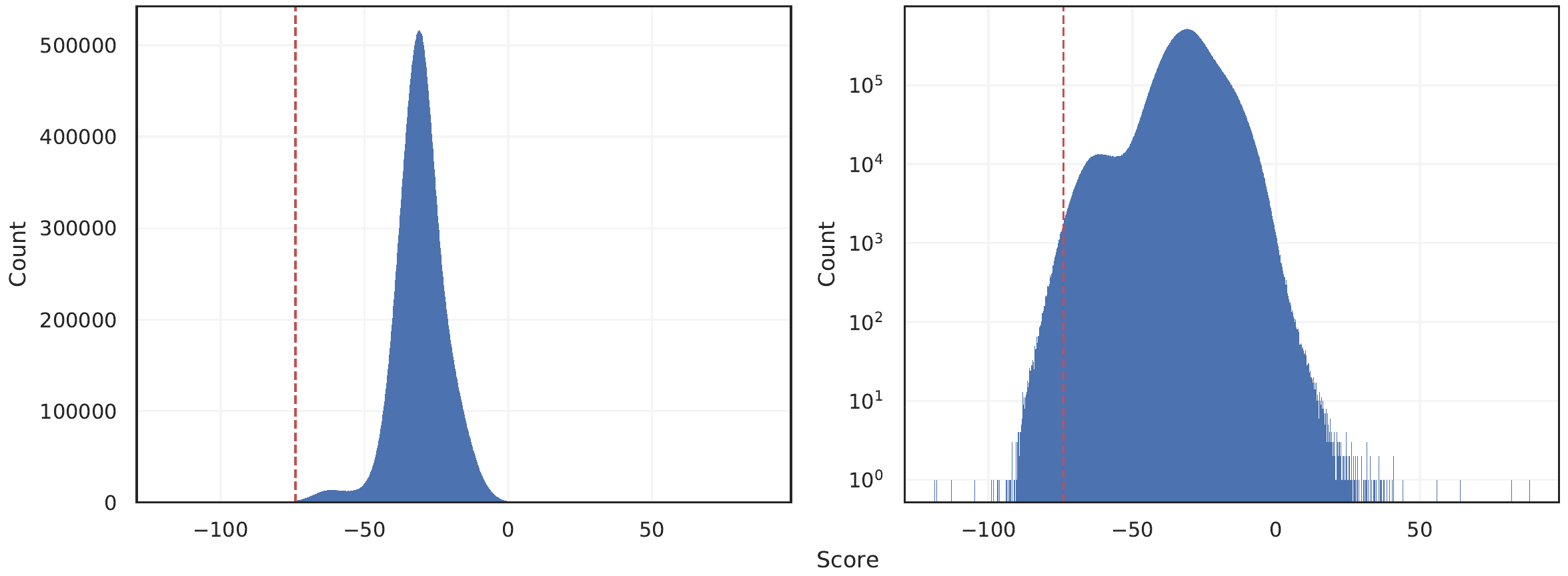}
    \caption{Distribution of docking scores in the AmpC dataset with a bin size of 0.1. Red, dashed line corresponds to the $k$\textsuperscript{th} best score ($k=50000$).}
    \label{fig:AmpC_score_hist}
\end{figure}

\FloatBarrier
\subsection{Library exploration across separate experiments}
\begin{figure}
    \centering
    \includegraphics[width=0.5\textwidth]{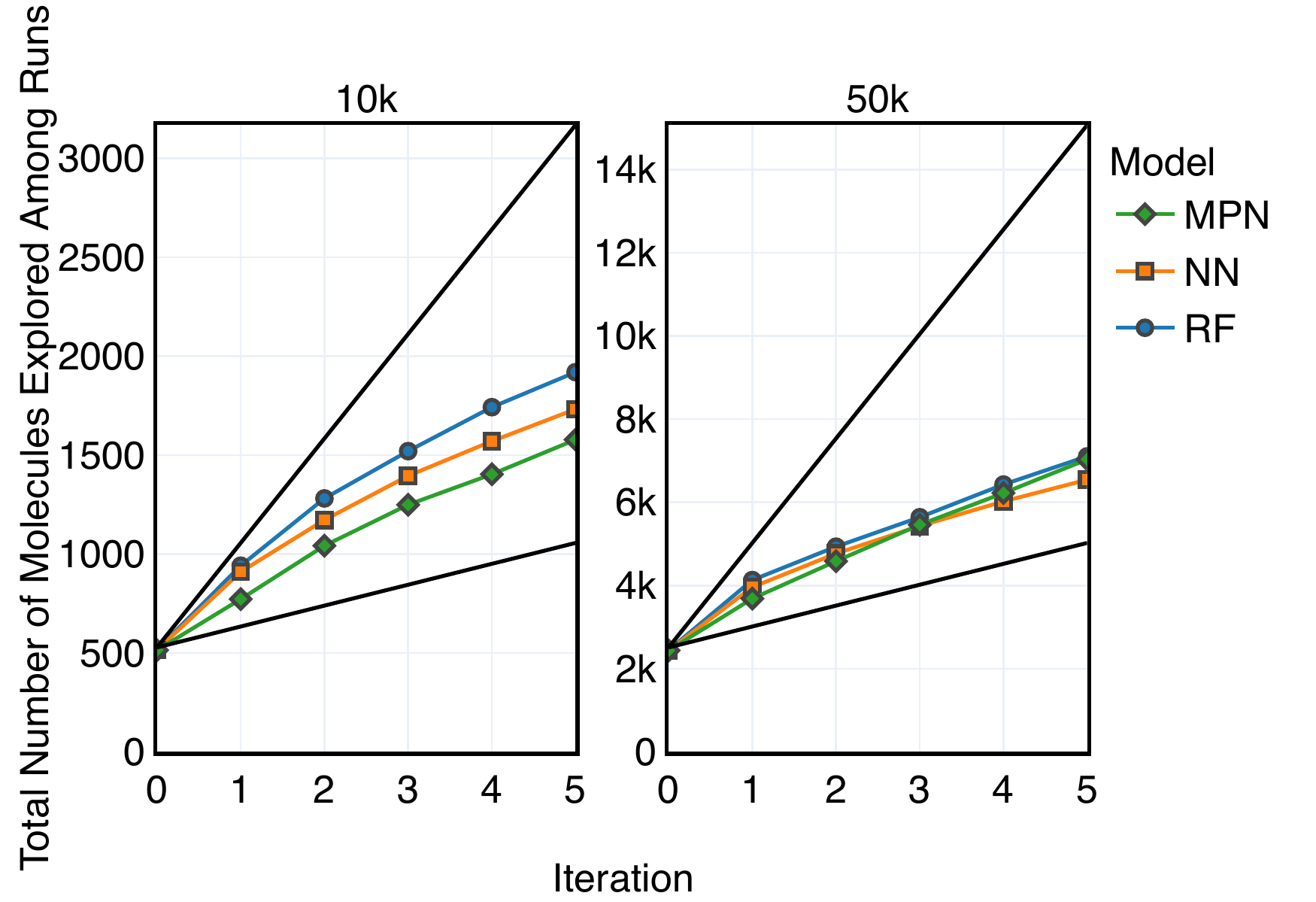}
    \caption{The total number of unique SMILES strings acquired across 5 greedy optimizations on the 10k and 50k libraries. The top black line is the theoretical maximum (i.e., repeated trials select distinct subsets of molecules to evaluate) the bottom black line is the theoretical minimum (i.e., repeated trials select identical subsets of molecules to evaluate).}
    \label{fig:10k50k_smis_greedy_union}
\end{figure}

\begin{figure}
    \centering
    \includegraphics[width=0.7\textwidth]{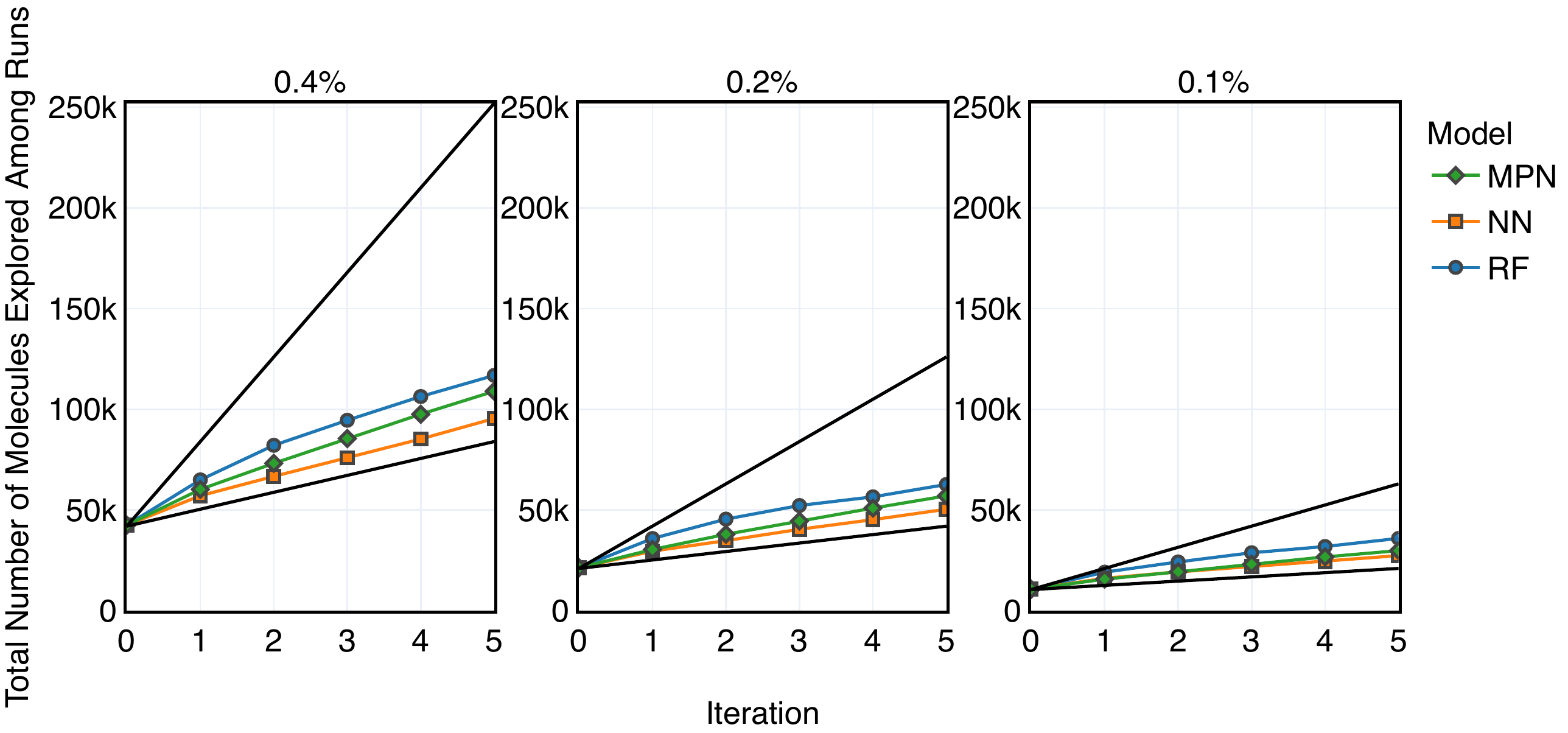}
    \caption{The total number of unique SMILES strings acquired across 5 greedy optimizations on the Enamine HTS library. The top black line is the theoretical maximum (i.e., repeated trials select distinct subsets of molecules to evaluate) the bottom black line is the theoretical minimum (i.e., repeated trials select identical subsets of molecules to evaluate).}
    \label{fig:HTS_smis_greedy_union}
\end{figure}

\FloatBarrier
\newpage
\subsection{Online training strategy}
\begin{figure}
    \centering
    \includegraphics[width=0.7\textwidth]{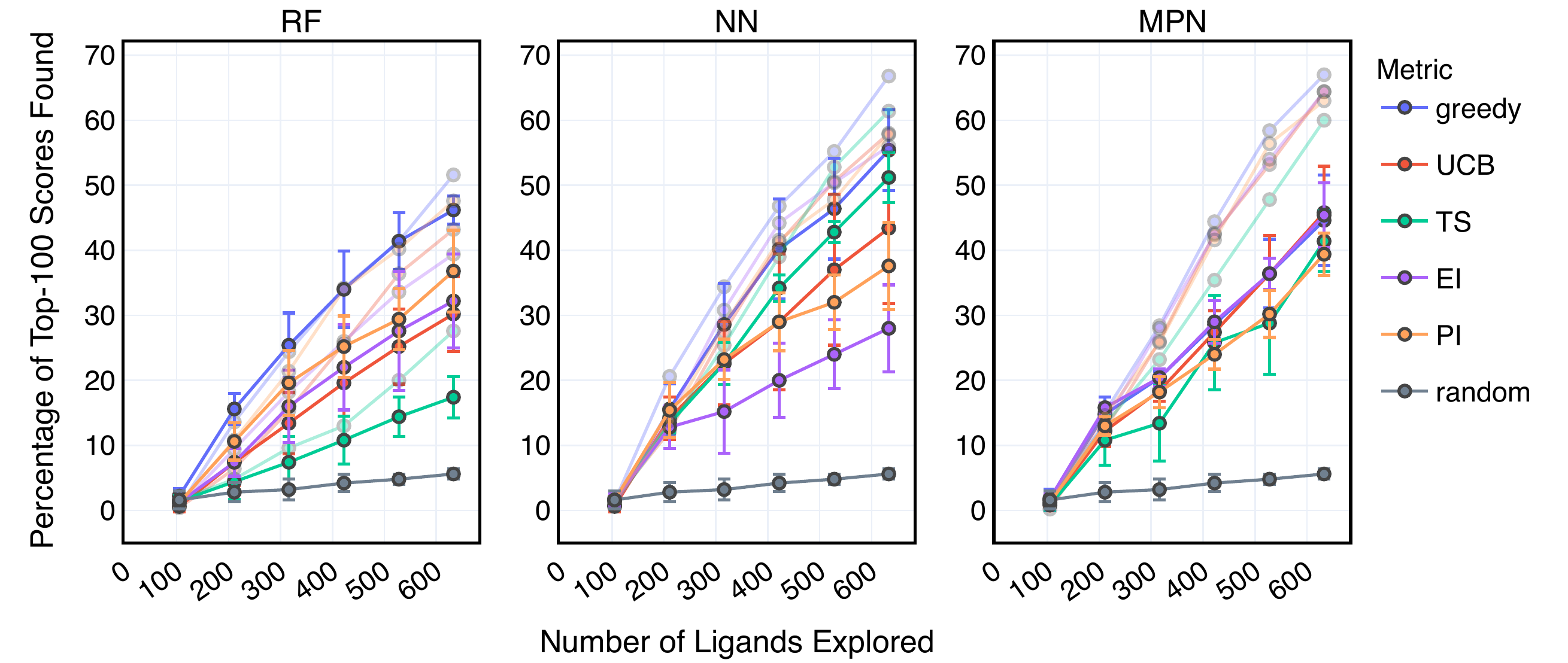}
    \caption{Bayesian optimization performance on Enamine 10k screening data as measured by the percentage of top-100 scores found as a function of the number of ligands evaluated. Each trace represents the performance of the given acquisition metric with the surrogate model architecture corresponding to the chart label. Full opacity: online model training. Faded: full model retraining. Each experiment began with a random 1\% acquisition (ca. 100 samples) and acquired 1\% more each iteration for five iterations. Error bars reflect $\pm$ one standard deviation across five runs.}
    \label{fig:10k_online}
\end{figure}

\begin{figure}[b!]
    \centering
    \includegraphics[width=0.7\textwidth]{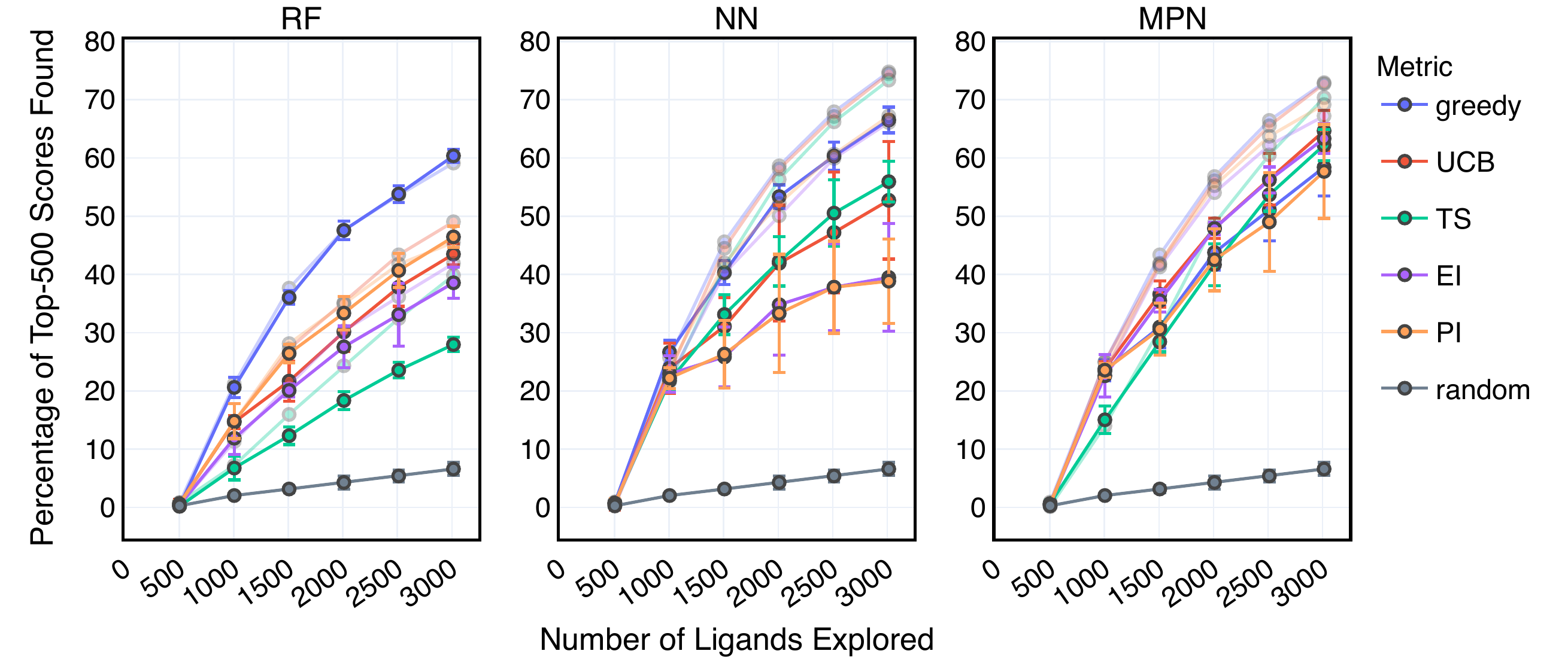}
    \caption{Bayesian optimization performance on Enamine 50k screening data as measured by the percentage of top-500 scores found as a function of the number of ligands evaluated. Each trace represents the performance of the given acquisition metric with the surrogate model architecture corresponding to the chart label. Full opacity: online model training. Faded: full model retraining. Each experiment began with a random 1\% acquisition (ca. 500 samples) and acquired 1\% more each iteration for five iterations. Error bars reflect $\pm$ one standard deviation across five runs.}
    \label{fig:50k_online}
\end{figure}

\begin{figure}
    \centering
    \includegraphics[width=0.7\textwidth]{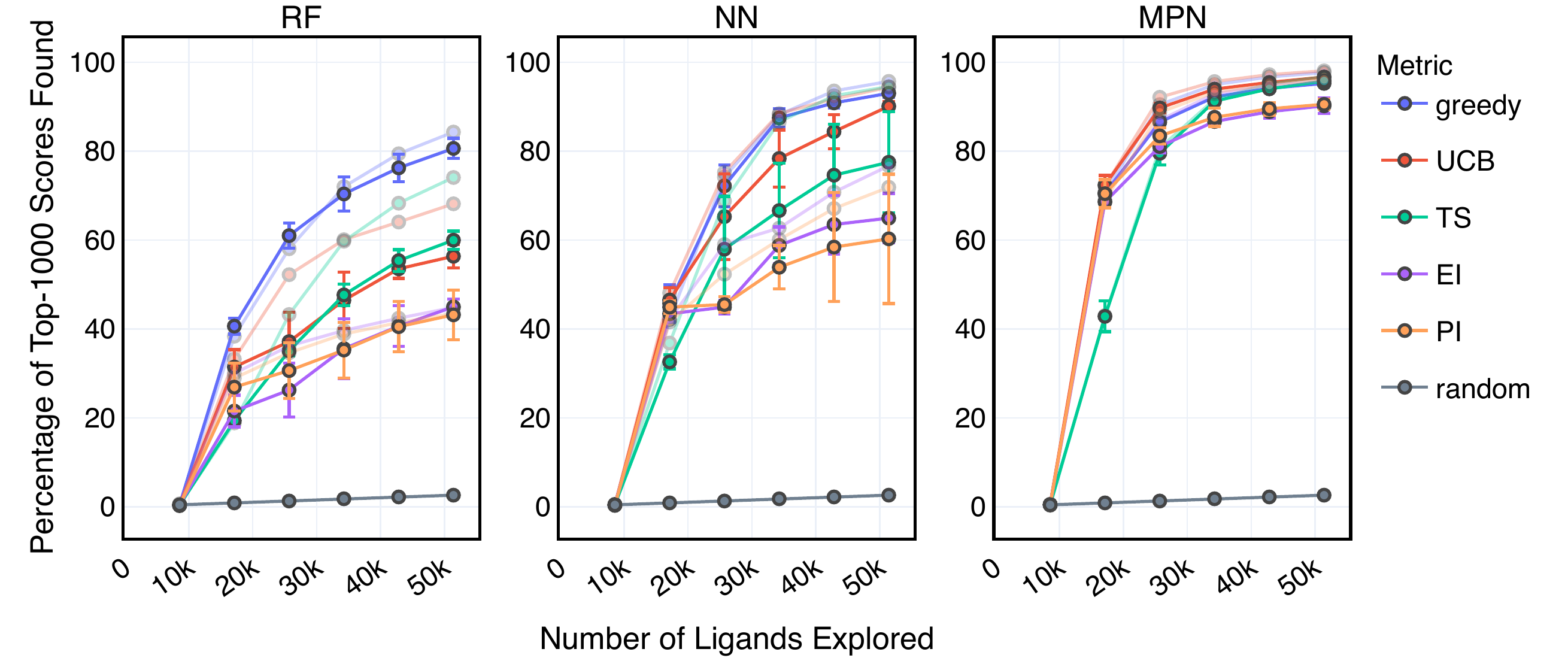}
    \caption{Bayesian optimization performance on Enamine HTS screening data as measured by the percentage of top-1000 scores found as a function of the number of ligands evaluated. Each trace represents the performance of the given acquisition metric with the surrogate model architecture corresponding to the chart label. Full opacity: online model training. Faded: full model retraining. Each experiment began with a random 0.4\% acquisition (ca. 8,400 samples) and acquired 0.4\% more each iteration for five iterations. Error bars reflect $\pm$ one standard deviation across five runs.}
    \label{fig:HTS_004_online}
\end{figure}

\begin{figure}
    \centering
    \includegraphics[width=0.7\textwidth]{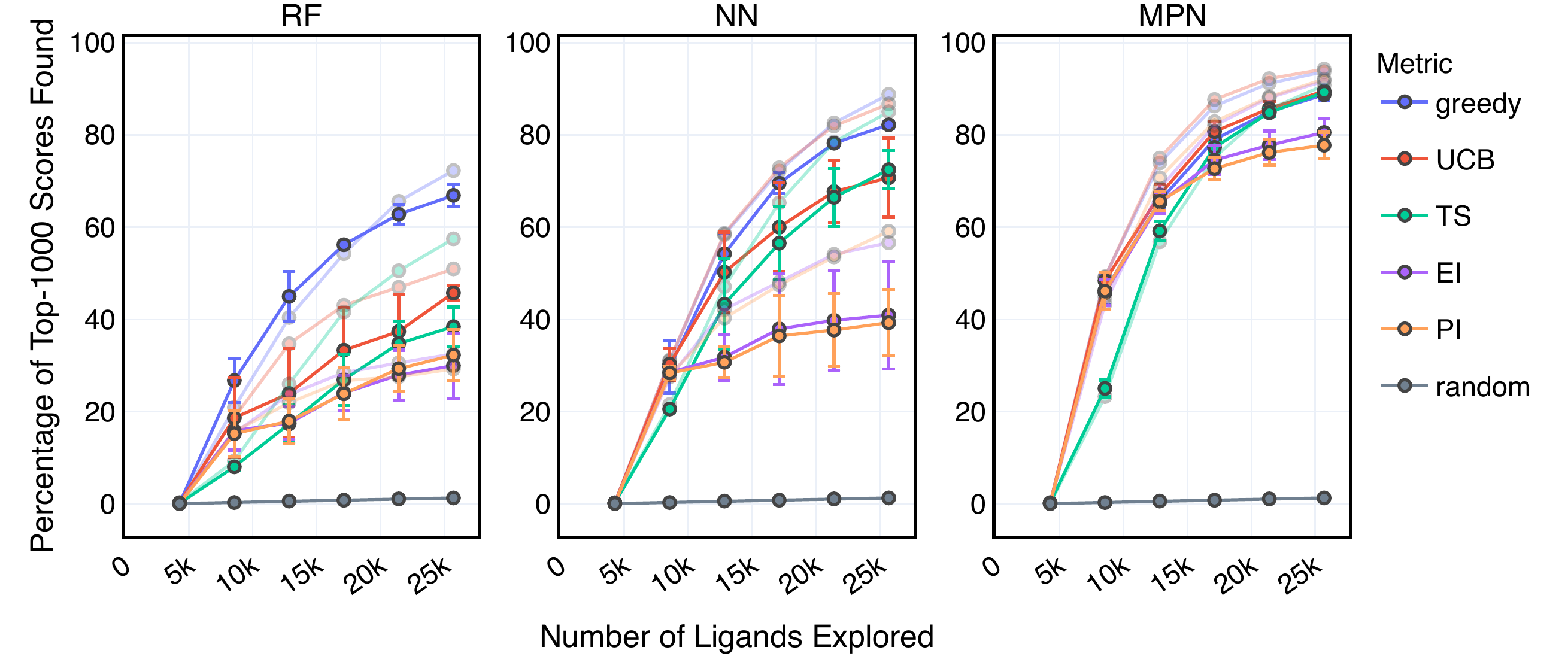}
    \caption{Bayesian optimization performance on Enamine HTS screening data as measured by the percentage of top-1000 scores found as function of the number of ligands evaluated. Each trace represents the performance of the given acquisition metric with the surrogate model architecture corresponding to the chart label. Full opacity: online model training. Faded: full model retraining. Each experiment began with a random 0.2\% acquisition (ca. 4,200 samples) and acquired 0.2\% more each iteration for five iterations. Error bars reflect $\pm$ one standard deviation across five runs.}
    \label{fig:HTS_002_online}
\end{figure}

\begin{figure}
    \centering
    \includegraphics[width=0.7\textwidth]{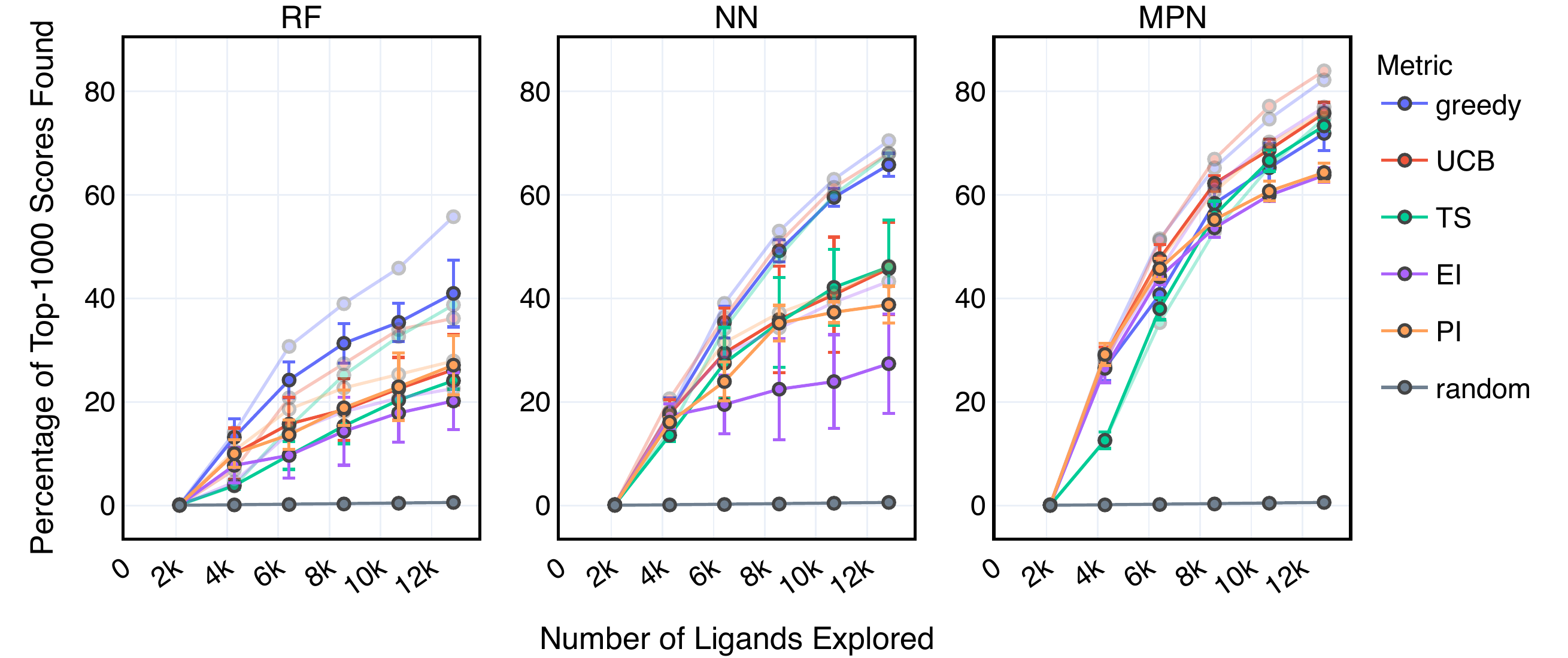}
    \caption{Bayesian optimization performance on Enamine HTS screening data as measured by the percentage of top-1000 scores found as function of the number of ligands evaluated. Each trace represents the performance of the given acquisition metric with the surrogate model architecture corresponding to the chart label. Full opacity: online model training. Faded: full model retraining. Each experiment began with a random 0.1\% acquisition (ca. 2,100 samples) and acquired 0.1\% more each iteration for five iterations. Error bars reflect $\pm$ one standard deviation across five runs.}
    \label{fig:HTS_001_online}
\end{figure}

\begin{figure}
    \centering
    \includegraphics[width=0.7\textwidth]{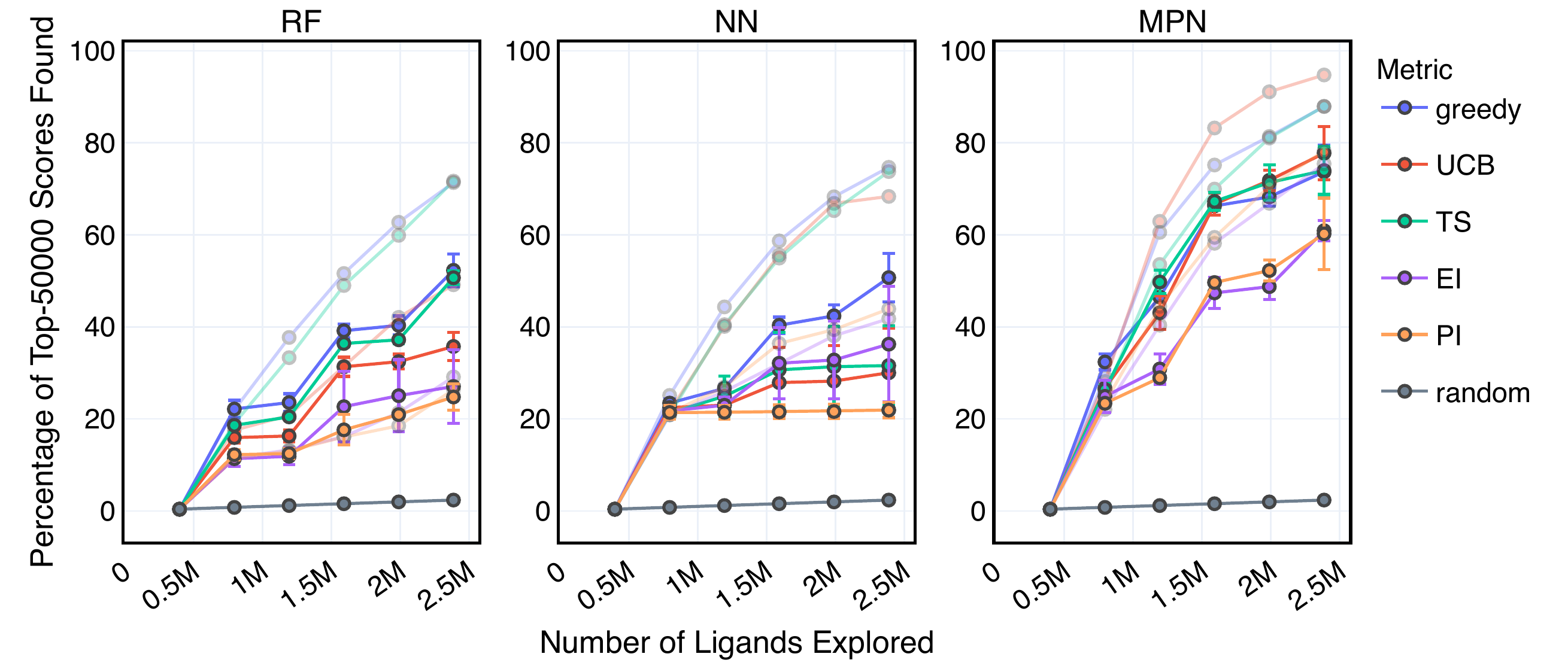}
    \caption{Bayesian optimization performance on AmpC screening data as measured by the percentage of top-50000 scores found as a function of the number of ligands evaluated. Each trace represents the performance of the given acquisition metric with the surrogate model architecture corresponding to the chart label. Full opacity: online model training. Faded: full model retraining. Each experiment began with a random 0.4\% acquisition (ca. 40,000 samples) and acquired 0.4\% more each iteration for five iterations. Error bars reflect $\pm$ one standard deviation across three runs.}
    \label{fig:AmpC_004_online}
\end{figure}

\begin{figure}
    \centering
    \includegraphics[width=0.7\textwidth]{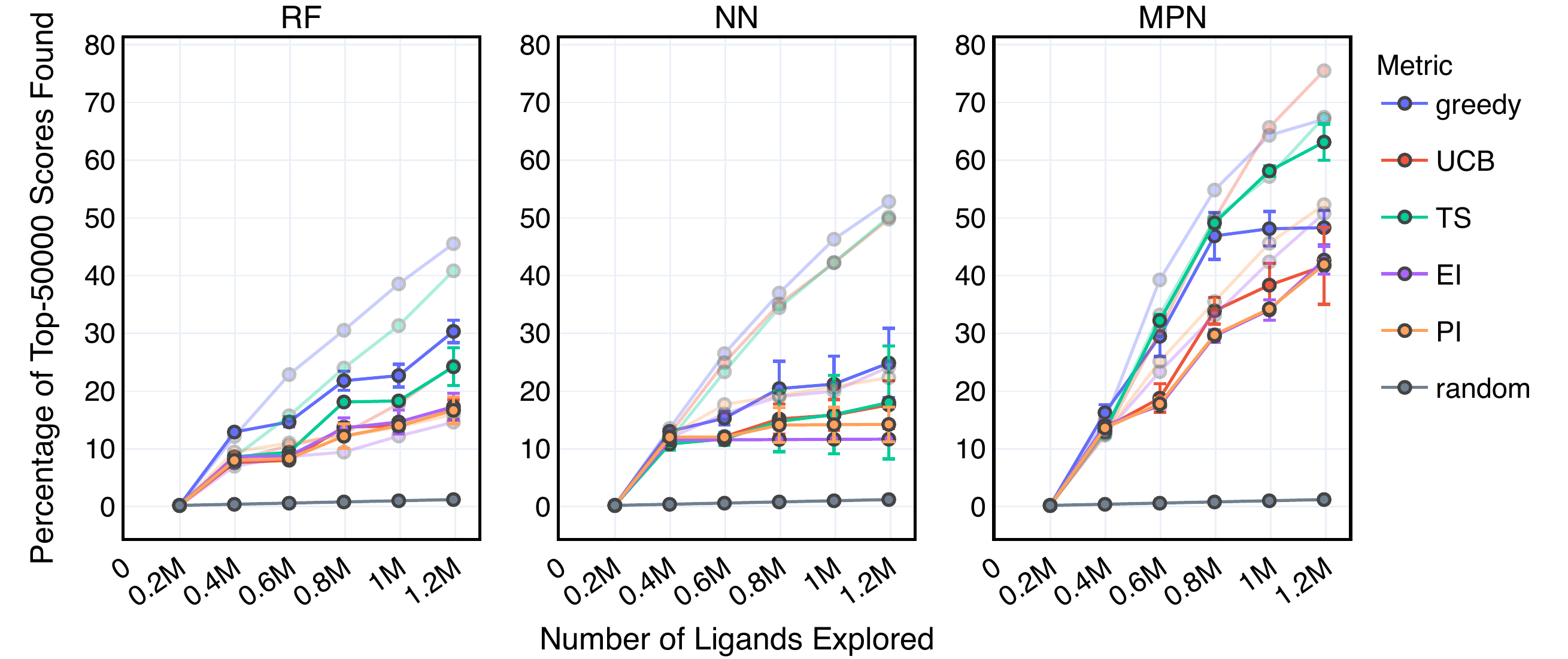}
    \caption{Bayesian optimization performance on AmpC screening data as measured by the percentage of top-50000 scores found as a function of the number of ligands evaluated. Each trace represents the performance of the given acquisition metric with the surrogate model architecture corresponding to the chart label. Full opacity: online model training. Faded: full model retraining. Each experiment began with a random 0.2\% acquisition (ca. 20,000 samples) and acquired 0.2\% more each iteration for five iterations. Error bars reflect $\pm$ one standard deviation across three runs.}
    \label{fig:AmpC_002_online}
\end{figure}

\begin{figure}[t!]
    \centering
    \includegraphics[width=0.7\textwidth]{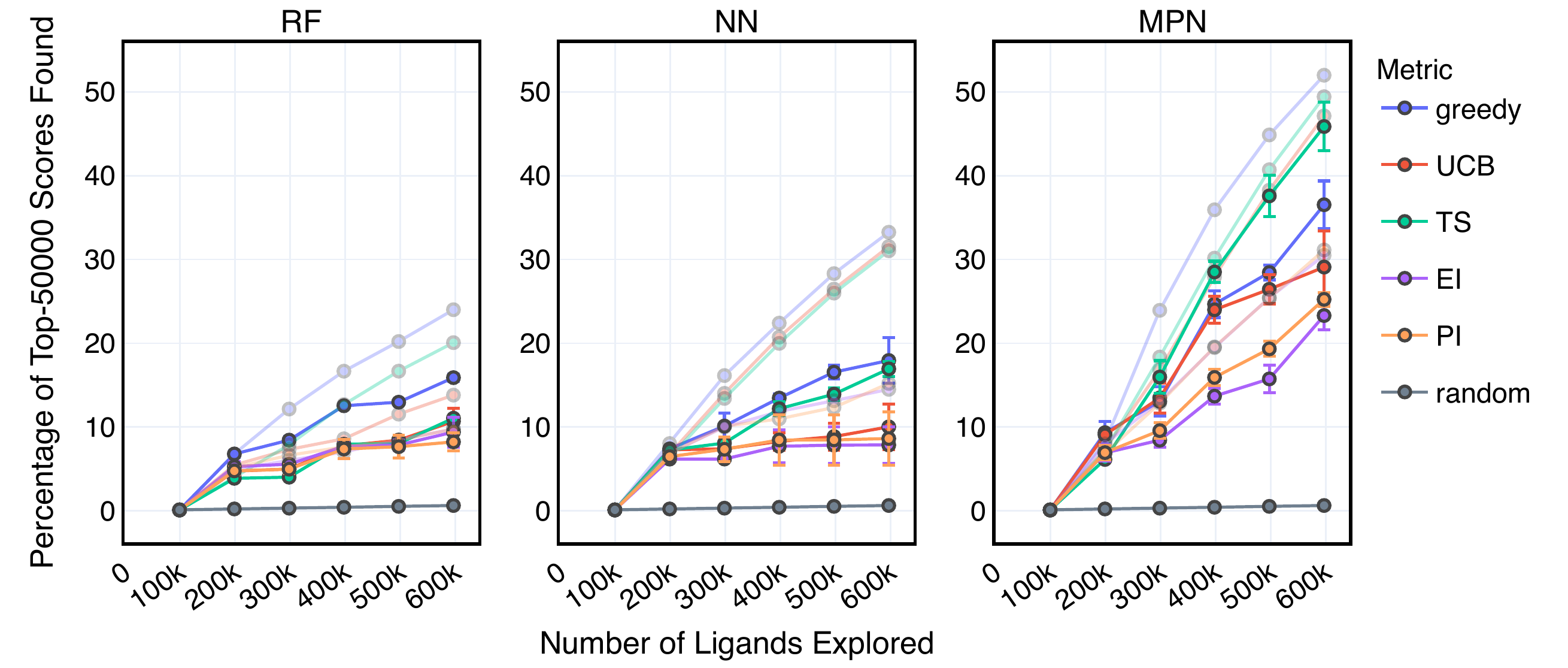}
    \caption{Bayesian optimization performance on AmpC screening data as measured by the percentage of top-50000 scores found as a function of the number of ligands evaluated. Each trace represents the performance of the given acquisition metric with the surrogate model architecture corresponding to the chart label. Full opacity: online model training. Faded: full model retraining. Each experiment began with a random 0.1\% acquisition (ca. 10,000 samples) and acquired 0.1\% more each iteration for five iterations. Error bars reflect $\pm$ one standard deviation across three runs.}
    \label{fig:AmpC_001_online}
\end{figure}

\FloatBarrier
\subsection{Bayesian Optimization Performance}
\begin{table}
    \caption{Final Bayesian optimization performance on Enamine 10k screening data with a 1.0\% batch size as measured by the given metric using the top-100 compounds found. Results are expressed as percentages and reflect the average (standard deviation) over five runs where higher is better.}
    \label{tbl:10k_results_final}
    \begin{tabular}{@{}cccccc@{}}
    \toprule
 Training & Model & Metric & Scores ($\pm$ s.d.) & SMILES ($\pm$ s.d.) & Average ($\pm$ s.d.) \\ \midrule
 \multirow{15}{*}{retrain} & \multirow{5}{*}{RF}  & Greedy & 51.6 (5.9) & 44.8 (5.8) & 98.21 (0.31) \\
 & & UCB & 43.2 (3.4) & 37.2 (3.1) & 97.58 (0.25) \\
 & & TS & 27.6 (1.9) & 22.6 (2.7) & 95.97 (0.34) \\
 & & EI & 39.4 (9.5) & 33.8 (9.1) & 97.16 (0.76) \\
 & & PI & 47.6 (4.2) & 41.4 (3.3) & 97.82 (0.24) \\ \cmidrule(l){2-6} 
 & \multirow{5}{*}{NN}  & Greedy & 66.8 (5.4) & 59.2 (6.1) & 98.97 (0.20) \\
 & & UCB & 58.0 (3.5) & 51.2 (3.4) & 98.59 (0.16) \\
 & & TS & 61.4 (3.9) & 54.6 (3.4) & 98.73 (0.19) \\
 & & EI & 56.0 (7.5) & 49.8 (6.9) & 98.42 (0.42) \\
 & & PI & 57.8 (2.4) & 51.6 (2.3) & 98.55 (0.15) \\ \cmidrule(l){2-6} 
 & \multirow{5}{*}{MPN} & Greedy & 67.0 (3.0) & 59.6 (2.3) & 98.94 (0.12) \\
 & & UCB & 64.4 (4.1) & 57.0 (3.2) & 98.80 (0.17) \\
 & & TS & 60.0 (4.1) & 52.2 (4.6) & 98.58 (0.20) \\
 & & EI & 64.4 (3.8) & 57.0 (3.6) & 98.80 (0.19) \\
 & & PI & 63.0 (3.2) & 55.2 (3.3) & 98.76 (0.16) \\ \midrule
 \multirow{15}{*}{online} & \multirow{5}{*}{RF} & Greedy & 46.2 (2.1) & 40.8 (3.7) & 97.86 (0.19) \\
 & & UCB & 30.2 (5.8) & 26.8 (5.3) & 96.46 (0.51) \\
 & & TS & 17.4 (3.2) & 15.2 (3.2) & 94.58 (0.31) \\
 & & EI & 32.2 (7.2) & 27.0 (5.8) & 96.53 (0.49) \\
 & & PI & 36.8 (6.3) & 31.4 (5.4) & 97.06 (0.58) \\ \cmidrule(l){2-6} 
 & \multirow{5}{*}{NN} & Greedy & 55.4 (6.2) & 49.8 (6.4) & 98.49 (0.30) \\
 & & UCB & 43.4 (11.6) & 38.4 (10.1) & 97.59 (0.82) \\
 & & TS & 51.2 (3.9) & 45.8 (3.5) & 98.16 (0.15) \\
 & & EI & 28.0 (6.7) & 24.6 (6.2) & 96.09 (1.12) \\
 & & PI & 37.6 (6.7) & 33.0 (5.4) & 96.93 (0.92) \\ \cmidrule(l){2-6} 
 & \multirow{5}{*}{MPN} & Greedy & 44.6 (6.9) & 37.2 (6.0) & 97.68 (0.50) \\
 & & UCB & 45.8 (7.1) & 38.8 (6.3) & 97.83 (0.47) \\
 & & TS & 41.4 (4.7) & 37.0 (5.2) & 97.44 (0.33) \\
 & & EI & 45.4 (5.0) & 38.8 (3.4) & 97.78 (0.27) \\
 & & PI & 39.4 (3.3) & 33.2 (2.9) & 97.40 (0.29) \\ \midrule
 \multicolumn{3}{c}{Random} & 5.6 (0.8) & 5.0 (0.9) & 91.41 (0.34) \\ \bottomrule
    \end{tabular}
    \label{tbl:E10k_results_final}
\end{table}

\begin{table}
    \caption{Final Bayesian optimization performance on Enamine 50k screening data with a 1.0\% batch size as measured by the given metric using the top-500 compounds found. Results are expressed as percentages and reflect the average (standard deviation) over five runs where higher is better.}
    \label{tbl:50k_results_final}
    \begin{tabular}{@{}cccccc@{}}
    \toprule
 Training & Model & Metric & Scores ($\pm$ s.d.) & SMILES ($\pm$ s.d.) & Average ($\pm$ s.d.) \\ \midrule
 \multirow{15}{*}{retrain} & \multirow{5}{*}{RF}  & Greedy & 59.1 (2.9) & 55.1 (3.0) & 98.74 (0.15) \\
 & & UCB & 49.0 (1.4) & 46.9 (1.3) & 98.16 (0.11) \\
 & & TS & 39.8 (2.9) & 37.6 (2.9) & 97.49 (0.23) \\
 & & EI & 41.9 (2.7) & 40.1 (2.7) & 97.62 (0.19) \\
 & & PI & 45.5 (2.4) & 43.4 (2.2) & 97.92 (0.15) \\ \cmidrule(l){2-6} 
 & \multirow{5}{*}{NN}  & Greedy & 74.8 (1.1) & 70.1 (1.1) & 99.39 (0.05) \\
 & & UCB & 74.4 (1.4) & 70.0 (1.2) & 99.39 (0.0 4) \\
 & & TS & 73.4 (2.3) & 68.9 (2.3) & 99.35 (0.07) \\
 & & EI & 66.1 (3.0) & 62.2 (2.9) & 99.08 (0.12) \\
 & & PI & 67.2 (4.0) & 63.1 (3.5) & 99.08 (0.16) \\ \cmidrule(l){2-6} 
 & \multirow{5}{*}{MPN} & Greedy & 72.9 (1.3) & 68.8 (1.0) & 99.34 (0.03) \\
 & & UCB & 72.7 (0.5) & 68.4 (0.4) & 99.33 (0.02) \\
 & & TS & 70.4 (0.7) & 66.0 (0.7) & 99.25 (0.05) \\
 & & EI & 67.2 (0.9) & 62.9 (0.9) & 99.12 (0.03) \\
 & & PI & 69.2 (1.7) & 64.9 (1.5) & 99.17 (0.07) \\ \midrule
 \multirow{15}{*}{online} & \multirow{5}{*}{RF}  & Greedy & 60.4 (1.2) & 56.4 (0.9) & 98.75 (0.04) \\
 & & UCB & 43.5 (1.8) & 41.4 (1.9) & 97.72 (0.15) \\
 & & TS & 28.0 (1.2) & 26.7 (1.3) & 96.30 (0.11) \\
 & & EI & 38.6 (2.7) & 37.1 (2.7) & 97.33 (0.25) \\
 & & PI & 46.4 (1.8) & 44.6 (1.4) & 98.00 (0.13) \\ \cmidrule(l){2-6} 
 & \multirow{5}{*}{NN}  & Greedy & 66.5 (2.2) & 62.8 (2.0) & 99.09 (0.10) \\
 & & UCB & 52.7 (10.1) & 49.9 (9.3) & 98.29 (0.54) \\
 & & TS & 55.9 (3.5) & 52.6 (3.3) & 98.56 (0.17) \\
 & & EI & 39.5 (9.2) & 37.1 (8.6) & 97.10 (1.05) \\
 & & PI & 38.8 (7.2) & 36.7 (6.8) & 97.25 (0.70) \\ \cmidrule(l){2-6} 
 & \multirow{5}{*}{MPN} & Greedy & 58.4 (4.9) & 54.8 (4.6) & 98.70 (0.32) \\
 & & UCB & 64.6 (3.5) & 60.6 (3.5) & 99.00 (0.15) \\
 & & TS & 62.2 (2.7) & 58.4 (2.3) & 98.94 (0.09) \\
 & & EI & 63.3 (2.6) & 59.2 (2.5) & 98.95 (0.11) \\
 & & PI & 57.7 (8.1) & 54.3 (7.6) & 98.64 (0.44) \\ \midrule
 \multicolumn{3}{c}{Random} & 6.6 (1.0) & 6.1 (1.2 & 91.36 (0.19) \\ \bottomrule
    \end{tabular}
\end{table}

\begin{table}
    \caption{Final Bayesian optimization performance on Enamine HTS screening data with a 0.4\% batch size as measured by the given metric using the top-1000 compounds found. Results are expressed as percentages and reflect the average (standard deviation) over five runs where higher is better.}
    \label{tbl:HTS_004_results_final}
    \begin{tabular}{@{}cccccc@{}}
    \toprule
 Training & Model & Metric & Scores ($\pm$ s.d.) & SMILES ($\pm$ s.d.) & Average ($\pm$ s.d.) \\ \midrule
 \multirow{15}{*}{retrain} & \multirow{5}{*}{RF}  & Greedy & 84.3 (1.1) & 79.8 (0.9) & 99.53 (0.02) \\
 & & UCB & 68.2 (2.7) & 65.2 (2.6) & 99.03 (0.10) \\
 & & TS & 74.1 (1.0) & 70.3 (1.2) & 99.26 (0.04) \\
 & & EI & 44.8 (4.0) & 43.2 (3.9) & 97.91 (0.26) \\
 & & PI & 43.5 (2.6) & 42.2 (2.7) & 97.80 (0.18) \\ \cmidrule(l){2-6} 
 & \multirow{5}{*}{NN}  & Greedy & 95.7 (0.0) & 93.2 (0.1) & 99.89 (0.00) \\
 & & UCB & 94.4 (0.5) & 91.5 (0.8) & 99.84 (0.02) \\
 & & TS & 94.5 (0.2) & 91.6 (0.3) & 99.85 (0.01) \\
 & & EI & 76.7 (2.3) & 72.9 (2.2) & 99.33 (0.05) \\
 & & PI & 71.9 (1.9) & 68.5 (1.8) & 99.17 (0.07) \\ \cmidrule(l){2-6} 
 & \multirow{5}{*}{MPN} & Greedy & 97.7 (0.2) & 94.8 (0.1) & 99.94 (0.01) \\
 & & UCB & 98.1 (0.4) & 94.9 (0.2) & 99.94 (0.01) \\
 & & TS & 96.7 (0.3) & 94.5 (0.1) & 99.92 (0.01) \\
 & & EI & 95.8 (0.6) & 93.7 (0.7) & 99.89 (0.02) \\
 & & PI & 96.2 (0.6) & 93.9 (0.7) & 99.90 (0.02) \\ \midrule
 \multirow{15}{*}{online}  & \multirow{5}{*}{RF}  & Greedy & 80.6 (2.3) & 76.5 (2.1) & 99.45 (0.05) \\
 & & UCB & 56.4 (2.6) & 54.0 (2.4) & 98.59 (0.14) \\
 & & TS & 60.0 (2.0) & 57.1 (1.8) & 98.69 (0.07) \\
 & & EI & 45.0 (1.8) & 43.5 (1.9) & 97.93 (0.10) \\
 & & PI & 43.2 (5.6) & 41.8 (5.3) & 97.80 (0.41) \\ \cmidrule(l){2-6} 
 & \multirow{5}{*}{NN}  & Greedy & 93.0 (0.8) & 89.8 (1.0) & 99.79 (0.03) \\
 & & UCB & 90.1 (1.1) & 86.1 (1.6) & 99.69 (0.03) \\
 & & TS & 77.5 (11.3) & 73.4 (10.5) & 99.28 (0.40) \\
 & & EI & 64.9 (5.6) & 61.5 (5.5) & 98.87 (0.24) \\
 & & PI & 60.3 (14.5) & 57.0 (13.8) & 98.58 (0.55) \\ \cmidrule(l){2-6} 
 & \multirow{5}{*}{MPN} & Greedy & 95.2 (0.4) & 93.2 (0.5) & 99.87 (0.02) \\
 & & UCB & 96.7 (0.4) & 94.5 (0.4) & 99.91 (0.01) \\
 & & TS & 95.8 (0.3) & 93.7 (0.5) & 99.88 (0.02) \\
 & & EI & 90.2 (1.7) & 85.7 (1.7) & 99.71 (0.05) \\
 & & PI & 90.5 (1.1) & 86.6 (1.3) & 99.70 (0.04) \\ \midrule
 \multicolumn{3}{c}{Random} & 2.6 (0.1) & 2.4 (0.1) & 90.09 (0.15) \\ \bottomrule
    \end{tabular}
\end{table}

\begin{table}
\caption{Final Bayesian optimization performance on Enamine HTS screening data with a 0.2\% batch size as measured by the given metric using the top-1000 compounds found. Results are expressed as percentages and reflect the average (standard deviation) over five runs where higher is better.}
\label{tbl:HTS_002_results_final}
\begin{tabular}{@{}cccccc@{}}
\toprule
 Training & Model & Metric & Scores ($\pm$ s.d.) & SMILES ($\pm$ s.d.) & Average ($\pm$ s.d.) \\ \midrule
 \multirow{15}{*}{retrain} & \multirow{5}{*}{RF} & Greedy & 72.3 (1.9) & 69.0 (1.9) & 99.23 (0.07) \\
 & & UCB & 51.0 (2.9) & 48.9 (2.9) & 98.25 (0.15) \\
 & & TS & 57.5 (1.4) & 54.8 (1.5) & 98.60 (0.05) \\
 & & EI & 32.6 (3.1) & 31.3 (3.0) & 96.86 (0.28) \\
 & & PI & 29.3 (5.2) & 28.3 (5.0) & 96.54 (0.52) \\ \cmidrule(l){2-6} 
 & \multirow{5}{*}{NN} & Greedy & 88.8 (0.8) & 83.9 (0.8) & 99.63 (0.03) \\
 & & UCB & 86.7 (0.5) & 82.1 (0.6) & 99.59 (0.01) \\
 & & TS & 85.0 (0.9) & 80.4 (0.9) & 99.53 (0.03) \\
 & & EI & 56.6 (4.3) & 54.0 (4.1) & 98.54 (0.23) \\
 & & PI & 59.1 (3.1) & 56.6 (3.0) & 98.67 (0.13) \\ \cmidrule(l){2-6} 
 & \multirow{5}{*}{MPN} & Greedy & 93.7 (0.9) & 90.7 (0.9) & 99.81 (0.04) \\
 & & UCB & 94.3 (0.4) & 91.6 (0.5) & 99.84 (0.01) \\
 & & TS & 90.8 (0.4) & 86.7 (0.6) & 99.71 (0.01) \\
 & & EI & 91.7 (0.7) & 87.4 (0.7) & 99.74 (0.01) \\
 & & PI & 92.1 (0.7) & 88.0 (0.9) & 99.76 (0.03) \\ \midrule
 \multirow{15}{*}{online} & \multirow{5}{*}{RF} & Greedy & 66.9 (2.4) & 64.0 (2.4) & 99.01 (0.11) \\
 & & UCB & 45.8 (1.6) & 44.1 (1.4) & 97.94 (0.09) \\
 & & TS & 38.5 (4.3) & 36.9 (4.0) & 97.39 (0.31) \\
 & & EI & 30.0 (7.1) & 29.0 (7.1) & 96.46 (0.75) \\
 & & PI & 32.3 (5.5) & 31.1 (5.2) & 96.79 (0.60) \\ \cmidrule(l){2-6} 
 & \multirow{5}{*}{NN} & Greedy & 82.2 (0.8) & 78.1 (0.6) & 99.47 (0.03) \\
 & & UCB & 70.7 (8.6) & 67.9 (8.3) & 99.13 (0.38) \\
 & & TS & 72.5 (4.2) & 68.9 (3.9) & 99.19 (0.14) \\
 & & EI & 40.9 (11.7) & 38.9 (11.1) & 97.49 (0.80) \\
 & & PI & 39.3 (7.1) & 37.4 (6.9) & 97.46 (0.48) \\ \cmidrule(l){2-6} 
 & \multirow{5}{*}{MPN} & Greedy & 88.7 (1.3) & 84.1 (1.5) & 99.62 (0.04) \\
 & & UCB & 89.5 (0.3) & 85.1 (0.6) & 99.67 (0.02) \\
 & & TS & 89.3 (0.3) & 84.9 (0.4) & 99.65 (0.01) \\
 & & EI & 80.5 (3.1) & 76.8 (2.9) & 99.48 (0.07) \\
 & & PI & 77.8 (2.8) & 74.3 (2.7) & 99.41 (0.08) \\ \midrule
 \multicolumn{3}{c}{Random} & 1.3 (0.4) & 1.3 (0.3) & 87.75 (0.14) \\ \bottomrule
\end{tabular}
\end{table}

\begin{table}
\caption{Final Bayesian optimization performance on Enamine HTS screening data with a 0.1\% batch size as measured by the given metric using the top-1000 compounds found. Results are expressed as percentages and reflect the average (standard deviation) over five runs where higher is better.}
\label{tbl:HTS_001_results_final}
\begin{tabular}{@{}cccccc@{}}
\toprule
Training & Model & Metric & Scores ($\pm$ s.d.) & SMILES ($\pm$ s.d.) & Average ($\pm$ s.d.) \\ \midrule
\multirow{15}{*}{retrain} & \multirow{5}{*}{RF} & Greedy & 55.8 (4.9) & 53.4 (4.6) & 98.54 (0.24) \\
 &  & UCB & 36.2 (4.2) & 34.9 (4.2) & 97.16 (0.42) \\
 &  & TS & 38.8 (2.5) & 37.1 (2.6) & 97.43 (0.18) \\
 &  & EI & 22.6 (2.7) & 21.9 (2.6) & 95.62 (0.37) \\
 &  & PI & 27.9 (2.7) & 27.1 (2.7) & 96.27 (0.37) \\ \cmidrule(l){2-6} 
 & \multirow{5}{*}{NN} & Greedy & 70.5 (1.8) & 66.9 (1.6) & 99.08 (0.09) \\
 &  & UCB & 68.0 (0.9) & 64.8 (1.2) & 99.04 (0.05) \\
 &  & TS & 68.0 (0.8) & 64.8 (0.8) & 99.05 (0.03) \\
 &  & EI & 43.3 (3.8) & 41.5 (3.5) & 97.74 (0.29) \\
 &  & PI & 46.3 (2.3) & 44.2 (2.2) & 97.94 (0.17) \\ \cmidrule(l){2-6} 
 & \multirow{5}{*}{MPN} & Greedy & 82.2 (0.9) & 78.1 (0.8) & 99.47 (0.03) \\
 &  & UCB & 84.0 (0.8) & 79.9 (0.7) & 99.54 (0.03) \\
 &  & TS & 74.8 (1.9) & 71.2 (1.8) & 99.27 (0.04) \\
 &  & EI & 77.0 (1.2) & 73.9 (1.2) & 99.42 (0.03) \\
 &  & PI & 76.1 (0.9) & 73.0 (0.7) & 99.41 (0.02) \\ \midrule
 \multirow{15}{*}{online} & \multirow{5}{*}{RF} & Greedy & 41.0 (6.5) & 39.5 (6.2) & 97.60 (0.49) \\
 &  & UCB & 26.2 (6.8) & 25.4 (6.6) & 95.84 (0.91) \\
 &  & TS & 24.1 (1.6) & 23.1 (1.5) & 95.63 (0.33) \\
 &  & EI & 20.2 (5.5) & 19.6 (5.2) & 94.90 (1.29) \\
 &  & PI & 27.1 (5.7) & 26.4 (5.6) & 96.05 (0.85) \\ \cmidrule(l){2-6} 
 & \multirow{5}{*}{NN} & Greedy & 65.8 (2.2) & 63.2 (2.0) & 98.96 (0.09) \\
 &  & UCB & 45.8 (8.9) & 43.9 (8.6) & 97.90 (0.62) \\
 &  & TS & 46.1 (9.1) & 44.3 (8.8) & 97.90 (0.62) \\
 &  & EI & 27.4 (9.6) & 26.2 (9.4) & 96.22 (0.98) \\
 &  & PI & 38.8 (3.6) & 37.3 (3.3) & 97.39 (0.30) \\ \cmidrule(l){2-6} 
 & \multirow{5}{*}{MPN} & Greedy & 71.9 (3.3) & 68.4 (3.5) & 99.17 (0.11) \\
 &  & UCB & 75.8 (2.1) & 72.4 (1.8) & 99.33 (0.05) \\
 &  & TS & 73.3 (1.0) & 70.0 (1.2) & 99.24 (0.03) \\
 &  & EI & 63.8 (1.4) & 61.3 (1.5) & 98.95 (0.06) \\
 &  & PI & 64.3 (1.8) & 61.9 (1.8) & 98.97 (0.09) \\ \midrule
 \multicolumn{3}{c}{Random} & 0.6 (0.2) & 0.5 (0.2) & 85.36 (0.11) \\ \bottomrule
\end{tabular}
\end{table}

\begin{table}
\caption{Final Bayesian optimization performance on AmpC screening data with a 0.4\% batch size as measured by the given metric using the top-50000 compounds found. Results are expressed as percentages and reflect the average (standard deviation) over three runs where higher is better.}
\label{tbl:AmpC_004_results_final}
\begin{tabular}{@{}cccccc@{}}
\toprule
 Training & Model & Metric & Scores ($\pm$ s.d.) & SMILES ($\pm$ s.d.) & Average ($\pm$ s.d.) \\ \midrule
 \multirow{15}{*}{retrain} & \multirow{5}{*}{RF} & Greedy & 71.4 (2.1) & 71.3 (2.1) & 98.79 (0.13) \\
 &  & UCB & 49.2 (7.7) & 49.1 (7.7) & 97.47 (0.58) \\
 &  & TS & 71.7 (1.9) & 71.6 (1.9) & 98.78 (0.10) \\
 &  & EI & 29.1 (4.4) & 29.1 (4.4) & 95.47 (0.59) \\
 &  & PI & 26.4 (4.7) & 26.4 (4.7) & 95.03 (0.71) \\ \cmidrule(l){2-6} 
 & \multirow{5}{*}{NN} & Greedy & 74.7 (1.4) & 74.6 (1.4) & 98.94 (0.08) \\
 &  & UCB & 68.4 (1.4) & 68.3 (1.4) & 98.65 (0.07) \\
 &  & TS & 73.8 (1.2) & 73.7 (1.2) & 98.92 (0.06) \\
 &  & EI & 41.8 (1.8) & 41.8 (1.8) & 96.90 (0.16) \\
 &  & PI & 43.9 (2.1) & 43.9 (2.1) & 97.08 (0.17) \\ \cmidrule(l){2-6} 
 & \multirow{5}{*}{MPN} & Greedy & 87.9 (2.3) & 87.9 (2.3) & 99.56 (0.09) \\
 &  & UCB & 94.8 (0.1) & 94.7 (0.1) & 99.83 (0.00) \\
 &  & TS & 87.9 (0.2) & 87.9 (0.2) & 99.56 (0.01) \\
 &  & EI & 75.4 (5.4) & 75.4 (5.4) & 99.17 (0.24) \\
 &  & PI & 78.3 (0.9) & 78.2 (0.9) & 99.31 (0.04) \\ \midrule
 \multirow{15}{*}{online} & \multirow{5}{*}{RF} & Greedy & 52.3 (3.6) & 52.3 (3.6) & 97.70 (0.26) \\
 &  & UCB & 35.8 (3.1) & 35.7 (3.1) & 96.29 (0.32) \\
 &  & TS & 50.7 (1.5) & 50.7 (1.5) & 97.59 (0.11) \\
 &  & EI & 27.1 (8.0) & 27.0 (8.0) & 95.05 (1.12) \\
 &  & PI & 24.8 (2.9) & 24.8 (2.9) & 94.81 (0.51) \\ \cmidrule(l){2-6} 
 & \multirow{5}{*}{NN} & Greedy & 50.7 (5.2) & 50.7 (5.2) & 97.59 (0.38) \\
 &  & UCB & 30.0 (9.7) & 30.0 (9.7) & 95.41 (1.16) \\
 &  & TS & 31.6 (8.6) & 31.6 (8.6) & 95.64 (1.11) \\
 &  & EI & 36.3 (12.6) & 36.2 (12.6) & 96.06 (1.38) \\
 &  & PI & 21.9 (1.6) & 21.9 (1.6) & 94.34 (0.30) \\ \cmidrule(l){2-6} 
 & \multirow{5}{*}{MPN} & Greedy & 73.7 (5.7) & 73.7 (5.7) & 98.95 (0.29) \\
 &  & UCB & 77.8 (5.8) & 77.7 (5.8) & 99.26 (0.25) \\
 &  & TS & 74.0 (5.2) & 73.9 (5.2) & 98.99 (0.25) \\
 &  & EI & 60.9 (2.2) & 60.9 (2.2) & 98.41 (0.12) \\
 &  & PI & 60.2 (7.8) & 60.2 (7.8) & 98.32 (0.51) \\ \midrule
 \multicolumn{3}{c}{Random} & 2.4 (0.1) & 2.4 (0.1) & 81.03 (0.04) \\ \bottomrule
\end{tabular}
\end{table}

\begin{table}
\caption{Final Bayesian optimization performance on AmpC screening data with a 0.2\% batch size as measured by the given metric using the top-50000 compounds found. Results are expressed as percentages and reflect the average (standard deviation) over three runs where higher is better.}
\label{tbl:AmpC_002_results_final}
\begin{tabular}{@{}cccccc@{}}
\toprule
 Training & Model & Metric & Scores ($\pm$ s.d.) & SMILES ($\pm$ s.d.) & Average ($\pm$ s.d.) \\ \midrule
 \multirow{15}{*}{retrain} & \multirow{5}{*}{RF} & Greedy & 45.5 (1.8) & 45.5 (1.8) & 97.19 (0.14) \\
 &  & UCB & 24.4 (2.0) & 24.4 (1.9) & 94.81 (0.40) \\
 &  & TS & 40.8 (1.9) & 40.8 (1.9) & 96.80 (0.17) \\
 &  & EI & 14.6 (2.7) & 14.6 (2.7) & 92.44 (0.85) \\
 &  & PI & 16.0 (1.6) & 16.0 (1.6) & 92.83 (0.44) \\ \cmidrule(l){2-6} 
 & \multirow{5}{*}{NN} & Greedy & 52.8 (0.5) & 52.8 (0.5) & 97.72 (0.03) \\
 &  & UCB & 49.8 (0.5) & 49.8 (0.5) & 97.52 (0.04) \\
 &  & TS & 50.1 (1.0) & 50.1 (1.0) & 97.53 (0.07) \\
 &  & EI & 24.2 (1.0) & 24.2 (1.0) & 94.75 (0.17) \\
 &  & PI & 22.3 (1.1) & 22.3 (1.1) & 94.42 (0.19) \\ \cmidrule(l){2-6} 
 & \multirow{5}{*}{MPN} & Greedy & 67.1 (2.1) & 67.0 (2.1) & 98.61 (0.09) \\
 &  & UCB & 75.5 (0.7) & 75.5 (0.7) & 99.16 (0.03) \\
 &  & TS & 67.4 (3.3) & 67.4 (3.3) & 98.68 (0.17) \\
 &  & EI & 50.8 (0.3) & 50.8 (0.3) & 97.72 (0.02) \\
 &  & PI & 52.3 (0.6) & 52.3 (0.6) & 97.84 (0.04) \\ \midrule
\multirow{15}{*}{online} & \multirow{5}{*}{RF} & Greedy & 30.4 (2.0) & 30.3 (2.0) & 95.67 (0.24) \\
 &  & UCB & 16.8 (1.8) & 16.8 (1.8) & 93.23 (0.46) \\
 &  & TS & 24.2 (3.3) & 24.2 (3.3) & 94.75 (0.54) \\
 &  & EI & 17.3 (2.3) & 17.3 (2.3) & 93.03 (0.66) \\
 &  & PI & 16.7 (2.3) & 16.7 (2.3) & 92.96 (0.63) \\ \cmidrule(l){2-6} 
 & \multirow{5}{*}{NN} & Greedy & 24.9 (6.0) & 24.8 (6.0) & 94.69 (1.09) \\
 &  & UCB & 17.6 (4.2) & 17.6 (4.2) & 93.29 (1.14) \\
 &  & TS & 18.0 (9.8) & 18.0 (9.8) & 92.83 (2.12) \\
 &  & EI & 11.7 (0.4) & 11.7 (0.4) & 91.48 (0.22) \\
 &  & PI & 14.2 (3.0) & 14.2 (3.0) & 92.26 (0.88) \\ \cmidrule(l){2-6} 
 & \multirow{5}{*}{MPN} & Greedy & 48.3 (3.0) & 48.3 (2.9) & 97.41 (0.24) \\
 &  & UCB & 41.7 (6.7) & 41.7 (6.7) & 96.78 (0.71) \\
 &  & TS & 63.1 (3.1) & 63.1 (3.1) & 98.46 (0.19) \\
 &  & EI & 42.7 (2.4) & 42.7 (2.4) & 96.98 (0.23) \\
 &  & PI & 41.9 (0.7) & 41.9 (0.7) & 96.91 (0.08) \\ \midrule
 \multicolumn{3}{c}{Random} & 1.2 ($<0.1$) & 1.2 ($<0.1$) & 72.23 (0.10) \\ \bottomrule
\end{tabular}
\end{table}

\begin{table}
\caption{Final Bayesian optimization performance on AmpC screening data with a 0.1\% batch size as measured by the given metric using the top-50000 compounds found. Results are expressed as percentages and reflect the average (standard deviation) over three runs where higher is better.}
\label{tbl:AmpC_001_results_final}
\begin{tabular}{@{}cccccc@{}}
\toprule
 Training & Model & Metric & Scores ($\pm$ s.d.) & SMILES ($\pm$ s.d.) & Average ($\pm$ s.d.) \\ \midrule
 \multirow{15}{*}{retrain} & \multirow{5}{*}{RF} & Greedy & 24.0 (2.2) & 24.0 (2.2) & 94.76 (0.35) \\
 &  & UCB & 13.8 (1.0) & 13.8 (1.0) & 92.01 (0.42) \\
 &  & TS & 20.1 (2.0) & 20.1 (2.0) & 94.08 (0.38) \\
 &  & EI & 9.8 (2.3) & 9.8 (2.3) & 90.03 (1.03) \\
 &  & PI & 9.8 (1.3) & 9.7 (1.3) & 90.44 (0.54) \\ \cmidrule(l){2-6} 
 & \multirow{5}{*}{NN} & Greedy & 33.3 (0.3) & 33.2 (0.3) & 96.00 (0.04) \\
 &  & UCB & 31.5 (0.6) & 31.5 (0.6) & 95.80 (0.07) \\
 &  & TS & 31.0 (0.8) & 31.0 (0.8) & 95.73 (0.10) \\
 &  & EI & 14.5 (0.8) & 14.5 (0.8) & 92.41 (0.30) \\
 &  & PI & 15.2 (1.4) & 15.2 (1.4) & 92.72 (0.43) \\ \cmidrule(l){2-6} 
 & \multirow{5}{*}{MPN} & Greedy & 52.0 (0.5) & 52.0 (0.5) & 97.68 (0.04) \\
 &  & UCB & 47.1 (0.5) & 47.1 (0.5) & 97.36 (0.05) \\
 &  & TS & 49.4 (0.3) & 49.4 (0.3) & 97.53 (0.03) \\
 &  & EI & 30.5 (1.8) & 30.5 (1.8) & 95.39 (0.29) \\
 &  & PI & 31.1 (1.3) & 31.1 (1.3) & 95.49 (0.17) \\ \midrule
 \multirow{15}{*}{online} & \multirow{5}{*}{RF} & Greedy & 15.9 (0.4) & 15.9 (0.4) & 93.01 (0.09) \\
 &  & UCB & 10.7 (1.6) & 10.7 (1.6) & 90.49 (0.96) \\
 &  & TS & 11.1 (0.1) & 11.0 (0.1) & 91.25 (0.10) \\
 &  & EI & 9.4 (1.8) & 9.4 (1.8) & 89.75 (1.22) \\
 &  & PI & 8.2 (1.1) & 8.2 (1.1) & 89.24 (0.59) \\ \cmidrule(l){2-6} 
 & \multirow{5}{*}{NN} & Greedy & 17.9 (2.7) & 17.9 (2.7) & 93.28 (0.67) \\
 &  & UCB & 10.0 (2.7) & 10.0 (2.7) & 90.16 (1.78) \\
 &  & TS & 16.9 (0.9) & 16.9 (0.9) & 93.16 (0.24) \\
 &  & EI & 7.9 (2.2) & 7.8 (2.2) & 89.01 (1.39) \\
 &  & PI & 8.6 (3.2) & 8.6 (3.2) & 89.25 (1.87) \\ \cmidrule(l){2-6} 
 & \multirow{5}{*}{MPN} & Greedy & 36.5 (2.8) & 36.5 (2.8) & 96.27 (0.33) \\
 &  & UCB & 29.1 (4.3) & 29.1 (4.3) & 95.02 (0.77) \\
 &  & TS & 45.9 (2.9) & 45.8 (2.9) & 97.21 (0.26) \\
 &  & EI & 23.3 (1.7) & 23.3 (1.7) & 94.15 (0.32) \\
 &  & PI & 25.2 (0.8) & 25.2 (0.8) & 94.50 (0.13) \\ \midrule
 \multicolumn{3}{c}{Random} & 0.6 ($\ll 0.1$) & 0.6 ($\ll 0.1$) & 64.44 (0.05) \\ \bottomrule
\end{tabular}
\end{table}

\FloatBarrier
\newpage
\subsection{Chemical Space Visualization}
\begin{figure}
    \centering
    \includegraphics[width=0.8\textwidth]{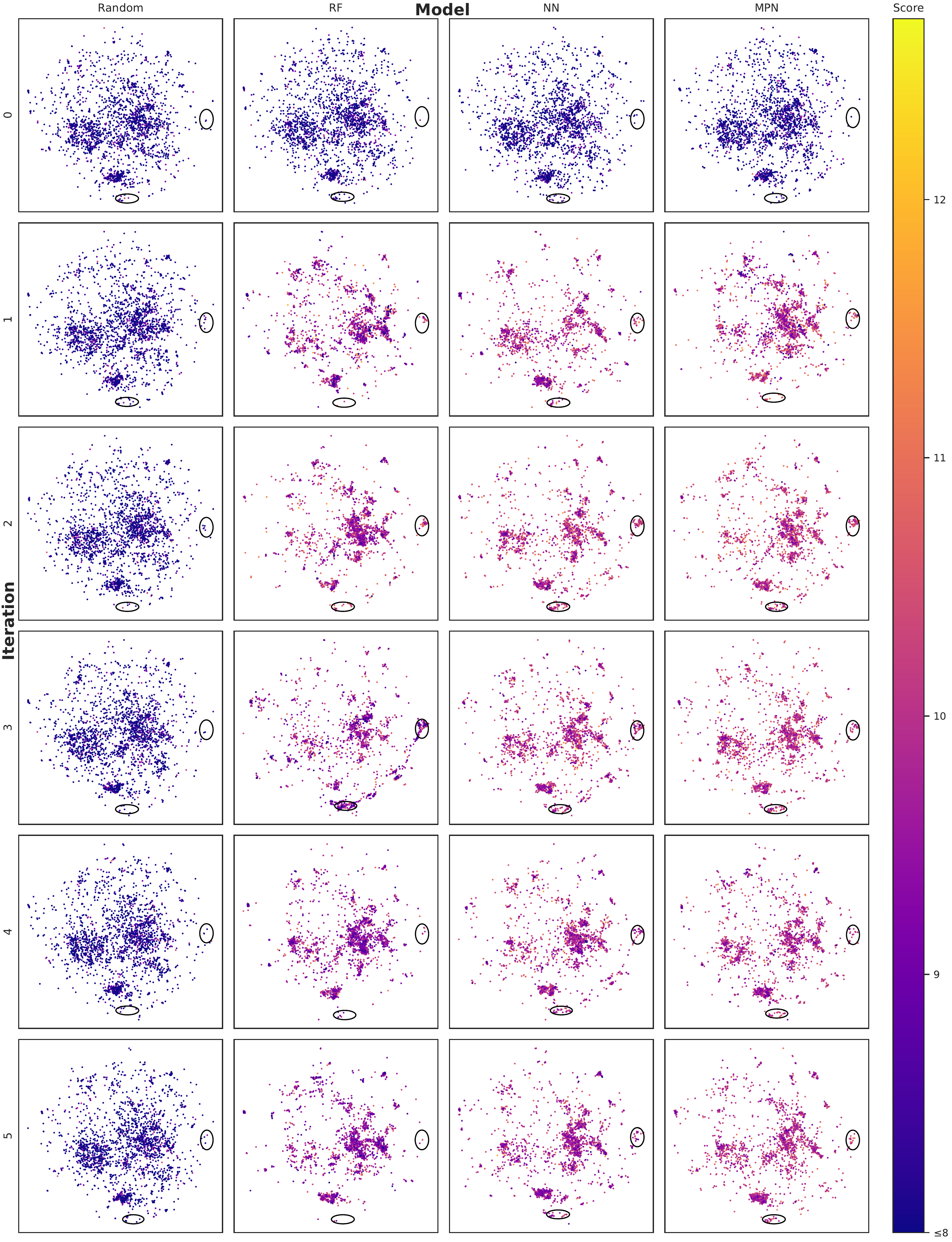}
    \caption{Visualization of the chemical space searched in the Enamine HTS library at the given iteration using a greedy acquisition metric, 0.1\% batch size, and specified surrogate model architecture. Points represent the 2-D UMAP embedding of the given molecule's 2048-bit Atom-pair fingerprint. The embedding was trained on a random 10\% subset of the full library. Circled regions indicate clusters of high-scoring compounds in sparse regions of chemical space. X- and y-axes are the first and second components of the 2-D UMAP embedding and range from -7.5 to 17.5. Color scale corresponds to the negative docking score (higher is better).}
    \label{fig:umap_si}
\end{figure}

\end{document}